\documentclass[conference]{IEEEtran}

\makeatletter
\newcommand\semihuge{\@setfontsize\semihuge{22.3}{22.6}}
\makeatother

\usepackage{algpseudocode}
\usepackage{algorithm}
\usepackage{algorithmicx}
\usepackage{multirow}
\usepackage{lipsum} 

\usepackage[dvips]{color}
\usepackage{comment}
\usepackage{todonotes}
\usepackage{epsf}
\usepackage{epsfig}
\usepackage{times}
\usepackage{epsfig}
\usepackage{graphicx}
\usepackage{bbold}
\usepackage{mathtools}
\usepackage{mathrsfs}
\usepackage{amssymb}
\usepackage{pdfpages}
\usepackage{epstopdf}
\usepackage{float}
\newfloat{algorithm}{t}{lop}

\usepackage{dsfont}
\usepackage{lettrine} 
\usepackage{amsmath,epsfig,amssymb,algorithm,algpseudocode,amsthm,cite,url}
\usepackage{subcaption}
\allowdisplaybreaks
\usepackage{csquotes}

\usepackage{verbatim}
\usepackage[english]{babel}
\usepackage{amsmath,amssymb}

\usepackage[justification=centering]{caption}
\usepackage{verbatim}

\IEEEoverridecommandlockouts	
\begin{document}
\title{\Huge A Tutorial on UAVs for Wireless Networks: Applications, Challenges, and Open Problems \vspace{-0.1cm}}

\author{\IEEEauthorblockN{Mohammad Mozaffari$^1$, Walid Saad$^1$, Mehdi Bennis$^2$, Young-Han Nam$^3$, and M\'erouane Debbah$^4$}\vspace{-0.15cm}\\
	\IEEEauthorblockA{
		\small $^1$ Wireless@VT, Electrical and Computer Engineering Department, Virginia Tech, VA, USA, Emails:\url{{mmozaff , walids}@vt.edu}.\\
		$^2$ CWC - Centre for Wireless Communications, University of Oulu, Finland, Email: \url{bennis@ee.oulu.fi}.\\
		$^3$ Standards \& 5G Mobility Innovations Lab, Samsung Research America, Richardson, TX, Email: \url{younghan.n@samsung.com}.\\
		$^4$ Mathematical and Algorithmic Sciences Lab, Huawei France R \& D, Paris, France, and CentraleSup´elec,\\   Universit´e Paris-Saclay, Gif-sur-Yvette, France, Email: \url{merouane.debbah@huawei.com}.
	}\vspace{-0.22cm}}
\maketitle\vspace{-0.1cm}
\vspace{0.1cm}
\begin{abstract}\vspace{-0.00cm}
	{\color{black}	
		The use of flying platforms such as unmanned aerial vehicles (UAVs), popularly known as drones, is rapidly growing. In particular, with their inherent attributes such as mobility, flexibility, and adaptive altitude, UAVs admit several key potential applications in wireless systems. On the one hand, UAVs can be used as aerial base stations to enhance coverage, capacity, reliability, and energy efficiency of wireless networks. On the other hand, UAVs can operate as flying mobile terminals within a cellular network. Such cellular-connected UAVs can enable several applications ranging from real-time video streaming to item delivery. In this paper, a comprehensive tutorial on the potential benefits and applications of UAVs in wireless communications is presented. Moreover, the important challenges and the fundamental tradeoffs in UAV-enabled wireless networks are thoroughly investigated. In particular, the key UAV challenges such as three-dimensional deployment, performance analysis, channel modeling, and energy efficiency are explored along with representative results. Then, open problems and potential research directions pertaining to UAV communications are introduced. Finally, various analytical frameworks and mathematical tools such as optimization theory, machine learning, stochastic geometry, transport theory, and game theory are described. The use of such tools for addressing unique UAV problems is also presented. In a nutshell, this tutorial provides key guidelines on how to analyze, optimize, and design UAV-based wireless communication systems.}
\end{abstract}

\section{{\color{black}Introduction and Overview on UAVs}}
Unmanned aerial vehicles (UAVs), commonly known as drones, have been the subject of concerted research over the past few years~\cite{HandbookUAV,austin2011unmanned, hanscomunmanned, Micro, stansbury2008survey}, owing to their autonomy, flexibility, and broad range of application domains. Indeed, UAVs have been considered as enablers of various applications that include military, surveillance and monitoring, telecommunications, delivery of medical supplies, and rescue operations \cite{HandbookUAV, hanscomunmanned}, and \cite{puri2005survey,IoTJournal, Irem,Bucaille, mozaffari2, HouraniOptimal,Mozaffari, Letter, zhang,bor, Rohde,yanmaz2018drone}. However, such conventional UAV-centric research has typically focused on issues of navigation, control, and autonomy, as the motivating applications were typically robotics or military oriented. In contrast, the communication challenges of UAVs have typically been either neglected or considered as part of the control and autonomy components.

\subsection{Motivation}
The unprecedented recent advances in drone technology \textcolor{black}{make} it possible to widely deploy UAVs, such as drones, small aircrafts, balloons, and airships for wireless communication purposes \cite{Bucaille,Sky,GoogleLoon,wu2018uav,wu2018common}. In particular, if properly deployed and operated, UAVs can provide reliable and cost-effective
wireless communication solutions for a variety of real-world scenarios. On the one hand, drones can be used as aerial base stations (BSs) that can deliver
reliable, cost-effective, and on-demand wireless communications to desired areas. On the other hand, drones can function as aerial user
equipments (UEs), known as \emph{cellular-connected UAVs}, in coexistence with ground users (e.g., delivery or surveillance drones).  This exciting new avenue for the use of UAVs warrants a rethinking of the research challenges with \emph{wireless communications and networking} being the primary focus, as opposed to control and navigation. 

In particular, when UAVs are used as flying, aerial base stations, they can support the connectivity of existing terrestrial wireless networks such as cellular and broadband networks.  { \color{black} Compared to conventional, terrestrial base stations,  the advantage of using UAVs as flying base stations is their ability to adjust their altitude, avoid obstacles, and  enhance the likelihood of  establishing line-of-sight (LoS) communication links to ground users }(see Tables \ref{Compare1} and \ref{Compare2} for a detailed comparison between UAVs and ground BSs). Indeed, owing to their inherent attributes such as mobility, flexibility, and adaptive altitude, UAV base stations can effectively complement existing cellular systems by providing additional capacity to hotspot areas and by delivering network coverage in hard to reach rural areas. Another important application of UAVs is in Internet of Things (IoT) scenarios \cite{al2015internet,PA00,IoT2014,ferdowsi2017deep, Ding1} whose devices often have small transmit power and may not be able to communicate over a long range. UAVs can also serve as wireless relays for improving connectivity and coverage of ground wireless devices and can also be used for surveillance scenarios, a key use case for the IoT. Last, but not least, in regions or countries where building a complete cellular infrastructure is expensive, deploying UAVs becomes highly beneficial as it removes the need for expensive towers and infrastructure deployment. 

From an industry perspective, key real-world example of recent projects that employ drones for wireless connectivity \textcolor{black}{includes} Google's Loon project. Within the scope of these practical deployments, UAVs are being used to deliver Internet access to developing countries and provide airborne global Internet connectivity. Moreover, Qualcomm and AT\&T are planning to deploy UAVs for enabling wide-scale wireless communications in the upcoming fifth generation (5G) wireless networks \cite{QualcomUAV}. Meanwhile, Amazon Prime Air and Google's Project Wing \cite{stewart2014google} initiatives are prominent examples of use cases for cellular-connected UAVs.

Despite such promising opportunities for drones, one must address a number of technical challenges in order to effectively use them
for each specific networking application. For instance, while using drone-BS, the key design considerations include performance
characterization, optimal 3D deployment of drones, wireless and computational resource allocation, flight time and trajectory optimization, and network planning. Meanwhile, in the drone-UE scenario, handover management, channel
modeling, low-latency control, 3D localization, and interference management are among the main challenges.
\subsection{UAV Classification}
Naturally, depending on the application and goals, one needs to use an appropriate type of UAV that can meet various requirements imposed by the \textcolor{black}{desired} quality-of-service (QoS), the nature of the environment, and federal regulations. In fact, to properly use UAVs for any specific wireless networking application, several factors such as the UAVs' capabilities and their flying altitudes must be taken into account. In general, UAVs can be  categorized, based on their altitudes, into high altitude platforms (HAPs) and low altitude platform (LAPs). HAPs have altitudes above 17\,km and are typically quasi-stationary \cite{HouraniModeling,zhang}. LAPs, on the other hand, can fly at altitudes of tens of meters up to a few kilometers, can quickly move, and \textcolor{black}{they} are flexible \cite{HouraniModeling}.

 We note that, according to US Federal aviation regulations, the maximum allowable altitude of LAP-drones that can freely fly without any permit is 400 feet\footnote{Hence, flying drones above 400 feet requires specific permissions from the Federal aviation administration (FAA).}\cite{FAA}. Compared to HAPs, the deployment of LAPs can be done more rapidly thus making them more appropriate for time-sensitive applications (e.g., emergency situations). Unlike HAPs, LAPs can be used for data collection from ground sensors. Moreover, LAPs can be readily recharged or replaced if needed.  In contrast, HAPs have longer endurance and they are designed for long term (e.g., up to few months) operations. Furthermore, HAP systems are typically preferred for providing and wide-scale wireless coverage for
large geographic areas \cite{zhang}. However, HAPs are costly and their deployment time is significantly longer than LAPs. 

\begin{figure}[t]
	\hspace{-0.1cm}	\begin{center}
		\includegraphics[width=9.01cm]{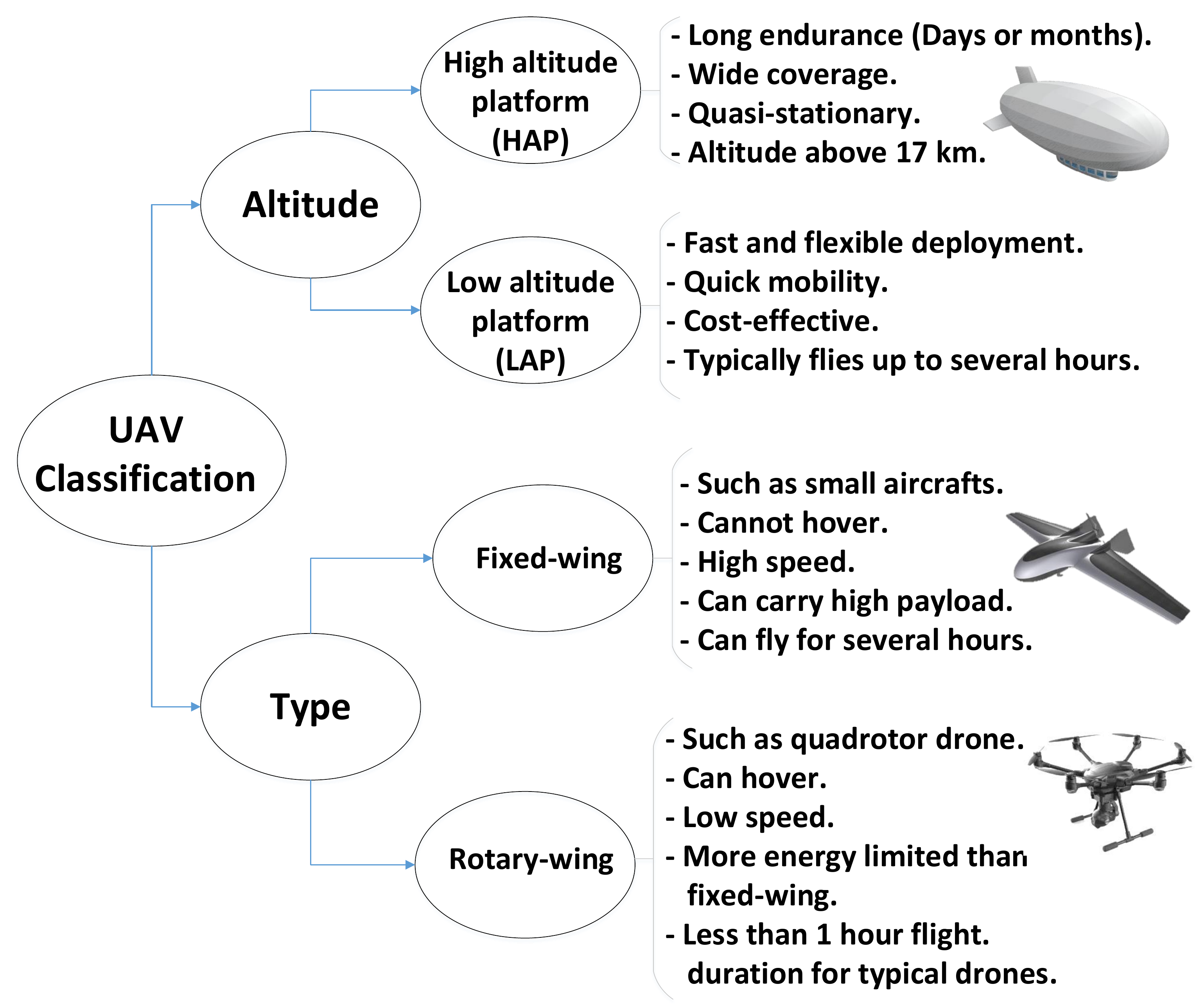}
		\vspace{-0.45cm}
		\caption{{\label{UAVClassification} UAV Classification.} }\vspace{-0.2cm}
	\end{center}
\end{figure}

 UAVs can also be categorized, based on type, into fixed-wing and rotary-wing UAVs. Compared to rotary-wing UAVs,  fixed-wing UAVs such as small aircrafts have more weights, higher speed, and they need to move forward in order to remain aloft. In contrast, rotary-wing UAVs such as quadrotor drones, can hover and remain stationary over a given area \cite{zhang}. In Figure \ref{UAVClassification}, we provide an overview on the different types of UAVs, their functions, and capabilities. {\color {black} We note that the flight time of a UAV depends on several factors such as energy source (e.g., battery, fuel, etc.,), type, weight, speed, and trajectory of the UAV.}

{\color{black}
\subsection{UAV Regulations}
Regulatory issues are important limiting factors facing the deployment of UAV-based communication systems. Despite the promising applications of UAVs in wireless networks, there are several concerns regarding privacy, public safety, security, collision avoidance, and data protection. In this regard, UAV regulations are being continuously developed to control the  operations of UAVs while considering various \textcolor{black}{factors} such as UAV type, spectrum, altitude, and speed of UAVs.  In general, five main criteria are often considered when developing UAV regulations \cite{fotouhi2018survey,stocker2017review}:  1) \emph{Applicability}: pertains to determining the scope (considering type, weight, and role of UAVs) where UAV regulations are applied, 2) \emph{Operational limitations}: related to restrictions on the locations of UAVs, 3) \emph{Administrative procedures}: specific legal procedures  could be needed to operate a UAV, 4) \emph{Technical requirements}: includes communications, control, and mechanical capabilities of drones, 5) \emph{Implementation of ethical constraints}: related to privacy protection.

UAV regulations vary between different countries and types of geographical areas (e.g., urban or rural). In the United States, regulations for UAV operations are issued by the federal aviation authority (FAA) and  national aeronautics and space administration (NASA). NASA is planning to develop UAV control frameworks in collaboration with federal communications commission (FCC) and FAA. FCC is currently investigating if new spectrum policy needs to be established for drone operations.

 In Table \ref{TableReg}, we list a number of UAV regulations for deployment of UAVs in various countries \cite{fotouhi2018survey}.
}

\begin{table} [!t] {\color{black}
	\normalsize
	\begin{center}
		\caption{\textcolor{black}{\small Regulations for the deployment of UAVs without any specific permit.}}
		\vspace{-0.1cm}
		\label{TableReg}
		\resizebox{9.2cm}{!}{
			\begin{tabular}{|c|c|c|c|}
				\hline
		\textbf{Country} & \textbf{Maximum altitude} & \textbf{Minimum distance to people}  & \textbf{Minimum distance to airport}\\ \hline \hline
				
				\large	US	&    \large 122\,m    &      \large N/A &  \large 8\,km \\ \hline
				
				\large Australia &\large120\,m	&    \large 30\,m     &      \large 5.5\,km     \\ \hline 
			\large	South
				Africa	&  \large  46\,m  &   \large50\,m& \large 10\,km \\ \hline
				\large UK	&  \large  122\,m     &    \large 50\,m &\large N/A    \\ \hline

			\large	Chile	&   \large 130\,m      &   \large 36\,m &\large N/A \\ \hline
				
		\end{tabular}}
		
	\end{center}}\vspace{-0.5cm}
\end{table}

\subsection{\textcolor{black}{Relevant Surveys on UAVs and Our Contributions} }
These exciting new opportunities for using  various types of UAVs for wireless networking purposes have spawned numerous recent research \textcolor{black}{activities} in the area~\cite{IoTJournal, Irem,Bucaille, HouraniOptimal,Mozaffari, Letter, zhang,AkramMagazin, mozaffari2,ALZ1,ALZ2,Qin, Azari, bor, VshalUAV, Lyu, Jeong, MozaffariFlightTime, Complition, Proactive}. {\color {black} These works also include a number of interesting surveys such as in \cite{zhang,bor,AkramMagazin,Flying1,Flying3,bekmezci2013flying,sahingoz2014networking,Low,Airborne,karapantazis2005broadband,sekander2018multi,fotouhi2018survey,hayat2016survey,gupta2016survey, LTE_Sky, Ismail_survey}.

The work in \cite{Flying1} introduced decentralized communication architectures for a multi-layer UAV ad hoc network. Furthermore, various routing protocols in flying ad-hoc networks are proposed along with open research problems.
In \cite{Flying3}, the authors provided an overview of flying ad-hoc networks while considering technological and social implications.  In particular, the work in \cite{Flying3} discussed the applications of flying ad-hoc networks, design considerations, communication protocols, and  privacy aspects.
In \cite{bekmezci2013flying}, a comprehensive review of  UAV-based flying ad hoc networks (FANETs) and their challenges \textcolor{black}{are} provided. Moreover, several FANET design challenges in terms of mobility, node density, topology change, radio propagation model, and power consumption are investigated. 
The survey in \cite{sahingoz2014networking} discussed the design challenges pertaining to the use of UAVs as relay
nodes in flying ad-hoc networks. 
The work in \cite{Low} provided a comprehensive survey on the potential use of UAVs for supporting IoT services. In particular, key challenges and requirements for designing UAV-assisted IoT networks are discussed in \cite{Low}. 
In \cite{Airborne}, the authors surveyed different mechanisms and protocols for developing airborne communication networks while considering low-altitude-platform communications, high-altitude-platform
communications, and integrated airborne communication systems.  The survey in \cite{karapantazis2005broadband} studied the use of HAPs for broadband communications. Moreover, it \textcolor{black} {described} key advantages of HAPs compared to terrestrial and satellite networks, suitable HAP airships, frequency bands, and possible HAP-based network architectures.
The authors in \cite{sekander2018multi} studied the challenges and advantages associated with a multi-tier drone network architecture. Moreover, this work investigated the performance of a multi-tier drone wireless system in terms of spectral efficiency.  In \cite{fotouhi2018survey} a survey on UAV-enabled cellular communications is provided with focus on  relevant 3rd generation partnership project (3GPP ) developments, standardization bodies for UAV users, vendor prototypes of UAV BSs, regulations, and cyber-security aspects of deploying UAVs in cellular networks.  The survey in \cite{hayat2016survey} presented the communications and networking requirements of UAVs for civil applications.  In \cite{gupta2016survey}, the authors conducted a survey on the key challenges in UAV-based wireless communication networks. In particular, the work in \cite{gupta2016survey} investigated issues pertaining to routing strategies in flying UAV networks, energy efficiency of UAVs, and seamless handover in UAV-enabled wireless networks.

While these surveys address important UAV communication problems, as listed in Table \ref{SurveysTable}, they mainly limit their discussions to cases in which UAVs are used as \emph{relay stations in ad-hoc networks \cite{ Flying1,Flying3,bekmezci2013flying,sahingoz2014networking, zhang}, rather than fully fledged flying base stations or drone-UEs} that can support complex ground networks, such as 5G cellular networks. 
Moreover, the surveys in \cite{zhang,bor,AkramMagazin,Low,Airborne,karapantazis2005broadband,sekander2018multi,fotouhi2018survey,hayat2016survey,gupta2016survey, LTE_Sky} remain restricted to isolated UAV topics and use cases in wireless networking. In addition, these surveys do not introduce potential analytical frameworks that are essentially needed for designing and analyzing UAV-based communication systems. More recently, some surveys such as  \cite{Ismail_survey} looked at channel models for UAVs, while overlooking broader networking problems. Clearly, the existing literature on wireless networking using UAVs is largely fragmented and, given the rapid emergence of the topic, in academia, industry, and government, there is a clear need for a unified and comprehensive overview on how UAVs can  be used as flying wireless base stations in emerging wireless, broadband, and beyond 5G scenarios.

\begin{table}[!t] {\color{black}
		\normalsize
	\begin{center}
		\caption{\textcolor{black}{\small Relevant surveys and magazines on UAV communications.}}
		\vspace{-0.1cm}
		\label{SurveysTable}
		\resizebox{9cm}{!}{
			\begin{tabular}{|c|p{9cm}|}
				\hline
				\textbf{References} &  	\hspace{5cm}\textbf{Focus}  \\ \hline \hline
				
				\cite{Flying1,Flying3,bekmezci2013flying, sahingoz2014networking}	&   Flying ad-hoc networks.  \\ \hline
				
				\cite{Low} & UAV in IoT	networks.       \\ \hline 
				
				\cite{Airborne}	&   Mechanisms for designing airborne communication networks.  \\ \hline
				
				\cite{karapantazis2005broadband}	&    Broadband communications with HAPs.      \\ \hline

				\cite{sekander2018multi} 	&    Network architecture for multi-tier drone.  \\ \hline

				\cite{fotouhi2018survey} 	&    3GPP developments, regulations, and cyber-security aspects of UAVs.  \\ \hline
				
				\cite{hayat2016survey}	&    Networking requirements of UAVs for civil applications.  \\ \hline

				\cite{zhang} 	& UAV relays for wireless communications. 
				\\ \hline			
				
				\cite{gupta2016survey}	&    Routing strategies, energy efficiency, and handover in UAV networks.  \\ \hline

				\cite{Ismail_survey} 	& Channel modeling for UAVs.   \\ \hline
				
				\cite{LTE_Sky}	& Interference and path loss study for UAVs.   \\ \hline

				\cite{bor,AkramMagazin} 	& UAV use cases. 
				\\ \hline
				
				This tutorial	& Opportunities, challenges, open problems, and mathematical tools for UAV base stations and cellular-connected drone-UEs.
				\\ \hline
		\end{tabular}}
		
	\end{center}}\vspace{-0.4cm}
\end{table}

The main contribution of this article is to provide the first holistic and comprehensive overview and tutorial on the use of UAVs for wireless communications and networking applications. To this end, the goal is to gather the state-of-the-art research contributions, from the largely fragmented and sparse literature on UAV-based wireless communications. Moreover,  this work presents  the major opportunities and challenges in deploying UAVs as flying wireless base stations that complement emerging wireless communication systems, or as cellular-connected UAV-UEs that use existing wireless infrastructure, with emphasis on application scenarios, challenges, representative results, open problems, and analytical techniques that will enable the real-world deployment of UAVs as aerial communication platforms. With the incessant growth in research revolving around the use of UAVs for wireless purposes,
this article constitutes one of the first comprehensives guides on how to fully exploit the potential of UAVs for wireless communications and networking.} To achieve this goal, we treat the following key topics:

\begin{itemize}
\item In Section II, we provide a comprehensive overview on potential applications of UAVs in a plethora of wireless networking scenarios. These applications will provide motivating examples and future use cases of UAVs, particularly in their role as flying base stations.
\item In Section III, we outline key research directions that will enable the applications identified in Section II. For each research direction, we provide an overview on the research challenges, the state of the art, and promising early results within these areas.
\item In Section IV, for each research direction identified in Section III, we provide an outline of challenging open problems that must be addressed, in order to fully exploit the potential of UAV-based wireless communications. This, in turn, will provide a roadmap for future research in this area.
\item In Section V, we then provide a summary on analytical frameworks that are expected to play an important role in the design of future UAV-based wireless networks \textcolor{black}{that can} enable network operators to leverage UAVs for various application scenarios.
\item The article is concluded in Section VI with additional insights on this fascinating area of research.
\end{itemize}

\section{{\color{black}Wireless Networking with UAVs: Motivating Application Use Cases}}\label{Opportunities}

In order to paint a clear picture on how UAVs can indeed be used as flying wireless base stations, in this section, we overview a number of prospective applications for such a wireless-centric UAV deployment. The applications are drawn from a variety of scenarios, that include imminent use cases, such as for public safety scenarios or hotspot coverage, as well as more ``futuristic'' applications such as the use of UAVs as caching apparatus or IoT enablers. Naturally, in all such applications, the UEs of the system can include cellular-connected UAV-UEs which we will also discuss. Note that this section restricts its attention to the application scenarios, while the challenges are left for a deeper treatment in Section~\ref{sec:Existing}.
{\color{black}
\subsection{UAV Aerial Base Station in 5G and Beyond}
Here, we discuss the key applications of UAV-mounted aerial base stations in 5G.}
\subsubsection{Coverage and Capacity Enhancement of Beyond 5G Wireless Cellular Networks}
The need for high-speed wireless access has been incessantly growing, fueled by the rapid proliferation of highly capable mobile devices such as smartphones, tablets, and more recently drone-UEs and IoT-style gadgets \cite{IoT2014}. As such, the capacity and coverage of existing wireless cellular networks \textcolor{black} {have} been extensively strained, which led to the emergence of a plethora of wireless technologies that seek to overcome this challenge. Such technologies, which include device-to-device~(D2D) communications, ultra dense small cell networks, and millimeter wave~(mmW) communications, are collectively viewed as the nexus of next-generation 5G cellular systems \cite{Samarakoon, Omid1, Omid2,Contract, ContextOmid}. However, despite their invaluable benefits, those solutions have limitations of their own. For instance, D2D communication will undoubtedly require better frequency planning and resource usage in cellular networks. Meanwhile, ultra dense small cell networks face many challenges in terms of backhaul, interference, and overall network modeling. Similarly, mmW communication is limited by blockage and high reliance on LoS communication to effectively deliver the promise of high-speed, low latency communications. These challenges will be further exacerbated in UAV-UEs scenarios.

We envision UAV-carried flying base stations as an inevitable complement for such a heterogeneous 5G environment, which \textcolor{black} {will allow overcoming} some of the challenges of the existing technologies. {\color{black}Deploying LAP-UAVs can be a cost-effective approach for providing wireless connectivity to  geographical areas with limited cellular infrastructure. Moreover, the use of UAV base stations becomes promising  when deploying
	small cells for the sole purpose of servicing  temporary
	events (e.g., sport events and festivals),
	is not economically viable, given
	the short period of time during which these events require
	wireless access.} Meanwhile, HAP-UAVs can provide a more long-term sustainable solution for coverage in such rural environments. { \color{black} Mobile UAVs can provide on-demand connectivity, high data rate wireless service, and traffic offloading opportunity \cite{Absolute, bor, OffloadingLyu} in hotspots and during temporary events such as football games or Presidential inaugurations. } In this regard, AT\&T and Verizon have already announced several plans to use flying drones to provide temporarily boosted Internet coverage for college football national championship and Super Bowl \cite{ATDrone}. Clearly, flying base stations can provide an important complement to ultra dense small cell networks.

\textcolor{black} {In addition}, {\color{black}  UAV-enabled mmW communications is a porpoising application of UAVs that can establish LoS communication links to users.} This, in turn, can be an attractive solution to provide high capacity wireless transmission, while leveraging the advantages of both UAVs and mmW links. Moreover, combining UAVs with mmW and potentially massive multiple input multiple output (MIMO) techniques can create a whole new sort of dynamic, flying cellular network \textcolor{black} {for providing} high capacity wireless services, if well planned and operated.

UAVs can also assist various terrestrial networks such as D2D and vehicular networks. For instance, owing to their mobility and LoS communications, drones can facilitate rapid information dissemination among ground devices. Furthermore, drones can potentially improve the reliability of wireless links in D2D and vehicle-to-vehicle (V2V) communications while exploiting transmit diversity. In particular, flying drones can help in broadcasting common information to ground devices thus reducing the interference in ground networks by decreasing the number of transmissions between devices. Moreover, UAV base stations can use air-to-air links to service other cellular-connected UAV-UEs, to alleviate the load on the terrestrial network.


For the aforementioned cellular networking scenarios, it is clear that the use of UAVs is quite natural due to their key features given in Tables \ref{Compare1} and \ref{Compare2} such as agility, mobility, flexibility, and adaptive altitude. In fact, by exploiting these unique features as well as establishing LoS communication links, UAVs can boost the performance of existing ground wireless networks in terms of coverage, capacity, delay, and overall quality-of-service.  Such scenarios are clearly  promising and one can see UAVs as being an integral part of beyond 5G cellular networks, as the technology matures further, and new operational scenarios emerge. Naturally, reaping these benefits will require overcoming numerous challenges, that we outline in Section~\ref{sec:Existing}.

 \begin{small}  
 	\begin{table}[!t]  { \color{black}\caption{UAV base station versus terrestrial base station.\label{Compare1}}\vspace{-0.3cm}
 		\begin{center} 
 			\begin{tabular}{ | p{4cm} |p{4cm} |} 
 				\hline
 				\textbf{UAV Base Stations}  & \textbf{Terrestrial Base Stations}
 				\\ \hline 
 				
 				$\bullet$\,\,\,Deployment is naturally three-dimensional.
 				& 	$\bullet$ Deployment is typically two-dimensional.
 				\\ \hline
 				$\bullet$	Short-term, frequently changing deployments.
 				& 	$\bullet$ Mostly long-term, permanent deployments.
 				\\ \hline
 				$\bullet$ Mostly unrestricted locations.
 				& 	$\bullet$ Few, selected locations.
 				\\
 				\hline
 				$\bullet$	Mobility dimension.	& 	$\bullet$ Fixed and static.
 				\\
 				\hline
 				
 			\end{tabular}
 		\end{center}}
 	\end{table}
 \end{small}

 \begin{small}   
 	\begin{table} [!t] { \color{black} \caption{UAV networks versus terrestrial networks. \label{Compare2}} \vspace{-0.3cm}
 		\begin{center} 
 			\begin{tabular}{ | p{4cm} |p{4cm} |} 
 				\hline
 				\textbf{UAV Networks
 				}  & \textbf{Terrestrial Networks
 				}
 				\\ \hline 
 				
 				$\bullet$ Spectrum is scarce.
 				& 	$\bullet$ Spectrum is scarce.
 				\\ \hline
 				$\bullet$ Elaborate and stringent energy constraints and models. 
 				& 	$\bullet$ Well-defined energy constraints and models.
 				\\
 				\hline
 				$\bullet$	Varying cell association.
 				& 	$\bullet$ Mainly static association.
 				\\
 				\hline
 				$\bullet$	Hover and flight time constraints.
 				
 				& 	$\bullet$ No timing constraints, BS always there.
 				\\
 				\hline
 			\end{tabular}
 		\end{center}}
 	\end{table}
 \end{small}


\subsubsection{UAVs as Flying Base Stations for Public Safety Scenarios}
Natural disasters such as floods, hurricanes, tornados, and severe snow storms often yield devastating consequences in many countries. During wide-scale natural disasters and unexpected events, the existing terrestrial communication networks can be damaged or even completely destroyed, thus becoming significantly overloaded, as evidenced by the recent aftermath of Hurricanes Sandy and Irma \cite{Gomez}. In particular, cellular base stations and ground communications infrastructure can be often compromised during natural disasters. In such scenarios, there is a vital need for public safety communications between first responders and victims for search and rescue operations. Consequently, a robust, fast, and capable emergency communication system is needed to enable effective communications during public safety operations. In public safety scenarios, such a reliable communication system will not only contribute to improving connectivity, but also to saving lives.

\begin{figure}[t]
	\begin{center}
		\vspace{-0.2cm}
		\includegraphics[width=8cm]{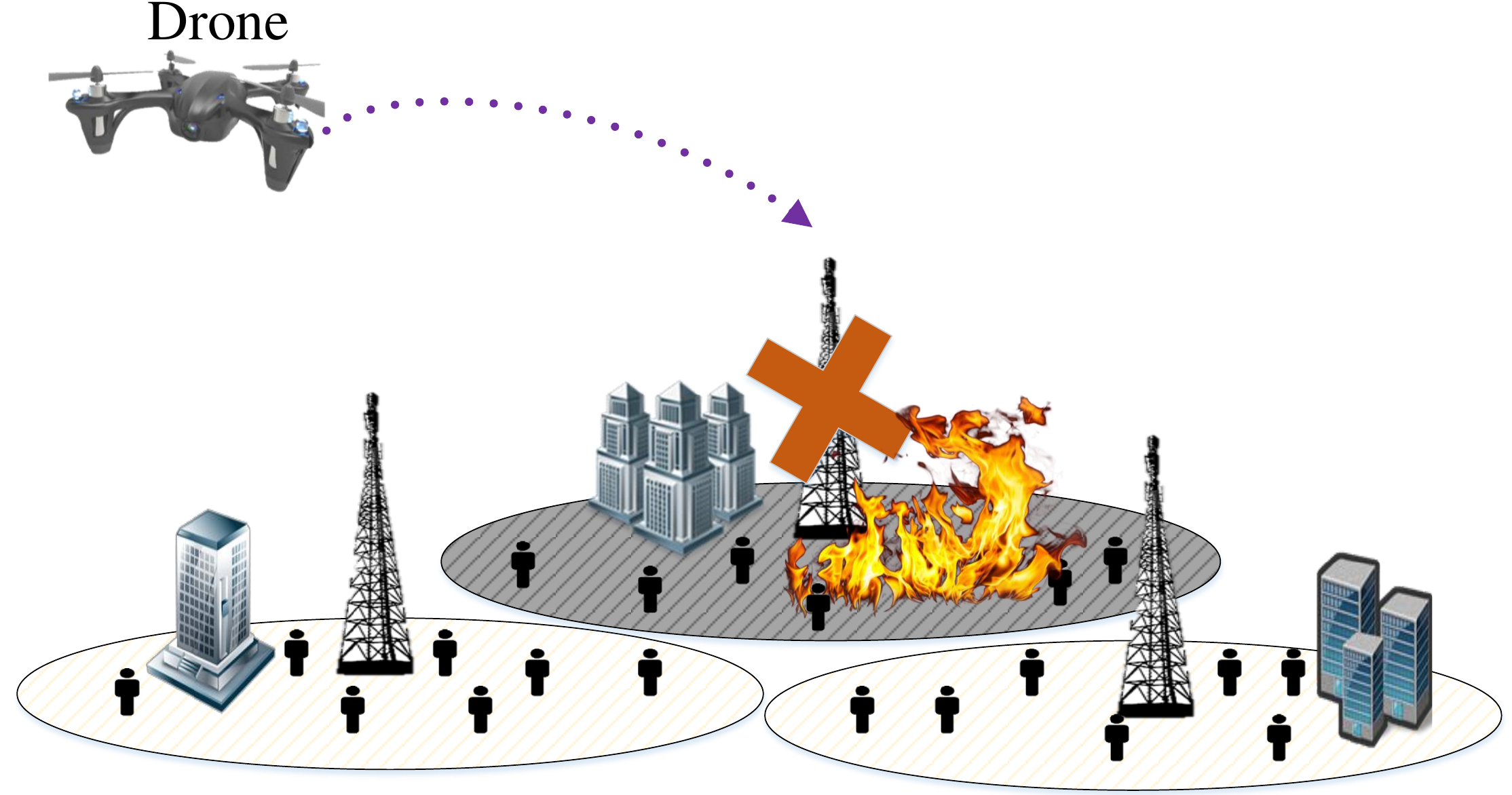}
		\vspace{-0.01cm}
		\caption{{Drone in public safety scenarios. }\label{Safety} }
	\end{center}\vspace{-0.4cm}
\end{figure}

In this regard, FirstNet in the United States was established to create a nationwide and high-speed broadband wireless network for public safety communications. The potential broadband wireless technologies for public safety scenarios include 4G long term evolution (LTE), WiFi, satellite communications,  and dedicated public safety systems such as TETRA and APCO25 \cite{PublicSafety}. However, these technologies may not provide flexibility,  low-latency services, and swift adaptation to the environment during natural disasters. In this regard, the use of UAV-based aerial networks \cite{Ismail}, as shown in Figure \ref{Safety}, is a promising solution to enable fast, flexible, and reliable wireless communications in public safety scenarios. Since UAVs do not require highly constrained and expensive infrastructure (e.g., cables), they can easily fly and dynamically change their positions to provide on-demand communications to ground users in emergency situations. In fact, due the unique features of UAVs such as mobility, flexible deployment, and rapid reconfiguration, they can effectively establish on-demand public safety communication networks. For instance, UAVs can be deployed as mobile aerial base stations in order to deliver broadband connectivity to areas with damaged terrestrial wireless infrastructure. Moreover, flying UAVs can continuously move to provide full coverage to a given area within a minimum possible time. Therefore, the use of UAV-mounted base stations can be an appropriate solution for providing fast and ubiquitous connectivity in  public safety scenarios.

\subsubsection{UAV-assisted Terrestrial Networks for Information Dissemination }
\begin{figure}[t]
	\begin{center}
		\vspace{-0.2cm}
		\includegraphics[width=8.5cm]{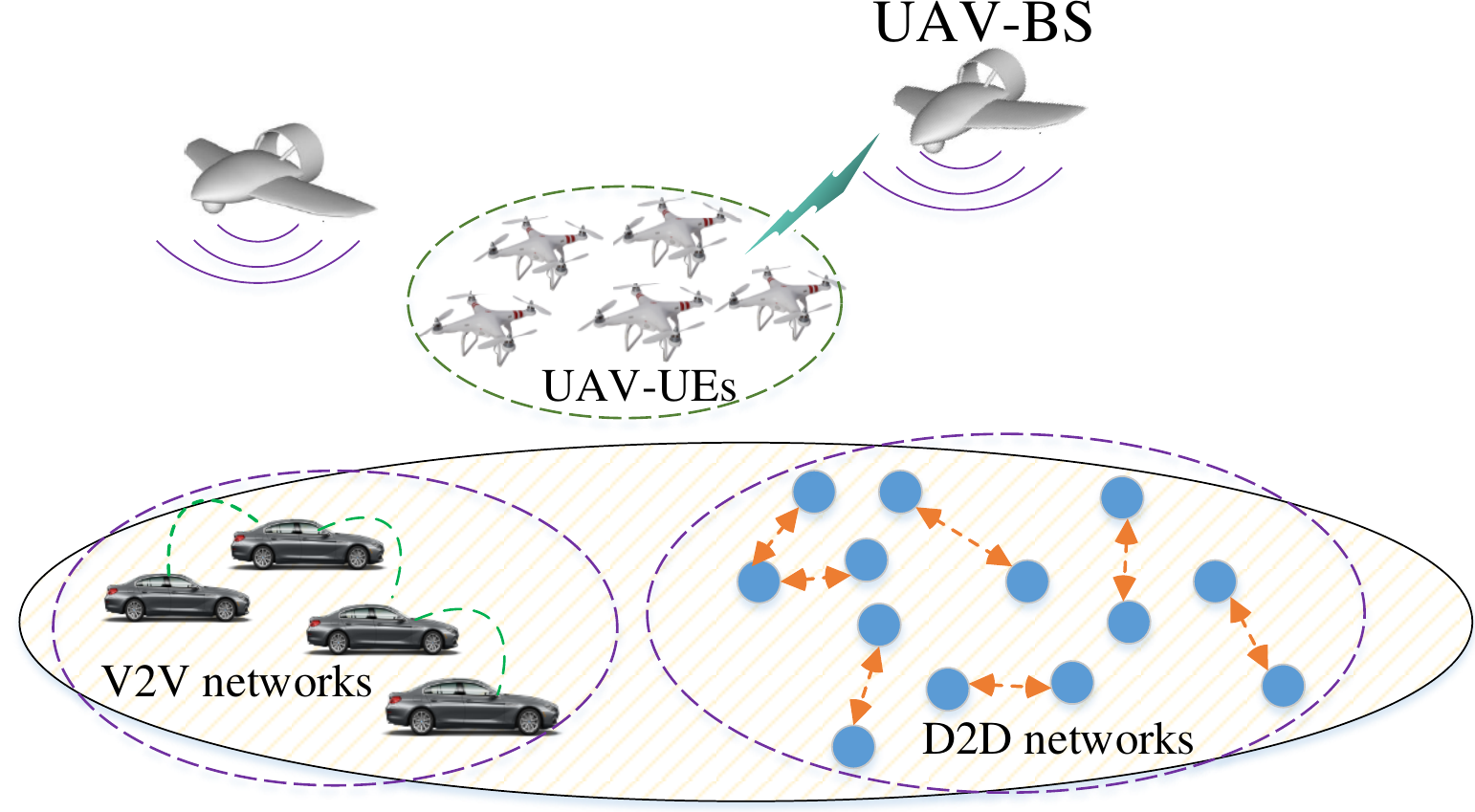}
		\vspace{-0.01cm}
		\caption{UAV-assisted terrestrial networks.}\label{UAVD2D} 
	\end{center}\vspace{-0.4cm}
\end{figure}

With their mobility and LoS opportunities, UAVs can support terrestrial networks for information dissemination and connectivity enhancement \cite{zhang, orsino2017effects}. For instance, as shown in Figure \ref{UAVD2D}, UAVs can be used as flying base stations to assist a D2D network or a mobile ad-hoc network in information dissemination among ground devices. While D2D networks can provide an effective solution for offloading cellular data traffic and improving network capacity and coverage, their performance is limited due to the short communication range of devices as well as potentially increasing interference. In this case, flying UAVs can facilitate rapid information dissemination by intelligently broadcasting common files among ground devices. For example, UAV-assisted  D2D networks allow the rapid spread of emergency or evacuation messages in public safety situations.

Likewise, drones can play a key role in vehicular networks (i.e., V2V communications) by spreading safety information across the vehicles. 
Drones can also enhance reliability and connectivity of D2D and V2V communication links. On the one hand, using drones can mitigate interference by reducing the number of required transmission links between ground devices. On the other hand, mobile drones can introduce transmit diversity opportunities thus boosting reliability and connectivity in D2D, ad-hoc, and V2V  networks.  One effective approach for employing such UAV-assisted terrestrial networks is to leverage clustering of ground users. Then, a UAV can directly communicate with the head of the clusters and the multi-hop communications are performed inside the clusters. In this case, the connectivity of terrestrial networks  can be significantly improved by adopting efficient clustering approaches and exploiting UAVs' mobility. \vspace{-0.05cm}

\subsubsection{3D MIMO and Millimeter Wave Communications}
\begin{figure}[t]
\begin{center}
\vspace{-0.2cm}
\includegraphics[width=7.5cm]{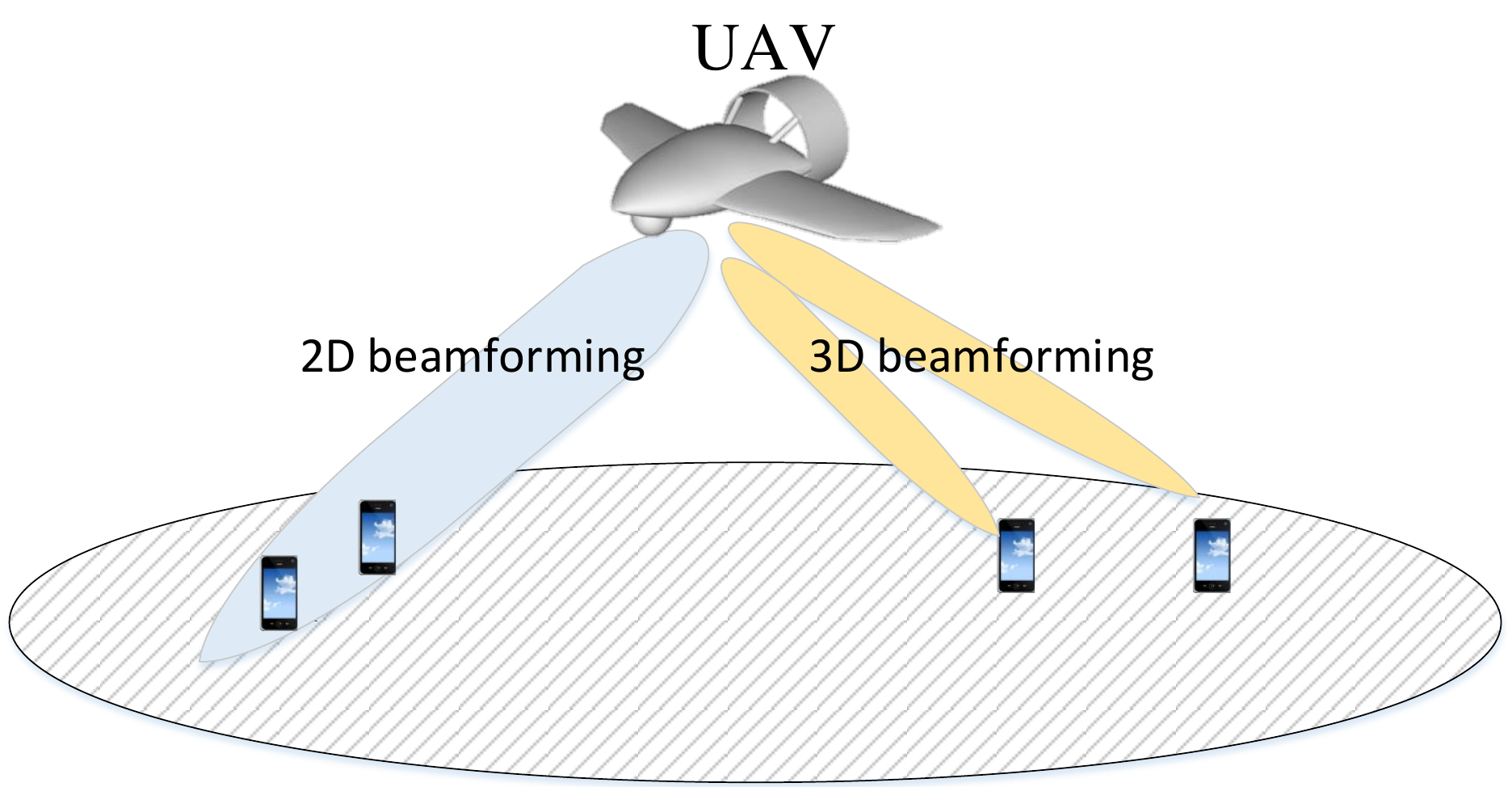}
\vspace{-0.01cm}
\caption{{3D beamforming using a drone.}\label{3DMIMO}}
\end{center}\vspace{-0.5cm}
\end{figure}

Due to their aerial positions and their ability to deploy on demand at specific locations, UAVs can be viewed as flying antenna systems that can be exploited for performing massive MIMO, 3D network MIMO, and mmW communications. For instance, in recent years, there has been considerable interest in the use of 3D MIMO, also known as full dimension MIMO, by exploiting both the vertical and horizontal dimensions in terrestrial cellular networks~\cite{Nam2013,3GPP36897, lee3D,namMIMO,sha,cheng,Li}. In particular, as shown in Figure \ref{3DMIMO}, 3D beamforming enables the creation of separate beams in the three-dimensional space at the same time, thus reducing inter-cell interference \cite{3GPP36777}.  
 Compared to the conventional two-dimensional MIMO,  3D MIMO solutions can yield higher overall system throughput and can support a higher number of users. In general, 3D MIMO is more suitable for scenarios in which the number of users is high and they are distributed in three dimensions with different elevation angles with respect to their serving base station \cite{zhang,Li}. Due to the high altitude of UAV-carried flying base stations, ground users can be easily distinguishable at different altitudes and elevation angles measured with respect to the UAV. Furthermore, LoS channel conditions in UAV-to-ground communications enable effective beamforming in both azimuth and elevation domains (i.e., in 3D). Therefore, UAV-BSs are suitable candidates for employing 3D MIMO. 

Furthermore, the use of a drone-based wireless antenna array, that we introduced in \cite{CommControl}, provides a unique opportunity for airborne beamforming.  A drone antenna array whose elements are single-antenna drones can  provide MIMO and beamforming opportunities to effectively service ground users in downlink and uplink scenarios. Compared to conventional
antenna array systems, a drone-based antenna array has the following advantages: 1) The number of antenna elements
(i.e., drones) is not limited by space constraints, 2) Beamforming gains can be increased
by dynamically adjusting the array element spacing,  and 3) The mobility and flexibility of drones allow effective mechanical beam-steering in any 3D direction. In addition, the use of a large number of small UAVs within an array formation can provide unique massive MIMO opportunists. Such UAV-based massive antenna array can form any arbitrary shape and effectively perform beamforming.

UAVs can also be a key enabler for mmW communications\footnote{It is worth noting that mmW communications have been  already adopted for satellite and HAPS communications \cite{3GPP38811}.} (e.g., see \cite{3GPP38811, bor,zhang,IsmailmmW}, and \cite{tork}). On the one hand, UAVs equipped with mmW capabilities can establish LoS connections to ground users \textcolor{black} {thus reducing} propagation loss while operating at high frequencies. On the other hand,
with the use of small-size
antennas (at mmW frequencies) on UAVs, one can exploit advanced MIMO techniques such as massive
MIMO in order to operate mmW communications. Meanwhile, swarms of UAVs can create reconfigurable antenna arrays in the sky \cite{CommControl}.  
 

\subsubsection{UAVs for IoT Communications} \label{IoTEnergy}
Wireless networking technologies are rapidly evolving into a massive IoT environment that must integrate a heterogeneous mix of devices ranging from conventional smartphones and tablets to vehicles, sensors, wearables, and naturally, drones. Realizing the much coveted applications of the IoT such as smart cities infrastructure management, healthcare, transportation, and energy management \cite{IoTVision,dawy,lien,IoT2014} requires effective wireless connectivity among a massive number of IoT devices that must reliably deliver their data, typically at high data rates or ultra low latency. The massive nature of the IoT requires a major rethinking to the way in which conventional wireless networks (e.g., cellular systems) operate.

 For instance, in an IoT environment, energy efficiency, ultra low latency, reliability, and high-speed uplink communications become major challenges that are not typically as critical in conventional cellular network use cases \cite{dawy}. In particular, IoT devices are highly battery limited and are typically unable to transmit over a long distance due to their energy constraints. For instance, in areas which experience
an intermittent or poor coverage by terrestrial wireless networks,
battery-limited IoT devices may not be able to transmit their
data to distant base stations due to their power constraints.
Furthermore, due to the various applications of IoT devices, they
might be deployed in environments with no terrestrial wireless
infrastructure such as mountains and desert areas.
 
 
In this regard, the use of mobile UAVs is a promising solution to a number of challenges associated with IoT networks.  
In  IoT-centric scenarios, UAVs can be deployed as flying base stations to provide reliable and energy-efficient uplink IoT communications (e.g., see \cite{mozaffari2,pang,IoTJournal}, and \cite{MehdiM2M}). In fact, due to the aerial nature of the UAVs
and their high altitude, they can be effectively deployed to reduce the shadowing and blockage effects as the major cause of signal attenuation in wireless links. As a result of such efficient placement of UAVs, the communication channel between IoT devices and UAVs can be significantly improved. Subsequently,  battery-limited IoT devices will need a significantly lower power to transmit their data to UAVs. In other words, UAVs can be placed based on the locations of IoT devices enabling those devices to successfully connect to the network using a minimum transmit power. Moreover, UAVs can also serve massive IoT systems by dynamically updating their locations based on the activation pattern of IoT devices.  This is in contrast to using ground small cell base stations which may need to be substantially expanded to service the anticipated number of devices in the IoT. Hence, by exploiting unique features of UAVs, the connectivity and energy efficiency of IoT networks can be significantly improved. 


\subsubsection{Cache-Enabled UAVs}
Caching at small base stations (SBSs) has emerged as a promising approach to improve users' throughput and to reduce the transmission delay \cite{ProactiveCaching2016,Tran2016Octopus,guo2015cooperative,bacstug2015cache,ye2016tradeoff}. However, 
 caching at static ground base stations may not be effective in serving mobile users in case of frequent handovers (e.g., as in ultra-dense networks with moving users). In this case, when a user moves to a new cell, its requested content may not be available at the new base station and, thus, the users cannot be served properly. To effectively service mobile users in such scenarios, each requested content needs to be cached at multiple base stations which is not efficient due to signaling overheads and additional storage usages. Hence, to enhance caching efficiency, there is a need to deploy flexible base stations that can track the users' mobility and effectively deliver the required contents. 

To this end, we envision futuristic scenarios in which UAVs, acting as flying base stations, can dynamically cache the popular contents, track the mobility pattern of the corresponding users and, then, effectively serve them \cite{Irem,mingzhe,Ding2}. In fact, the use of cache-enabled UAVs is a promising solution for traffic offloading in wireless networks. By leveraging user-centric
information, such as content request distribution and mobility patterns, cache-enabled UAVs can be optimally moved and deployed to deliver desired services to users.  {\color{black}  Another advantage of deploying cache-enabled
	UAVs is that the caching  complexity can be
	reduced compared to a conventional static SBSs case. For instance, whenever a
	mobile user moves to a new cell, its requested content needs to be stored at the new base station. However, cache-enabled drones can track the mobility pattern of  users and, consequently, the content stored at the drones will no longer require such additional caching at SBSs. \textcolor{black}{In practice, in a cache-enabled UAV system, a central cloud processor can utilize various user-centric information including users' mobility patterns and their content request distribution to manage the UAV deployment. In fact, such user-enteric information can be learned by a cloud center using any previous available users' data. Then, the cloud center can effectively determine the locations and mobility paths of cache-enabled UAVs to serve ground users \footnote{Caching with UAVs can also be an important use-case for future flying taxis \cite{Ramy}.}.} This, in turn,  can reduce the overall overhead of updating the cache content.} While performing caching with SBSs, content requests of a mobile user may need to be dynamically stored at different SBSs. However, cache-enabled UAVs can track the mobility pattern of users and avoid frequently updating the content requests of mobile users. Therefore, ground users can be effectively served by exploiting mobile cache-enabled UAVs that predict mobility patterns and content request information of users.

{\color{black}
\subsection{Cellular-Connected Drones as User Equipments}}
Naturally, drones can act as users of the wireless infrastructure. In particular, drone-users can be used for package delivery, surveillance, remote sensing, and virtual reality applications. Indeed, cellular-connected UAVs will be a key enabler of the IoT. For instance, for delivery purposes, drones are used for Amazon's prime
air drone delivery service, and autonomous delivery of emergency drugs \cite{bamburry2015drones}. The key advantage of drone-users is their ability to swiftly move and optimize their path to quickly complete their missions. To properly use drones as user equipments (i.e., cellular-connected drone-UEs \cite{3GPP36777}), there is a need for reliable and low-latency  communication between drones and ground BSs.  In fact, to support a large-scale deployment of drones, a reliable wireless communication infrastructure is needed to effectively control the drones' operations while supporting the traffic stemming from their application services \cite{mozaffari2018beyond}.

Beyond their need for ultra low latency and reliability, when used for surveillance purposes, drone-UEs will require high-speed uplink connectivity from the terrestrial network and from other UAV-BSs. In this regard, current cellular networks may not be able to fully support drone-UEs as they were designed for ground users whose operations, mobility, and traffic characteristics are substantially different from the drone-UEs. There are a number of key differences between drone-UEs and terrestrial users. First, drone-UEs typically experience different channel conditions due to nearly LoS communications between ground BSs and flying drones. In this case, one of the main challenges for supporting drone-UEs is significant LoS interference caused by ground BSs\footnote{One approach for mitigating such LoS interference is to utilize full-dimensional MIMO in BS-to-drone communications \cite{3GPP36777}.}. Second, unlike terrestrial users, the on-board energy of drone-UEs is highly limited. Third, drone-UEs are in general more dynamic than ground users as they can continuously fly in any direction. Therefore, incorporating cellular-connected drone-UEs in wireless networks will introduce new technical challenges and design considerations. \vspace{0.15cm}



{\color{black}
	\subsection{Flying Ad-hoc Networks with UAVs }
	
	One of the key use cases of UAVs is in flying ad-hoc networks (FANETs) in which multiple UAVs communicate in an ad-hoc manner. With their mobility, lack of central control, and self-organizing nature, FANETs can expand the connectivity and communication range at geographical areas with limited cellular infrastructure \cite{Flying3}.  Meanwhile, FANETs play important roles in various applications such as traffic monitoring, remote sensing, border surveillance, disaster management, agricultural management, wildfire management, and relay networks \cite{Flying3, bekmezci2013flying,sahingoz2014networking}.  In particular, a relaying network of UAVs maintains reliable communication links between a remote transmitters and receivers that cannot directly communicate due to obstacles or their long separation distance.
	
	Compared to a single UAV, a FANET with multiple small UAVs has the following advantages \cite{bekmezci2013flying}:

	\begin{itemize}
		\item Scalability: The operational coverage of FANETs can be easily increased by adding new UAVs and adopting efficient dynamic routing schemes.  
		
		\item Cost: The deployment and maintenance cost of small UAVs is lower than the cost of a large UAV with complex hardware and heavy payload.
		
		\item Survivability: In FANETs, if one UAV becomes inoperational (due to weather conditions or any failure in the UAV system), FANETs’ missions can still proceed with rest of flying UAVs. Such flexibility does not exist in a single UAV system.
	\end{itemize}
}

{\color{black}
\subsection{Other Potential UAV Use Cases}}
\subsubsection{UAVs as Flying Backhaul for Terrestrial Networks}
Wired backhauling  is a common approach for connecting base stations to a core network in terrestrial networks. However, wired connections can be expensive and infeasible due to geographical constraints, especially when dealing with ultra dense cellular networks \cite{densification, Ge, MmWave}. While wireless backhauling is a viable and cost-effective solution, it suffers from blockage and interference that degrade the performance of the radio access network \cite{WirelessBackhaul}. In this case, UAVs can play a key role in enabling cost-effective, reliable, and high speed wireless backhaul connectivity for ground networks \cite{ursula}. In particular, UAVs can be optimally placed to avoid obstacles and establish LoS and reliable communication links. Moreover, the use of UAVs with mmW capabilities can establish high data rate wireless backhaul connections that are needed to cope with high traffic demands in congested areas. UAVs can also create a reconfigurable network in the sky and provide multi-hop LoS wireless backhauling opportunities. Clearly, such flexible UAV-based backhaul networks can significantly improve the reliability, capacity, and operation cost of backhauling in terrestrial networks.



\subsubsection{Smart Cities}
Realizing a global vision of smart and connected communities and cities is a daunting technological challenge. Smart cities will effectively have to integrate many of the previously mentioned technologies and services including an IoT environment (with its numerous services), a reliable wireless cellular network, resilience to calamities, and huge amounts of data \cite{ferdowsi2017colonel}. To this end, UAVs can provide several wireless application use cases in smart cities. On the one hand, they can be used as data collection devices that can gather vast amounts of data  across various geographical areas within a city and deliver them to central cloud units for big data analytics purposes. On the other hand, UAV base stations can be used to simply enhance the coverage of the cellular network in a city or to respond to specific emergencies. UAVs can also be used to sense the radio environment maps \cite{GesbertMap} across a city, in order to assist network operators in their network and frequency planning efforts. 
 Another key application of UAVs in smart cities is their ability to act as mobile cloud computing systems \cite{Jeong}. In this regards,  a UAV-mounted
cloudlet can provide fog computing and offloading opportunities for devices that are unable to perform computationally heavy tasks. We note that, within smart cities, drones may need to temporarily position themselves on buildings for specific purposes (e.g., recharge). In such case, there is a need for on-demand site renting management to accommodate drones' operation. Overall, UAVs will be an integral part of smart cities, from both  wireless and operational perspectives.


{\color{black}
	
	\subsection{Summary of Lessons Learned}
	The key lessons learned from Section II are listed as follows:
	\begin{itemize}
		\item Flying UAVs can play several roles in wireless networks. In particular, UAVs can be used as aerial base stations, user equipments in cellular networks, or mobile relay in flying ad-hoc networks. Moreover, they have promising applications in wireless backhauling and smart cities.
	
	\item UAV base stations can significantly improve the coverage and capacity of wireless networks. Furthermore, they can be deployed to enable connectivity in public safety information dissemination scenarios. UAVs can also facilitate millimeter wave communications and reliable energy efficient IoT communications. Meanwhile, the deployment of cache-enabled UAV base stations is a promising solution for traffic offloading in wireless networks.  
	
	\item Drones can also act as flying users within a cellular network in various applications such as package delivery and virtual reality.  Cellular-connected drones can freely move and optimize their route so as	to quickly complete their missions and deliver their tasks. Such cellular-connected drones  require reliable and low-latency communications with ground base stations.

	\item Self-organizing and flexible flying ad-hoc networks of UAVs can provide coverage expansion for geographical areas with limited wireless infrastructure. 
		
	\end{itemize}
	
}

Clearly, the aforementioned applications are only a selected sample of potential use cases of UAVs as flying wireless platforms. If realized, such applications will have far reaching technological and societal impacts. However, in order to truly deploy such UAV-centric applications, one must overcome numerous technical challenges, as outlined in the next section.

\section{Research Directions, Challenges, and State-of-the-Art}\label{sec:Existing}
In this section, inspired by the aforementioned applications, we present a comprehensive overview on the key research directions that must be pursued for practically deploying UAVs as flying wireless platforms. For each research direction, we first outline the key challenges, and then we discuss the state of the art, while also providing an overview on recent results.

\begin{figure}[!t] 
	\begin{center}
		\vspace{-0.2cm}
		\includegraphics[width=9.1cm]{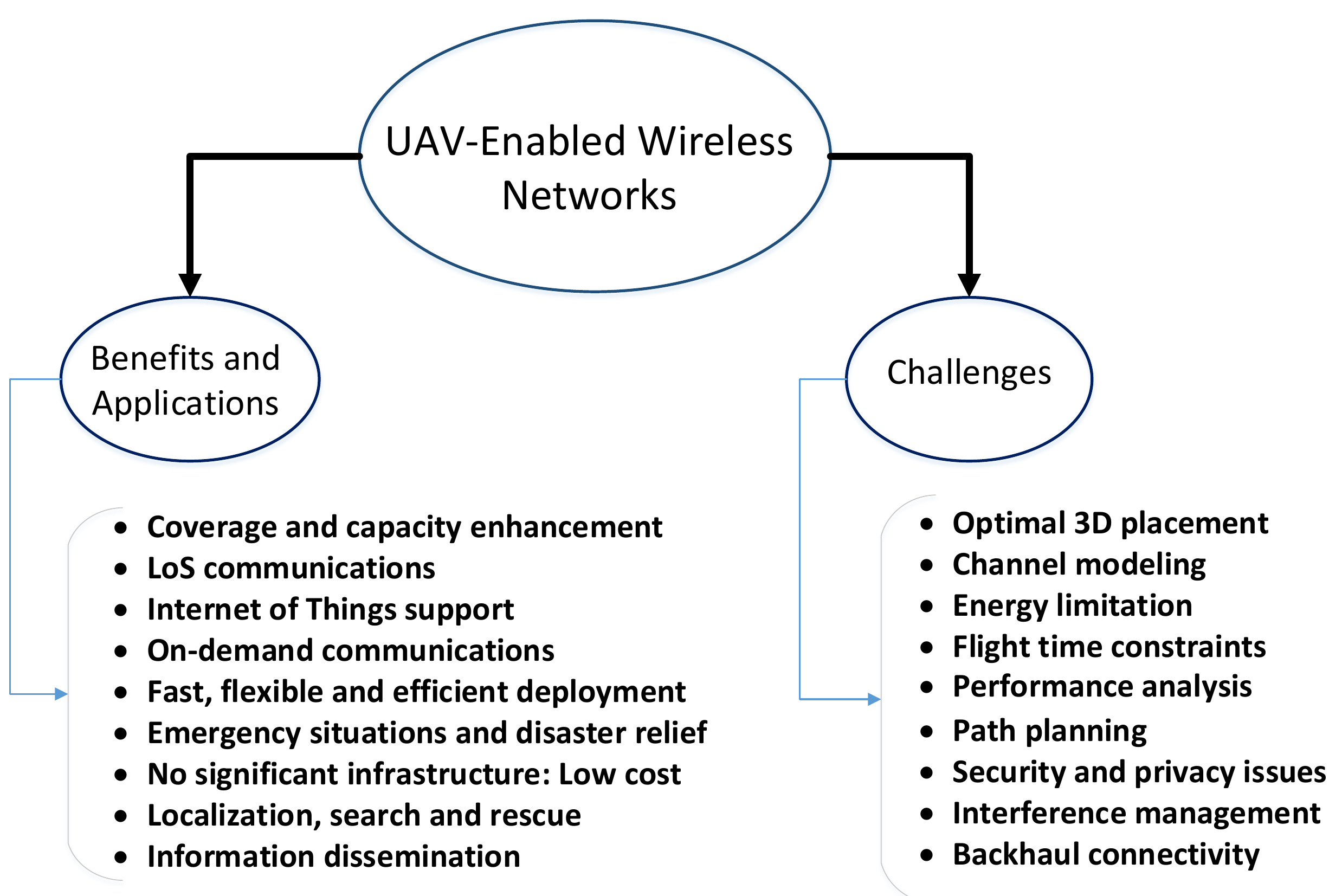}
		\vspace{-0.4cm}
		\caption{{Opportunities, applications, and challenges of UAV-enabled wireless networks.} \label{Benefits}}
	\end{center}\vspace{-0.6cm}
\end{figure}

\subsection{Air-to-Ground Channel Modeling}

\subsubsection{Challenges}
Wireless signal propagation is affected by the medium between the transmitter
and the receiver. The air-to-ground (A2G) channel characteristics significantly differ from classical ground communication
channels  which, in turn, can determine the performance of UAV-based wireless communications in terms of coverage
and capacity \cite{zaj,Zheng,Holis,HouraniModeling}. Also, compared to air-to-air communication links that experience dominant LoS, A2G channels are more susceptible to blockage. Clearly, the optimal design and deployment of drone-based communication systems \textcolor{black} {require} using an accurate A2G channel model. While the ray-tracing technique is a reasonable approach for channel modeling,  it lacks sufficient accuracy, particularly at low frequency operations \cite{yun2015ray}.  An accurate A2G channel modeling is important especially when using UAVs in applications such as coverage enhancement, cellular-connected UAV-UEs, and IoT communications. 

The A2G channel characteristics significantly differ from ground communication channels \cite{3GPP36777}. In particular, any movement or vibration by the UAVs can affect the channel
characteristics. Moreover, the A2G channel is highly dependent on the altitude and type of the UAV, elevation angle, and type of the propagation environment. Therefore, finding a
generic channel model for UAV-to-ground communications needs comprehensive simulations and measurements in various environments. In addition, the effects of a  UAV's altitude, antennas' movements, and shadowing caused by the UAV's body  must be captured in channel modeling. Clearly, capturing such factors is challenging in A2G
channel modeling.\vspace{0.02cm}

\subsubsection{State of the Art}
Now, we discuss a number of recent studies on A2G channel modeling. The work in \cite{Matolak} presented an overview of existing research related to A2G channel modeling. In \cite{Matolak2017}, the authors provided both simulation and measurement results for path loss, delay spread, and fading in A2G communications. In \cite{Ismail_survey}, the authors provided a comprehensive survey on A2G propagation while describing large-scale and small-scale fading models. 
In \cite{zaj} and \cite{Zheng}, the authors performed thorough path loss modeling for high altitude A2G communications. {\color{black}  As discussed in  \cite{zaj}, \cite{bor}, and \cite{zhang}, by efficiently deploying UAVs, their A2G communication links can experience a better channel quality (and a higher likelihood of LoS connections) compared to fixed terrestrial base stations.} The authors in \cite{Holis} presented a channel propagation model for high altitude platforms and ground users communications in an urban area.  In \cite{Holis}, based on empirical results, the statistical characteristics of the channel are modeled as a function of the elevation angle. In particular, the authors in  \cite{Holis} considered LoS and NLoS links between the HAP and ground users and derived the probability of occurrence associated with each link. In \cite{FengModelling}, the likelihood of LoS links for A2G communication was derived as a function of elevation angle and average height of buildings in urban environments.   In addition, there are some measurement-based studies on UAV-to-ground channel modeling such as \cite{Channel3D,UAVChannel2,UAVChannel3,Bettstetter} that identified some of the key channel characteristics. These works provide some insights on the A2G channel characteristics that can be used to find a more generic channel model.\vspace{0.02cm} 

\subsubsection{Representative Result}
One of the most widely adopted A2G path loss model for low altitude platforms is presented in \cite{ HouraniModeling} and, thus, we explain it in more detail. As shown in \cite{ HouraniModeling}, the path loss between a UAV and a ground device depends on the locations of the UAV and the ground device as well as the type of propagation environment (e.g., rural, suburban, urban, high-rise urban). In this case, depending on the environment,  A2G communication links can be either LoS or NLoS. Note that, without any additional information about the exact locations, heights, and number of the obstacles, one must consider the randomness associated with the LoS and NLoS links. As a result, many of the existing literature on UAV communication (e.g., \cite{bor,Irem,Kalantari, Jacob, Azari, Azari2, Haya, Gomez, jia, mingzhe,ursula}) adopted the probabilistic path loss model given in \cite{HouraniOptimal}, and \cite{HouraniModeling}. As discussed in these works, the LoS and non-LoS (NLoS) links can be considered separately with
different probabilities of occurrence. The probability
of occurrence is a function of the environment, density
and height of buildings, and elevation angle between UAV and ground device. The common probabilistic LoS model is based on the general geometrical statistics of various environments  provided by the International Telecommunication
Union (ITU-R) \cite{ITUR}. In particular, for various types of environments, the ITU-R provides some environmental-dependent parameters to determine the density, number, and hight of the buildings (or obstacles). For instance, according to \cite{ITUR}, the buildings' heights can be modeled using a Rayleigh distribution as:
\begin{equation}
	f({h_B}) = \frac{{{h_B}}}{{{\gamma ^2}}}\exp \left( {\frac{{ - {h_B}}}{{2{\gamma ^2}}}} \right),
\end{equation}
where $h_B$ is the height of buildings in meters, and $\gamma$ is a environmental-dependent parameter \cite{HouraniOptimal}.
Clearly, due to the randomness (uncertainty) associated with the height of buildings (from a UAV perspective), one must consider a probabilistic LoS model while designing  UAV-based communication systems. Therefore, using the statistical parameters provided by ITU-R, other works such as \cite{HouraniOptimal} and \cite{HouraniModeling} derived an expression for the LoS probability, which is given by \cite{HouraniModeling, Irem,Kalantari, Jacob, Azari, Azari2, Haya, Gomez, jia}:\vspace{-0.2cm}
\begin{equation}\label{PLoS}
	{P_{{\rm{LoS}}}} = \frac{1}{{1 + C\exp ( - B\left[ {\theta  - C} \right])}},
\end{equation}
where $C$  and $B$  are constant values that depend on the environment (rural, urban, dense urban, or others) and $\theta$  is the elevation angle in degrees.  Clearly, ${\theta} = \frac{{180}}{\pi } \times {\sin ^{ - 1}}\left( {{\textstyle{{{h}} \over { {{d}} }}}} \right)$, with $h$ being the UAV's altitude, and $d$ is the distance between the UAV and a given ground user. In this case,  the NLoS probability will be ${{P}_{{\text{NLoS}}}} = 1 - {{P}_{{\text{LoS}}}}$. We note that the probabilistic path loss model in (\ref{PLoS}) is an example of existing A2G channel models such as the one proposed by the 3GPP \cite{3GPP36777}.

Equation  (\ref{PLoS}) captures the probability of having LoS connection between the aerial base station and ground users is an increasing function of elevation angle. According to this equation, by increasing the elevation angle between the receiver and the transmitter, the blockage effect decreases and the communication link becomes more LoS. 

It is worth noting that the small-scale fading in A2G communications can be characterized by Rician fading channel model \cite{Matolak2017}. The Rician $K$-factor that represents the strength of LoS component is a function of elevation angle and the UAV's altitude.    
\begin{figure}[t]
	\begin{center}
		\vspace{-0.2cm}
		\includegraphics[width=7cm]{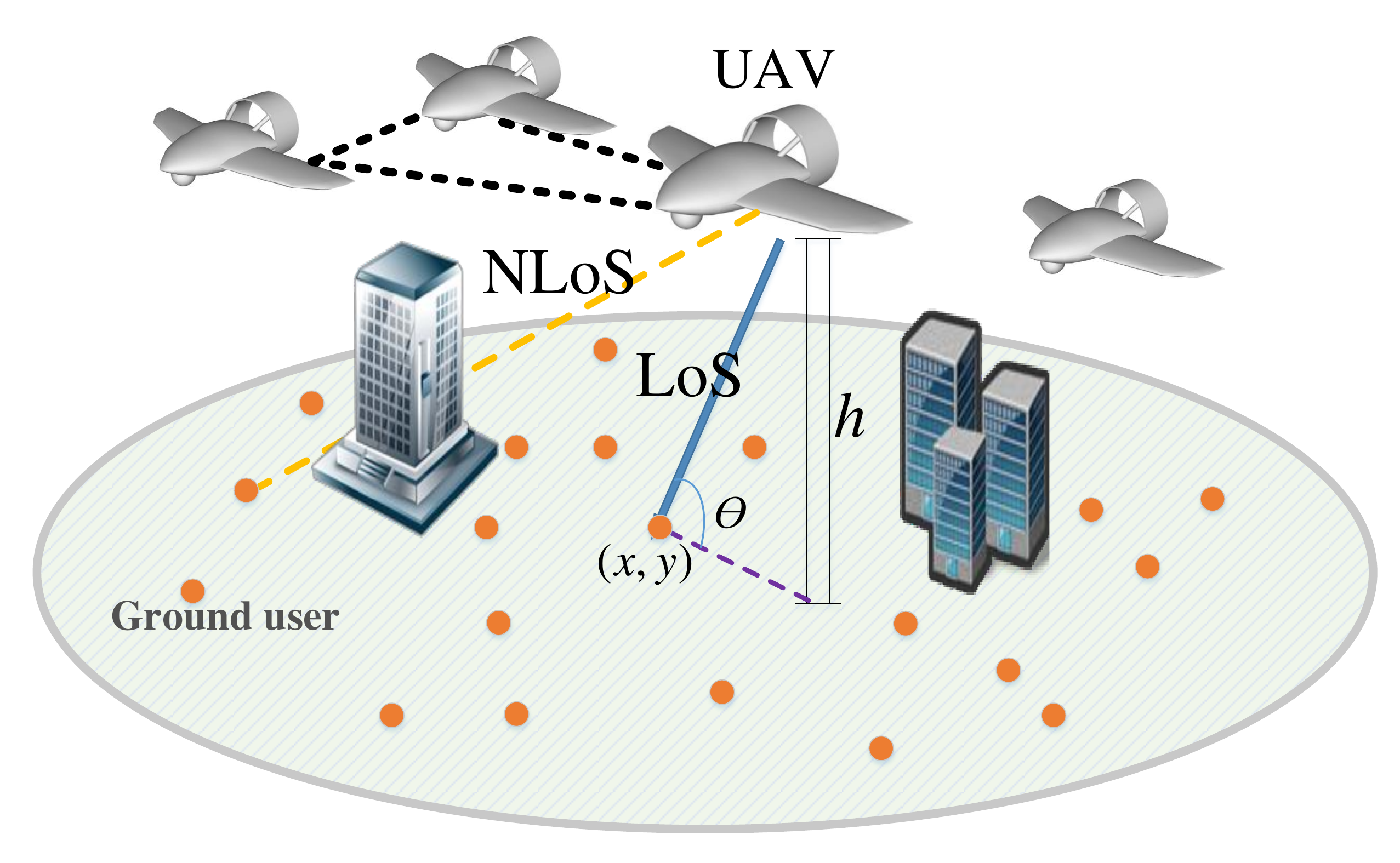}
		\vspace{-0.1cm}
		\caption{{UAV communication.}\label{ChannleModeling} }
	\end{center}\vspace{-0.6cm}
\end{figure}


\subsection{Optimal Deployment of UAVs as Flying Base Stations}\label{1-2}

\subsubsection{Challenges}
The three dimensional deployment of UAVs is one of the key challenges in UAV-based communications. In fact, as mentioned in Tables \ref{Compare1} and \ref{Compare2}, the adjustable height of UAVs and their potential mobility provide additional degrees of freedom for an efficient deployment. As a result, optimal deployment of UAVs has received significant attention \cite{HouraniOptimal,Mozaffari,Letter,Kalantari,Irem,Azari,IoTJournal,kosmerl,ALZ1,ALZ2, Kalantari2}. In fact, deployment is a key design consideration while using UAVs for coverage and capacity maximization, public safety, smart cities, caching, and IoT applications. 
The optimal 3D placement of UAVs is a challenging task as it depends on many factors such as deployment environment (e.g., geographical area), locations of ground users, and UAV-to-ground channel characteristics which itself is a function of a UAV's altitude.
In addition, simultaneously deploying multiple UAVs becomes more challenging due to the impact of inter-cell interference on the system performance. In fact, the deployment of UAVs is significantly more challenging than that of ground base stations, as done in conventional cellular network planning. Unlike terrestrial base stations UAVs needs to be deployed in a continuous 3D space while considering the impact of altitude on the A2G channel characteristics. Moreover,
while deploying UAVs, their \textcolor{black}{flight time} and energy constraints must be also taken into account, as they directly impact the network performance.\vspace{0.02cm}

\subsubsection{State of the Art}
Recently, the deployment problem of UAVs in wireless networks has been extensively studied in the literature. {\color{black}For instance, in \cite{IoTJournal}, the optimal deployment and mobility of multiple UAVs for energy-efficient data collection from IoT devices was investigated.} In \cite{HouraniOptimal}, the authors derived the optimal altitude enabling a single UAV to achieve a maximum coverage radius. In this work, the deterministic coverage range is determined by comparing the average path loss with a specified threshold. As shown in \cite{HouraniOptimal}, for very low altitudes, due to the shadowing effect, the probability of LoS connections between transmitter and receiver decreases and, consequently, the coverage radius decreases. On the other hand, at very high altitudes, LoS links exist with a high probability. However, due to the large distance between transmitter and receiver, the path loss increases and consequently the coverage performance decreases. Therefore, to find the optimal UAV's altitude, the impact of both distance and LoS probability should be considered simultaneously.

In \cite{Mozaffari}, we extended the results of \cite{HouraniOptimal} to the case of two, interfering UAVs. In \cite{Letter}, we investigated the optimal 3D placement of multiple UAVs, that use directional antennas, to maximize total coverage area. The work in \cite{Azari} analyzed the impact of a UAV's altitude on the sum-rate maximization of a UAV-assisted terrestrial wireless network.  In \cite{bor}, the authors investigated the 3D placement of drones with the goal of maximizing the number of ground users which are covered by the drone. In \cite{Kalantari}, the authors studied the efficient deployment of aerial base stations to maximize the coverage performance. Furthermore, the authors in \cite{Kalantari} determined the minimum number of drones needed for serving all the ground users within a given area.  In \cite{kosmerl}, the authors used evolutionary algorithms to find the optimal placement of LAPs and portable base stations for disaster relief scenarios. In this work, by deploying the UAVs at the optimal locations, the number of base stations required to completely cover the desired area was minimized. The work in \cite{E} proposed a framework for a cooperative deployment and task allocation of UAVs that service ground users. In \cite{E}, the problem of joint deployment and task allocation was addressed  by exploiting the concepts of coalitional game theory and queueing theory.    

Moreover, the deployment of UAVs for supplementing existing cellular infrastructure was discussed in \cite{Daniel}.  In this work, a general view of the potential integration of UAVs with cellular networks was presented. 
In \cite{zhan2006}, the authors investigated the optimal deployment of a UAV that acts as a wireless relay between the transmitter and the receiver. The optimal location of the UAV was determined by maximizing the average rate while ensuring that the bit error rate will not exceed a specified threshold. As shown in \cite{zhan2006}, a UAV should be placed closer to the ground device (transmitter or receiver) which has a poor link quality to the UAV. The authors in \cite{de} studied the use of UAV relays to enhance the connectivity of a ground wireless network. In this work, flying UAVs are optimally deployed to guarantee the message delivery of sensors to destinations. The work in \cite{orfanus} investigated the deployment of multiple UAVs as wireless relays in order to provide service for ground sensors. In particular, this work addressed the tradeoff between connectivity among the UAVs and maximizing the area covered by the UAVs. 


\subsubsection{Representative Results}
\begin{figure}[t]
	\begin{center}
		\vspace{-0.2cm}
		\includegraphics[width=7.7cm]{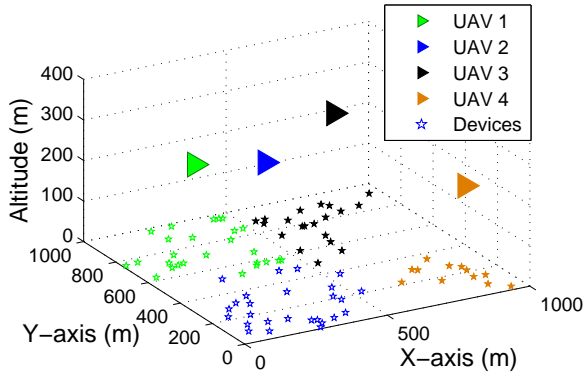}
		\vspace{-0.1cm}
		\caption{{ Optimal 3D locations of UAVs \cite{IoTJournal}.}}\label{Locations}
	\end{center}\vspace{-0.2cm}
\end{figure}

In \cite{IoTJournal}, we proposed a framework for dynamic deployment and mobility of UAVs to enable reliable and energy-efficient IoT communications.   In Figure \ref{Locations}, we show a representative result on the optimal 3D placement of UAVs, taken from \cite{IoTJournal}. In this case, four UAVs are deployed to collect data (in the uplink) from IoT devices which are uniformly distributed within a geographical area of size $1 \textrm{km} \times 1 \textrm{km}$.  Here, using tools from optimization theory and facility location problems, we derived the optimal 3D positions of the UAVs as well as the device-UAV associations such that the total uplink transmit power of devices is minimized while ensuring reliable communications. As a result, the devices are able to send their data to the associated UAVs while using a minimum
total transmit power. This result shows that UAVs can be optimally deployed to enable reliable and energy-efficient uplink communications in IoT networks. 

\begin{figure}[t]
	\begin{center}
		\vspace{-0.2cm}
		\includegraphics[width=7.7cm]{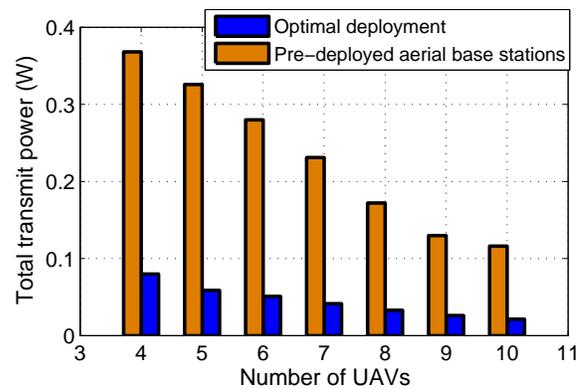}
		\vspace{-0.1cm}
		\caption{{Total transmit power of devices vs. number of UAVs (for 80 IoT devices).}}\label{Power_UAV}
	\end{center}\vspace{-0.5cm}
\end{figure}

Figure \ref{Power_UAV} shows the average transmit power of devices in the optimal deployment scenario with a case in which aerial base stations are pre-deployed (i.e., without optimizing the UAVs' locations). As we can see, the average transmit power of devices can be reduced by 78\% by optimally deploying the UAVs. Figure \ref{Power_UAV} also shows that the uplink transmit power decreases while increasing the number of UAVs. Clearly, the energy efficiency of the IoT network is significantly improved  by exploiting the flexibility of drones and optimizing their locations. 

Next, we discuss another key result on the deployment of multiple UAVs for maximizing wireless coverage. In our work in \cite{Letter}, we consider multiple UAV-BSs that must provide a downlink wireless service to a circular geographical area of radius 5\,km. We assume that the UAVs are symmetric and have the same transmit power and altitude. In the considered model, each UAV uses a directional antenna with a certain beamwidth, and UAVs \textcolor{black}{operate} at the same frequency band. Our goal is to optimally deploy the UAVs in 3D space such that their total coverage area is maximized while avoiding mutual interference between the UAVs. To this end, we tackle our problem by  exploiting circle packing theory \cite{gaspar2000upper}. Our results provide rigorous guidelines on how to optimally adjust the location and, in particular, the altitude of UAVs, based on the antenna beamwidth, size of the area, and the number of UAVs.

\begin{figure}[t]
	\begin{center}
		\vspace{-0.2cm}
		\includegraphics[width=7.7cm]{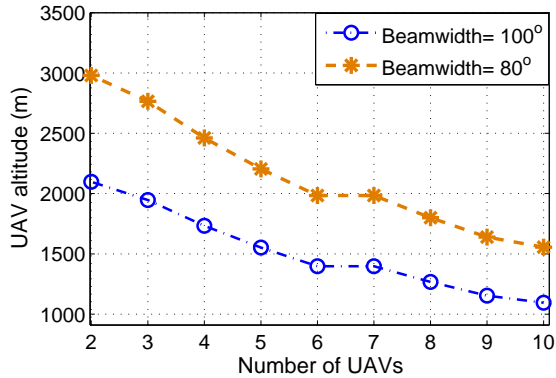}
		\vspace{-0.1cm}
		\caption{{Each UAV's altitude for various number of UAVs.}}\label{Figer_Letter}
	\end{center}\vspace{-0.5cm}
\end{figure}

In Figure \ref{Figer_Letter}, we show a representative result from \cite{Letter}. In particular, Figure \ref{Figer_Letter} shows how the optimal UAVs' altitude varies by changing the
number of UAVs. Intuitively, to avoid interference, the height of UAVs must be decreased
as the number of UAVs increases. In this case, for a higher number
of UAVs, the coverage radius of each UAV must be decreased by reducing its altitude to avoid overlapping (or interference) between their coverage regions. For instance,
by increasing the number of UAVs from 3 to 6, the optimal
altitude decreases from 2000\,m to 1300\,m. This figure also shows that the UAVs must be placed at lower altitudes when they use directional antennas with higher antenna beamwidths.

\subsection{Trajectory Optimization }\label{1-3}
Optimal path planning for UAVs is another important challenge in UAV-based communication systems. In particular, optimizing the trajectory of UAVs is crucial while using them for smart cities, drone-UE, and caching scenarios. The trajectory of a UAV is significantly affected by  different factors  such as flight time, energy constraints, ground users' demands, and collision avoidance. 

Naturally, optimizing the flight path of UAVs is challenging as it requires
considering many physical constraints and parameters. For instance, while finding the trajectories of UAVs for performance optimization, one needs to consider various key factors such as channel variation due to the mobility, UAV's dynamics, energy consumption of UAVs, and flight constraints. Furthermore, solving a continuous UAV trajectory optimization problem is known to be analytically challenging as it involves finding an infinite number of optimization variables (i.e. UAV's locations) \cite{zhang}. In addition, trajectory optimization in UAV-enabled wireless networks requires capturing coupling between mobility and various QoS metrics in wireless communication. \vspace{0.02cm}

\subsubsection{State of the Art}
Trajectory optimization for UAVs has been primarily studied from a robotics/control perspective \cite{dou, Rucco, CooperativePath, FlightDemonstrations,Autonomous,chandler2000uav}. More recently, there has been a number of works that study the interplay between the trajectory of a UAV and its wireless communication performance. The work in \cite{Qin} jointly optimized user scheduling and UAV trajectory
for maximizing the minimum average rate among ground users.
In \cite{Jiang}, the authors investigated the optimal trajectory of UAVs equipped with multiple antennas for maximizing sum-rate in uplink communications. The work in \cite{zengThroughput} maximized the throughput of a relay-based UAV system by jointly optimizing the UAV's trajectory as well as the source/relay transmit power.  In \cite{fran}, a UAV path planning algorithm for photographic sensing of a given geographical area was proposed. The algorithm of \cite{fran} led to a minimum total energy consumption for the UAV while covering the entire survey area. To this end, the authors in \cite{fran} computed the optimal set of waypoints and the optimal speed of the UAV in the path between the waypoints. In \cite{gro}, considering collision avoidance, no-fly zones, and altitude constraints, the optimal paths of UAVs that minimize the fuel consumption were computed using the mixed integer linear programming.\\
\indent Moreover, the authors in \cite{tis} investigated the path planning problem for UAVs in the search and localization applications using camera measurements. In this work, path planning was analyzed by maximizing the likelihood of target detection.
In \cite{H}, the authors investigated how to optimally move UAVs for improving connectivity of ad-hoc networks assuming that the drones have complete information on the location of devices. The work in \cite{Qin} studied the joint user scheduling and UAV trajectory design to maximize the minimum rate of ground users in a multi-UAV enabled wireless network. In addition, there are some works that studied the UAV trajectory optimization for localization purposes. For instance, the work in \cite{dou} investigated path planning for multiple UAVs for localization of a passive emitter.  
In this work, using the angle of arrival  and time difference of arrival information, the set of waypoints which leads to a minimum localization error was determined. However, the work in \cite{dou} was limited to localization and did not directly address any wireless communication problem.  Other works on UAV navigation and cooperative control are found in \cite{Rucco, CooperativePath, FlightDemonstrations,Autonomous,chandler2000uav}.\\
\indent In fact, prior studies on UAV trajectory optimization focused on three aspects: control and navigation, localization \cite{MozaffariGPS}, and wireless communications. In particular, in the existing works on UAV communications, trajectory optimization was performed with respect to energy consumption, rate, and reliability.   \vspace{0.02cm}
\subsubsection{Representative Result}
One representative result on  trajectory optimization can be found in our work in \cite{IoTJournal}. In particular, we considered a drone-assisted IoT network scenario in which 5 drones are used to collect data from ground IoT devices. A set of 500 IoT devices are uniformly distributed within a geographical size of 1\,km $\times$ 1\,km. We considered a time-varying IoT network in which the set of active IoT devices changes over time, based on a beta distribution \cite{3GPP}. Hence, to effectively serve the IoT devices, the drones must update their locations according to the locations of active devices. In this model, we consider some pre-defined time slots during which the drones collect data from active IoT devices. At the end of each time slot (i.e., update time), the drones' update their locations based on the activation pattern of IoT devices. Given such a time-varying network, our goal is to find the optimal trajectory of drones such that they can update their locations with a minimum energy consumption. Therefore, while serving IoT devices, the drones move within optimal paths so as to minimize their mobility energy consumption.

\begin{figure}[!t]
	\begin{center}
		\vspace{-0.2cm}
		\includegraphics[width=8.1cm]{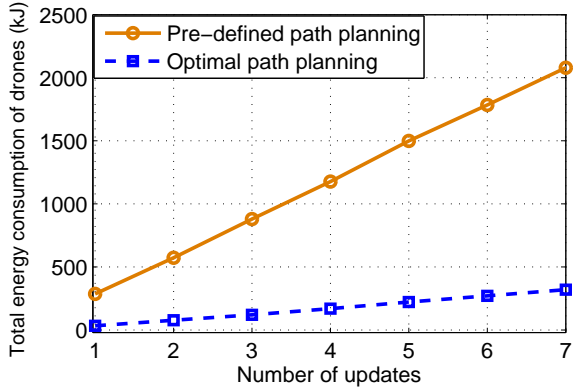}
		\vspace{-0.1cm}
		\caption{ \small Total energy consumption of drones on mobility vs. number of updates.}\vspace{-0.3cm}
		\label{Energy_Drone}
	\end{center}
\end{figure}

Figure \ref{Energy_Drone} shows the total energy consumption of drones as a function  of the number of updates. As expected, a higher number of updates
requires more mobility of the drones thus more energy consumption. We compare the performance of the optimal path planning with a case that drones update their locations following pre-defined paths. As we can see, by using  optimal path planning, the average total energy consumption of drones decreases by 74\% compared to the non-optimal case.

In fact, to effectively use UAVs in wireless networks, the trajectory of UAVs needs to be optimized with
respect to wireless metrics such as throughput and coverage as well as energy constraints of
UAVs. While jointly optimizing trajectory and communication is a challenging task, it can significantly improve the performance of UAV-enabled wireless networks.

\subsection{Performance Analysis of UAV-Enabled Wireless Networks}
\subsubsection{Challenges}
A fundamental analysis of the performance of UAV-enabled wireless systems is required in order to evaluate the impact of each design parameter on the overall system performance \cite{mozaffari2, SudheeshLetter}. In particular, the performance of the UAV systems must analyzed in terms of the key QoS metrics such as coverage probability, throughput, delay, or reliability (e.g., for cellular-connected drones). Such performance evaluations can also reveal the inherent tradeoffs that one faces when designing UAV-based systems. 

Clearly, while designing UAV-based communication systems, a fundamental performance analysis needs to be done in order to evaluate the impact of design parameters on the overall system
performance. Naturally, devising a fundamental analysis of the wireless performance of a UAV-based wireless system will substantially differ from conventional ground networks due to the altitude
and potential mobility of UAVs as well as their different channel characteristics. The stringent energy limitations of UAVs also introduce unique challenges. The limited available on-board energy of UAVs which leads to the short flight duration is a major factor impacting the performance
of wireless communications using UAVs. Indeed, analyzing the performance of a complex heterogeneous aerial-terrestrial wireless network that is composed of flying and ground base stations is a challenging task. In fact,  there is a need for a comprehensive performance analysis of UAV-enabled wireless networks while capturing various aspects of UAVs including mobility, and specific A2G channel characteristics in coexistence with terrestrial networks. Moreover, performance characterization of cellular-connected drone networks with flying users and base stations has its own complexity due to the  mobile and highly dynamic nature of the network. \vspace{0.02cm}  




\subsubsection{State of the Art}
Prior to our seminal work in this area in \cite{mozaffari2}, most of the existing works focused on performance analysis of UAVs acting as relays, or in ad-hoc networks \cite{Performance2011,guo,H,RelayCommunications}. For instance, the work in \cite{Performance2011} evaluated the performance of a UAV ad-hoc network
in terms of achievable transmission rate and end-to-end delay. In \cite{guo}, the authors studied the use of macro UAV relays to enhance the throughput of the cellular networks. The work in \cite{H}, derived the probability of successful connectivity among ground devices in a UAV-assisted ad-hoc network. In \cite{RelayCommunications}, the authors analyzed the performance
of UAVs acting as relays for ground devices in a wireless network. In particular, the authors derived closed-form expressions for signal-to-noise-ratio (SNR) distribution and ergodic capacity of  UAV-ground devices links. In contrast, in \cite{mozaffari2}, we considered the use of UAVs as stand-alone aerial base stations. In particular, we investigated the downlink coverage and rate performance of a single UAV that co-exists with a device-to-device communication network.

Following our work in \cite{mozaffari2}, the authors in \cite{VishnuJournal} derived an exact expression for downlink coverage probability for ground receivers which are served by multiple UAVs. In particular, using tools from stochastic geometry, the work in \cite{VishnuJournal} provided the coverage analysis in a finite UAV network considering a Nakagami-$m$ fading channel for UAV-to-user communications. In \cite{Haya}, the performance of a single drone-based communication system in terms of outage probability, bit error rate, and outage capacity was investigated. The work in \cite{Spectrum} analyzed the coverage and throughput  for a network with UAVs and underlaid traditional cellular networks. In this work, using 3D and 2D Poisson point processes (PPP), the downlink coverage probability and rate expressions were derived. In \cite{mumtaz}, the authors evaluated the performance of using UAVs for  overload and outage compensation in cellular networks. Clearly, such fundamental performance analysis is needed to provide various key design insights for UAV communication systems.\vspace{0.02cm}

\subsubsection{Representative Result}
As per our work in \cite{mozaffari2}, we considered a circular area with in which a number of users are spatiality distributed according to a PPP \cite{haenggi}, and a UAV-mounted aerial base station is used to serve a subset of those users. In the considered network, there are two types of users: downlink users and D2D users. Here, we consider the downlink scenario for the UAV while the D2D users operate in an underlay fashion. Moreover, we assume that a D2D receiver connects to its corresponding D2D transmitter located at a fixed distance away from it \cite{lee}. Hence, a D2D receiver receives its desired signal from the D2D transmitter pair, and interference from the UAV and other D2D transmitters. The received signals at a downlink user include the desired signal from the UAV and interference from all the D2D transmitters.

\begin{figure}[t]
	\begin{center}
		\vspace{-0.2cm}
		\includegraphics[width=8.0cm]{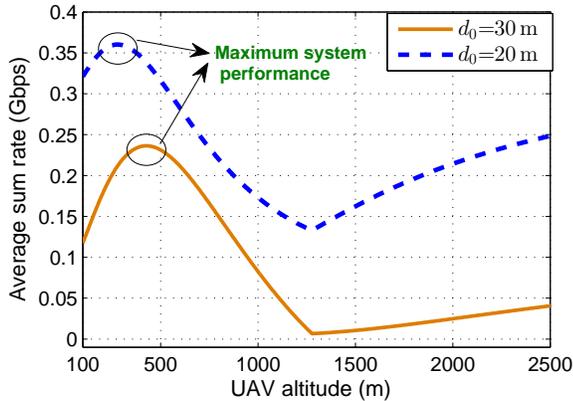}
		\vspace{-0.1cm}
		\caption{{Average sum-rate in a UAV-D2D network vs. UAVs altitude.}}\label{ASR_vs_H_D}
	\end{center}\vspace{-0.7cm}
\end{figure}

For this UAV-D2D network, we derived tractable analytical expressions for the coverage and rate analysis for both static and mobile UAV scenarios (see \cite{mozaffari2}). In Figure \ref{ASR_vs_H_D}, we show the average sum-rate versus the UAV altitude for different values of the fixed distance, $d_0$, between a D2D transmitter/receiver pair. As we can see from this figure, the average sum-rate is maximized when  the UAV's altitude are around 300\,m for $d_0$ = 20\,m. From Figure \ref{ASR_vs_H_D}, we can see that for altitude above 1300\,m, the average sum-rate starts increasing. This is due to the fact that, as the UAV's altitude exceeds a certain value, downlink users cannot be served while the interference on D2D users decreases thus increasing the sum-rate. Moreover, for altitudes within a range $300\,\textnormal{m}$ to $1300\,\textnormal{m}$, the sum-rate performance decreases due to the impact of LoS interference from the UAV on the D2D users. Note that, the optimal UAV's altitude depends on $d_0$, as shown in Figure \ref{ASR_vs_H_D}. For instance, the sum-rate is maximized at a 400\,m altitude when $d_0=30$\,m. 

We note that, in the literature, there are also additional insightful results on the performance of UAV communication systems. For instance, the work in \cite{VishnuJournal} showed the downlink coverage probability varies as a function of SIR threshold in a network of multiple UAV-BSs. In \cite{Azari}, the authors presented the impact of the UAV's altitude on the minimum required transmit power of UAV that ensures ground coverage. In \cite{Spectrum}, the network throughput of a UAV-assisted cellular network is determined as a function of the number of base stations. 


\subsection{Cellular Network Planning and Provisioning with UAVs}
\subsubsection{Challenges}
Network planning involves addressing a number of key problems such as base station positioning, traffic estimation, frequency allocation, cell association, backhaul management, signaling, and interference mitigation. Network planning with UAVs is particularly important when UAVs are used for coverage and capacity enhancement. 
In a UAV-assisted cellular network, network planning becomes more challenging due to the various properties of UAVs including mobility, LoS interference, energy constraints, and wireless backhaul connectivity. For example, joint radio and backhaul designs and deployment are needed during network planning with UAVs \cite{XX}. Furthermore, network planning in presence of flying drone-UEs requires new considerations. On the one hand, LoS interference stemming from a potentially massive number of drone-UEs in uplink significantly impacts network planning. On the other hand,  ground base stations must be equipped with appropriate types of antennas (considering e.g.,  radiation pattern and beam tilting) so as to serve drone-UEs in downlink. Another difference between network planning for traditional cellular networks and UAV systems is the amount of signaling and overhead. Unlike static terrestrial networks, in the UAV case, there is a need for dynamic signaling to continuously track the location and number of UAVs in the network. Such dynamic signaling may also be needed to register the various UAVs as users or base stations in the cellular system. Clearly, handling such signaling and overhead must be taken into account in cellular network planning with UAVs. 

{\color{black}
Backhaul connectivity for flying UAVs is another key challenge in designing UAV communication systems. Due to aerial nature of done base stations, wireless backhauling needs to be employed for connecting them to a core network. WiFi and satellite technologies are promising solutions for wireless backhauling \cite{AkramMagazin}. Satellite links can provide  wider backhaul coverage compared to WiFi. However, WiFi links have the advantages of lower cost and lower latency compared to the satellite backhauling. Other promising solutions for wireless backhauling are millimeter wave  and free space optical communications (FSO) with ground stations \cite{bor,horwath2007experimental,fidler,HalimBackhaul}.  Aerial base stations can adjust their altitude, avoid obstacles, and establish LoS communication links to ground stations. Such LoS opportunity is a key requirement for millimeter wave and FSO  communications that can provide  high capacity wireless backhauling services. We note that wireless backhauling for UAVs is still a challenging problem in UAV communications and further studies need to be done to find an efficient backhauling solution.}

\subsubsection{State of the Art}
Recent studies on UAV communications have addressed various problems pertaining to network planning. For example, in \cite{Vishal}, the authors investigated the optimal user-UAV assignment for capacity enhancement in UAV-assisted heterogeneous wireless networks. In \cite{Kalantari}, the authors jointly optimized the locations and number of UAVs for maximizing wireless coverage. The work in \cite{OTUAV} optimized the deployment and cell association of UAVs for meeting the users' rate requirements while using a minimum UAVs' transmit power. In \cite{Letter_OT}, a delay-optimal cell planning was proposed for a UAV-assisted cellular network. The work in \cite{FarajLetter}  proposed a novel approach for strategic placement of multiple UAV-BSs in a large-scale network. In \cite{Kalantari2}, the authors proposed a backhaul aware optimal drone-BS placement algorithm that maximizes the number of the served users as well as the sum-rate for the users.  The work in \cite{Boris_Back} provided an analytical expression for the probability of backhaul connectivity for UAVs that can use either an LTE or a millimeter wave backhaul. In \cite{ursula}, a framework for the use of UAVs as an aerial backhaul network for ground base stations was proposed. In fact, the previous studies on UAV network planning primarily analyzed problems related to user association, 3D placement, backhaul connectivity, and optimizing the number of UAVs that must be deployed in the network. Also, there does not exist any concrete work focusing on the signaling challenges.
\subsubsection{Representative Result}
\begin{figure}[!t]
	\begin{center}
		\vspace{-0.2cm}
		\includegraphics[width=8.0cm]{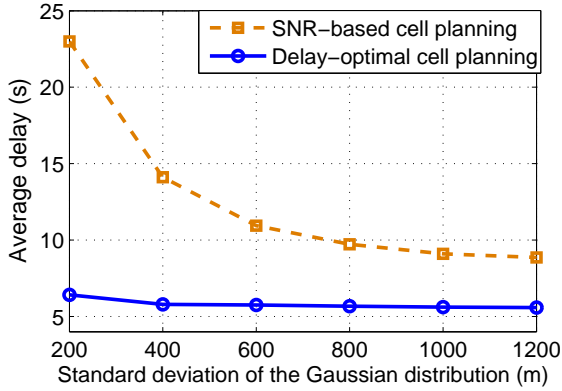}
		\vspace{-0.15cm}
		\caption{ \small Average network delay per 1Mb data transmission.}\vspace{-0.2cm}
		\label{Delay}
	\end{center}
\end{figure}

In terms of network planning, in \cite{Letter_OT}, we studied the problem of optimal cell association for delay minimization in a UAV-assisted cellular network. In particular, we considered a geographical area of size $4\,\text{km}\times 4 \,\text{km}$ in which 4 UAVs (as aerial base stations) and 2 ground macro base stations are deployed according to a traditional grid-based deployment. Within this area, ground users are distributed according to a truncated Gaussian distribution with a standard deviation $\sigma_o$, which is suitable to model a hotspot area. Here, our main performance metric is transmission delay, which is the time needed for transmitting a given number of bits to ground users. Our goal is to provide an optimal cell planning (e.g., cell association) for which the average network delay is minimized. \\
\indent In Figure \ref{Delay}, we compare the delay performance of our proposed cell association with the classical SNR-based association. For users' spatial distribution, we consider a truncated Gaussian distribution with a center (1300\,m, 1300\,m), and a standard deviation $\sigma_o$ that varies from 200\,m to 1200\,m. Lower values of $\sigma_o$ correspond to cases in which users are more congested around a hotspot center. This figure shows that the proposed cell association significantly outperforms the SNR-based association and yields up to a 72\% lower average delay. This is due to the fact that, in the proposed approach, the impact of network congestion is taken into consideration. 
In fact, unlike the SNR-based cell association, the proposed approach avoid creating highly loaded cells that cause delay in the network. Hence, compared to the SNR-based association case, our approach is more robust against network congestion, and it significantly reduces the average network delay.\\ 
\indent Clearly, the performance of UAV-enabled wireless networks significantly depends on the network planning. In general, network planning impacts several key metrics of UAV networks  such as throughput, delay (as also shown in Figure \ref{Delay}), operational cost, and energy consumption. 

\subsection{Resource Management and Energy Efficiency}\label{1-3}
\subsubsection{Challenges}
Resource management and energy efficiency require significant attention when operating UAVs in key scenarios such as IoT, public safety, and UAV-assisted cellular wireless networks.  
While resource management is a major challenge for cellular networks \cite{Talebi1, mumtaz, Talebi2}, UAVs introduce unique challenges due to: 1) Interplay between the UAVs' flight time, energy, path plan, and
 spectral efficiency, 2) Stringent energy and flight limitations for UAVs, 3) LoS interference stemming from A2G and
 air-to-air links, and 4) Unique mobility of UAVs. Hence, there is a need for optimizing and managing
 resource allocation in complex  UAV-assisted wireless networks operating
 over heterogeneous spectrum bands and co-existing with ground networks. In fact, resource management
 and spectrum sharing \cite{PantisanoSpectrum} processes must properly handle the inherent dynamics of wireless networks such as time-varying interference, varying traffic patterns, mobility, and energy constraints of the UAVs. 
 
 Naturally, flying drones have a limited amount of on-board energy which must be used for transmission, mobility, control,
 data processing, and payloads purposes \cite{Uragun}. Consequently, the flight duration of drones is typically short and insufficient for providing a long-term, continuous wireless coverage. The energy consumption of the UAV also depends on the role/mission of the UAV, weather conditions, and the navigation path. Such energy constraints, in turn, lead to limited flight and hover time durations. Hence, while designing UAV communication systems, the energy and flight constraints of UAVs need to be explicitly taken into account. Therefore, the energy efficiency of UAVs requires careful consideration as it significantly impact the performance of UAV-communication systems. In fact, the limited on-board energy of UAVs is a key constraint for deployment and mobility of UAVs in various applications.


 


\subsubsection{State of the Art}
 Energy efficiency and resource management in UAV-based wireless communication systems have been studied from various perspectives. For instance, the work in \cite{ZhangEnergy} provided an analytical framework for minimizing the energy consumption of a fixed-wing UAV by determining the optimal trajectory of the UAV. In \cite{Cooperative}, the authors proposed an energy-efficient scheduling framework for cooperative UAVs communications. In \cite{Zorbas}, the authors studied the energy efficiency of drones in target tracking scenarios by adjusting the number of active drones. Energy harvesting from vibrations and solar sources for small UAVs was investigated in \cite{ant}. The work in \cite{ZhangLetter} proposed a framework for optimizing transmission times in user-UAV communications that maximizes the minimum throughput of the users. The authors in \cite{shar} studied the use of antenna array on UAVs for improving the SNR and consequently for reducing the required transmit power. The work in \cite{Ceran} investigated an optimal resource allocation scheme for an energy harvesting flying access point. In \cite{MozaffariFlightTime}, the problem of bandwidth and flight time optimization of UAVs that \textcolor{black}{service} ground users was studied. The work in \cite{Mingzhe_LTE} proposed a resource allocation framework for enabling cache-enabled UAVs to effectively service users over licensed and unlicensed bands.

 Clearly, the performance of UAV communication systems is significantly affected by battery lifetime of UAVs. {\color{black}
	The flight time (i.e., battery lifetime) of a UAV depends on several factors such as the energy source (e.g., battery, fuel, etc.,), type, weight, speed, and trajectory of the UAV.   In Table \ref{Battery}, we provide some examples for the battery lifetime of various types of UAVs \cite{fotouhi2018survey}.

	\begin{table}[!t] {\color{black}
			\normalsize
			\begin{center}
				\caption{\small Battery lifetime of UAVs.}
				\vspace{-0.1cm}
				\label{Battery}
				\resizebox{9cm}{!}{
					\begin{tabular}{|c|c|c|c|}
						\hline
						\textbf{Size} & \textbf{Weight} & \textbf{Example}  & \textbf{Battery lifetime}\\ \hline \hline
						
						Micro	&     $<100$\,g   &      Kogan Nano Drone &  6-8\,min \\ \hline
						
						Very small & 100\,g--2\,kg	&   Parrot Disco     &      45\,min     \\ \hline 
						
						Small	&    2\,kg--25\,kg  &   DJI Spreading Wings & 18\,min \\ \hline
						
						Medium	&    25\,kg--150\,kg     &     Scout B-330 UAV
						helicopter & 180\,min    \\ \hline

						Large	&    $>150$\,kg      &   Predator B & 1800\,min \\ \hline
						
				\end{tabular}}
				
		\end{center}}\vspace{-0.5cm}
	\end{table}

In general, the total energy consumption of a UAV is composed of two main components \cite{ZhangEnergy,zeng2018energy,fotouhi2018survey}: 1) Communication related energy, and 2) Propulsion energy. The  related energy.  The communication related energy is used for various communication functions such as signal transmission, computations, and signal processing. The propulsion energy pertains to the  mechanical energy consumption for movement and hovering of UAVs. Typically, the  propulsion energy consumption is significantly more than the communication-related energy consumption. Next, we provide some baseline propulsion energy consumption models for fixed-wing and rotary-wing UAVs in a forward flight with speed $V$.

For a fixed-wing UAV, the propulsion energy consumption during a flight time $T$ is given by \cite{ZhangEnergy}: 
	\begin{equation}
	E = T\left( {{a_1}{V^3} + \frac{{{a_2}}}{V}} \right),
	\end{equation}
	where $a_1$ and $a_2$ are constants that depend on several factors such as UAV's weight, wing area, and air density \cite{ZhangEnergy}.

	For a rotary-wing UAV, the propulsion energy consumption during a flight time $T$ is given by \cite{zeng2018energy}: 
	\begin{align}
	E = T\Bigl[& {c_1}\left( {1 + \frac{{3{V^2}}}{{{q^2}}}} \right) + {c_2}{{\left( {\sqrt {1 + \frac{{{V^4}}}{{4v_o^4}}}  - \frac{{{V^2}}}{{2v_o^2}}} \right)}^{1/2}} \nonumber\\ 
	&+ \frac{1}{2}{d_o}\rho s A{V^3} \Bigr],
	\end{align}
	where $c_1$ and $c_2$ are constants which depend on drone's weight, rotor's speed, rotor disc area, blade angular velocity, and air density. $q$ is the tip speed of the rotor, $d_o$ is the fuselage drag ratio, $v_o$ is the mean rotor speed, $\rho$ is air density, $s$ is the rotor solidity,  and $A$ is the rotor disc area.
}

\subsubsection{Representative Result}
In \cite{MozaffariFlightTime}, we studied the resource management problem with a focus on optimal bandwidth allocation in UAV-enabled wireless networks. In particular, we considered a scenario in which 5 UAVs are deployed as aerial base stations over a rectangular area of size 1\,km $\times$ 1\,km in order to provide service for 50 ground users. These UAVs must fly (or hover) over the area until all the users receive their desired service (in terms of number of bits) in the downlink. Our goal is to optimally share the total available bandwidth between the users such that the total flight time that the UAVs need to service the users is minimized. Note that the flight time is directly related to the energy consumption of UAVs. Hence, minimizing the flight time of UAVs will effectively improve their energy-efficiency.

\begin{figure}[!t]
	\begin{center}
		\vspace{-0.2cm}
		\includegraphics[width=8.0cm]{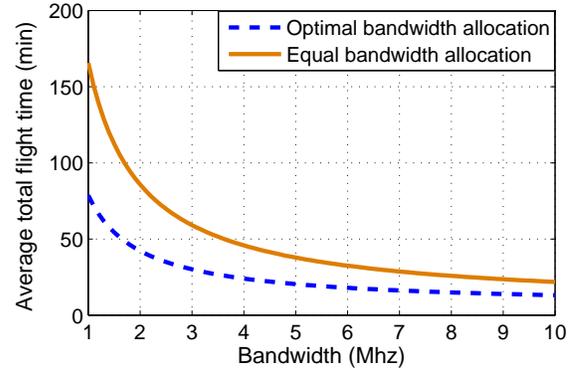}
		\vspace{-0.1cm}
		\caption{ \small Average flight time vs. bandwidth.}\vspace{-0.3cm}
		\label{Hover_BW}
	\end{center}
\end{figure}

Figure \ref{Hover_BW} shows the average total flight time of UAVs versus the transmission bandwidth. Here, the total flight time represents the time needed to provide service to all ground users, each of which requires a 100\,Mb data. We consider two bandwidth allocation schemes, the optimal bandwidth allocation, and an equal bandwidth allocation. Clearly, by increasing the bandwidth, the total flight time that the UAVs require to service their users decreases. {\color{black} Naturally, a higher bandwidth can provide a higher
transmission rate and, thus, users can be served within a shorter time duration.} From Figure \ref{Hover_BW}, we can observe that the optimal bandwidth allocation scheme can lead to a 51\% shorter flight time
compared to the equal bandwidth allocation case. This is because, by optimally allocating the bandwidth to each user based on its load and location, the total flight time of UAVs can be minimized.
  
\begin{figure}[!t]
	\begin{center}
		\vspace{-0.2cm}
		\includegraphics[width=8.0cm]{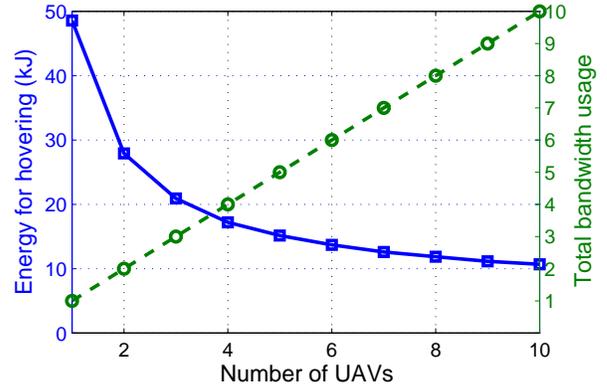}
		\vspace{-0.1cm}
		\caption{ \small UAV energy consumption (due to hover time) and spectrum
			tradeoff.}\vspace{-0.4cm}
		\label{Hover_Energy}
	\end{center}
\end{figure}

In Figure \ref{Hover_Energy}, we show  the total hovering energy consumption of the UAVs as a function of number of UAVs. This
result corresponds to the interference-free scenario in which the UAVs operate on different
frequency bands. Hence, the total bandwidth usage linearly increases by increasing the number
of UAVs. Clearly, the total energy consumption decreases as the number of UAVs
increases. A higher number of UAVs corresponds to a higher number of cell partitions. Therefore,
the size of each cell partition decreases and the users will have a shorter distance to the UAVs.
Increasing the number of UAVs leads to a higher transmission rate thus shorter hover time and energy consumption. For instance, Figure \ref{Hover_Energy} shows that when the number of UAVs
increases from 2 ot 6, the total energy consumption of UAVs  decreases by 53\%. Nevertheless, deploying more
UAVs in interference-free scenario requires using more bandwidth. Hence, there is a
fundamental tradeoff between the energy consumption of UAVs for hovering and bandwidth efficiency.

In summary, to efficiently employ UAVs for wireless networking applications, one must efficiently manage the use of available resources such as energy, bandwidth, and time. In fact, the performance of UAV-communication systems is significantly affected by resource allocation strategies and energy constraints of UAVs.

\subsection{Drone-UEs in Wireless Networks}
\subsubsection{Challenges}
Beyond the use of drones as aerial base stations, they can also act as flying users as part of cellular networks. In particular, drone-UEs play key roles in air delivery applications, such as Amazon prime air and in surveillance applications. Another important application of drone-UEs is virtual reality (VR) \cite{chen2017virtual, JacobVR,chen2017echo} where drones capture any desired information about a specific area and transmit it to remote VR users. However, current cellular networks have been primarily designed for supporting terrestrial devices whose characteristics are significantly different from drone-UEs. Naturally,
classical wireless challenges such as performance analysis, interference management,
mobility management, and energy and spectrum efficiency, will be further exacerbated by the
use of drone-UEs due to their relatively high altitude, stringent
on-board energy limitations, dynamic roles, potentially massive deployment, and
their nearly unconstrained mobility. In particular,  incorporating drone-UEs in cellular networks introduces unique challenges such as uplink interference management due to massive deployment of drone-UEs, ground-to-air channel modeling for BSs-to-drones communications, and designing suitable BS's antennas that can support high altitude (i.e., high elevation angle) drones.   In addition, drone-UEs will require ultra-reliable, low
latency communications (URLLC) so as to swiftly control their operations, and ensure
their safe and effective navigation. Clearly, such a need for URLLC also leads to new
wireless networking challenges.

Furthermore, there is a need for effective handover management mechanisms to deploy an aerial network of flying drone-UEs and drone-BSs. Handover is a key process in wireless networks  in which user association changes in order to maintain the connectivity of mobile users.  Meanwhile, handover management will result in signaling overhead in wireless networks \cite{zhang2011signalling}. Such handover signaling depends on the size of the network, network mobility (user and BS movements), locations of users and base stations, and handover rate \cite{zhang2011signalling,godor2015survey,Handover1}. In  UAV-based communication systems, handover management needs to be done in order to reduce the handover signaling and also to properly provide connectivity for flying UAVs in beyond visual LoS (BVLoS) scenarios. Handover management in UAV communications is significantly more challenging than traditional cellular networks due to the highly dynamic nature of drone-UEs and drone-BSs. In particular, efficient handover mechanisms must be designed to accommodate 3D movements of both drone-UEs a drone-BSs, while ensuring low-latency communications and control when serving drone-UEs. This handover design for flying devices must be done jointly with existing handover mechanisms for mobile ground users, such as vehicles.

Moreover, for drone-UEs, all of the aforementioned challenges must also take into account the fact that ground base stations will have their antennas downtilted to maximize coverage of ground users. As a result, it is imperative to understand the impact of antenna tilt on the performance of UAV-UEs, while also studying how one can overcome this limitation via adaptive beamforming or new UAV-UE aware design of ground base stations.

\subsubsection{State of the Art}
While the use case of UAV-BSs has been widely studied in the literature, there are only a handful of studies on drone-UEs scenarios. For example, the work in \cite{Coexistence} analyzed the coexistence of aerial and ground users in cellular networks. In particular, the authors in \cite{Coexistence} proposed a framework
for characterizing the downlink coverage performance in a network that includes drone-UEs and terrestrial-UEs. In \cite{azari2017reshaping}, the authors derived an exact expression for coverage probability of drone-UEs which are served by ground BSs. The work in \cite{LTE_Sky} analyzed the impact of both drone-BSs and drone-UEs on uplink and downlink performance of an LTE network. In \cite{lin2017sky}, the authors studied the feasibility of wireless connectivity for drone-UEs via LTE networks. Moreover, in \cite{lin2017sky}, propagation characteristics of BSs-to-drones communications was studied  using measurements
and ray tracing simulations. {\color{black}The work in \cite{challita2018cellular} developed an interference-aware path planning scheme for drone-UEs that yields a minimum communication latency of drones as well as their interference on ground users. In \cite{garcia2018essential}, the authors studied the potential use of massive MIMO for supporting drone-UEs with cellular networks. In particular, the work in \cite{garcia2018essential} studied the uplink and downlink performance of drone-UEs in coexistence with ground users, while utilizing massive MIMO in cellular networks. Finally, in \cite{Ramy}, we studied how various network parameters, such as downtilted antenna patterns and network structure, impact the performance of drone-UEs with caching capability.} \vspace{0.02cm} 

\subsubsection{Representative Result}
\begin{figure}[!t]
	\begin{center}
		\vspace{-0.2cm}
		\includegraphics[width=8.0cm]{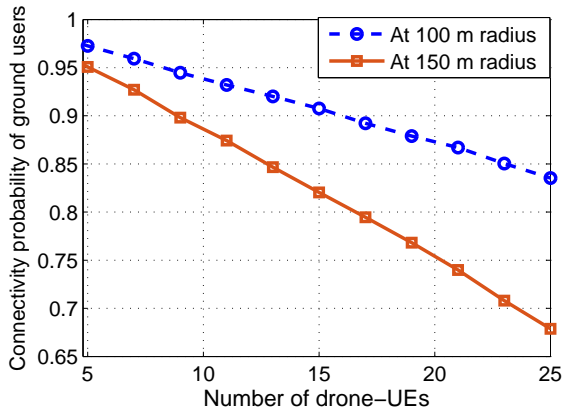}
		\vspace{-0.25cm}
		\caption{ \small Impact of drone-UEs on connectivity of ground users.}\vspace{-0.6cm}
		\label{Drone_UE}
	\end{center}
\end{figure}

Here, we show how uplink interference stemming from drone-UEs impact the connectivity of ground users. We consider a number of flying drone-UEs which are uniformly deployed on a disk of radius 1000\,m at an altitude 100\,m over a given geographical area.  Meanwhile, ground users attempt to connect to a ground base station located at the center of the area. Figure \ref{Drone_UE} shows the uplink connectivity probability of ground users (at a given radius from the base station) as the number of drone-UEs varies. Clearly, the connectivity of ground users decreases as the number of drones increases. This is due to the  dominant LoS interference caused by the drone-UEs. For instance, the connectivity probability at a 150\,m radius  decreases by 18\% when the number of drone-UEs increases from 5 to 15. Our result in Figure \ref{Drone_UE} highlights the need for adopting effective interference management techniques in drone-UEs scenarios \cite{mozaffari2,IoTJournal,OmidMatching,PantisanoInterf, ZhaoInterf}. \vspace{-0.02cm}

{\color{black}
\subsection{Summary of Lessons Learned}
In summary, the main lessons learned from this section include:
\begin{itemize}
	\item Despite  promising roles of UAVs in wireless networks, a number of design challenges need to be studied. In fact, each role has its own challenges and opportunities. For instance, for flying base stations, one prominent challenge is to maximize network performance under unique UAV features and constraints such as flight time, air-to-ground channel models, and mobility. The key challenges for cellular-connected UAV-UEs include co-existence with ground networks, mobility and handover management, and interference mitigation. Meanwhile, in flying ad-hoc networks, routing and path planning for UAVs are among important design challenges.


\item The design of UAV-enabled wireless networks is  affected by channel models used for air-to-ground  air-to-air communications.  Channel modeling in UAV communications is an important research direction and can be done  using various approaches such as ray-tracing technique, extensive measurements, and machine learning.

\item Optimizing the 3D locations of drones is a key design consideration as it significantly impacts the performance drone-enabled wireless networks.  Drone deployment is particularly of important in use cases for coverage and capacity enhancement, public safety, IoT applications, and caching. While optimizing the drones' positions, various factors such as A2G channel, users' locations, transmit power, and obstacles must be taken into account.

\item In order to optimize the trajectory of UAVs, several constraints and parameters must be considered. The UAV's trajectory is determined based on the users' QoS requirements, the UAV's energy consumption, type of the UAV, as well as shape and locations of obstacles in the environment.

\item Performance evaluation of a UAV-enabled wireless network is needed in order to capture key network design tradeoffs.  The performance of UAV communication systems can be analyzed in terms of various metrics such as coverage probability, area spectral efficiency, reliability, and latency. These metrics can be linked to unique UAV parameters such as its altitude, trajectory, and hover time. 

\item Network planning in a UAV-assisted wireless networks requires addressing various problems pertaining to aerial and terrestrial base station deployment, frequency planning, interference management, and user association. Network planning must be efficiently done so as to maximize the overall UAV system performance in terms of coverage, capacity, and operational costs.

\item Given the limited on-board energy of drones, the energy efficiency aspects of drone-based communication systems require careful consideration. In fact, the flight time and transmit power constraints of drones will significantly impact the performance of drone-enabled wireless networks. A drone's energy consumption can be minimized by developing energy-efficient deployment, path planning, and drone communication designs.

\item The use of flying UAV-UEs in a cellular-connected UAVs scenario introduces new challenges. For instance, traditional cellular networks with downtilted base station antennas  that have been primary designed for serving ground users, may not be able to effectively support connectivity and low-latency requirements of  UAV-UEs. In fact, there is need for designing an efficient cellular-connected UAV systems that can support ultra-reliable and low latency communications requirements, mobility and handover management, and seamless connectivity for flying UAV-UEs.
	
\end{itemize}
}

\begin{table*}\caption{Challenges, open problems, and tools for designing UAV-enabled wireless networks.\label{BigTable}}\vspace{-0.3cm}
	\begin{center}{\color{black}
		\begin{tabular}{ |l|l|l|l| } 
			\hline
			\textbf{Research Direction} & \textbf{Key References} & \textbf{Challenges and Open Problems} & \textbf{Mathematical Tools and Techniques} \\ \hline
			\multirow{3}{*}{Channel Modeling} & 	\multirow{3}{2.8cm}{\centering\cite{Zheng,Holis, FengPath, FengModelling,dan,HouraniModeling,HouraniOptimal, ITUR,Channel3D,UAVChannel2,UAVChannel3,Matolak,Matolak2017,Ismail_survey}}  &$\bullet$ Air-to-ground path loss. & $\bullet$ Ray-tracing techniques. \\
			&	&$\bullet$ Air-to-air channel modeling. &  $\bullet$  Machine learning.\\
			&  &$\bullet$ Small scale fading. & $\bullet$  Extensive measurements.	
			\\ \hline
			\multirow{3}{2.5cm}{Deployment} & \multirow{3}{2.8cm}{\centering\cite{Letter, Azari2, Kalantari,ALZ2,ALZ1, HouraniOptimal,Mozaffari,kosmerl,Daniel,zhan2006,de,Rohde,bor}}& $\bullet$ Deployment in presence of terrestrial networks.& $\bullet$ Centralized optimization theory.\\
			&& $\bullet$ Energy-aware deployment.&$\bullet$ Facility location theory.\\
			&&	$\bullet$ Joint 3D deployment and resource allocation.& 
			\\  \hline
			\multirow{3}{2.5cm}{Performance Analysis} &\multirow{3}{2.8cm}{ \centering \cite{VishnuJournal,Mozaffari,mozaffari2,Spectrum, Haya,RelayCommunications,H,mumtaz,guo}}& $\bullet$ Analyzing heterogeneous aerial-terrestrial networks. & $\bullet$ Probability theory.  \\
			&& $\bullet$ Performance analysis under mobility considerations. & $\bullet$ Stochastic geometry. \\
			&& $\bullet$ Capturing spatial and temporal correlations.& $\bullet$ Information theory\\ 
			\hline
			
			\multirow{3}{2.5cm}{Cellular Network Planning with UAVs} & 	\multirow{3}{3cm}{\centering \hspace{0.2cm} \cite{Vishal,Kalantari2,OTUAV, Boris_Back,Kalantari}}& $\bullet$ Backhaul-aware cell planning. & $\bullet$ Centralized optimization theory.\\
			&& $\bullet$ Optimizing number of UAVs.& $\bullet$ Facility location theory.\\
			&& $\bullet$ Traffic-based cell association.& $\bullet$ Optimal transport theory.\\
			&& $\bullet$ Analysis of signaling and overhead.&\\
			\hline
			
			\multirow{3}{2.8cm}{Resource Management and Energy Efficiency} & \multirow{3}{3cm}{\centering \hspace{0.4cm}\cite{ZhangEnergy, Cooperative, Uragun,Zorbas,ant,vin,Ceran,shar}}& $\bullet$ Bandwidth and flight time optimization.& $\bullet$ Centralized optimization theory.\\
			&& $\bullet$ Joint trajectory and transmit power optimization. & $\bullet$ Optimal transport theory.\\
			&& $\bullet$ Spectrum sharing with cellular networks.& $\bullet$ Game theory and machine learning.\\
			&& $\bullet$ Multi-dimensional resource management.&\\
			\hline
			
			\multirow{3}{2.5cm}{Trajectory Optimization} & \multirow{3}{3cm}{\centering\cite{zengThroughput, Qin, fran,gro,dou,tis,H,Jiang,Rucco, CooperativePath, FlightDemonstrations,Autonomous,chandler2000uav,WangTrajectory}} & $\bullet$ Energy-efficient trajectory optimization.&  $\bullet$ Centralized optimization theory. \\
			&& $\bullet$ Joint trajectory and delay optimization.&  $\bullet$ Machine learning.\\
			&& $\bullet$ Reliable communication with path planning.&\\
			\hline
			
			\multirow{4}{2.3cm}{Cellular Connected UAV-UEs} & \multirow{4}{3cm}{\centering\cite{Coexistence,azari2017reshaping,LTE_Sky,lin2017sky,challita2018cellular}} & $\bullet$ \textcolor{black}{Effective connectivity with downtilted ground base stations.}&  $\bullet$ Centralized optimization theory. \\
			&& $\bullet$ Interference management.&  $\bullet$ Machine learning.\\
			&& $\bullet$ Handover management.& $\bullet$ Optimal transport theory.\\
			&& $\bullet$ Ground-to-air channel modeling. & $\bullet$ Game theory.\\
			&& $\bullet$ Ultra reliable, low latency communication and control.& $\bullet$ Stochastic geometry.\\
			\hline		
			
		\end{tabular}}
	\end{center}
\end{table*}

\section{Open Problems and Future Opportunities for UAV-based Wireless Communication and Networking}\label{sec:Open}
In the previous section, we have outlined the general research directions and challenges of wireless communications with UAVs. The next natural step is to discuss open research problems in each one of the covered areas, in order to shed light on future opportunities, as done in this section. Despite a considerable number of studies on UAV communications, there are still many key open problems that must be investigated. 

\subsection{UAV Channel Modeling}
For air-to-ground channel modeling, there are several key open problems. First and foremost, there is a need for more realistic channel models that stem from real-world measurements \cite{Ismail_survey}. While efforts in this regard already started, most of them remain limited to a single UAV or to very specific environments. A broader campaign of channel measurements that can cut across urban and rural areas, as well as various operational environments (e.g., weather conditions) is needed. Such experimental work can complement the existing, mostly ray tracing simulation based results. Moreover, the simulation results can also be expanded to model small-scale fading A2G communications. In addition, as UAVs become more commonly used as flying base stations, drone-UEs, or even for backhaul support, one must have more insights on air-to-air channel modeling. In particular, there is a need for an accurate UAV-to-UAV channel model that can capture time-variation of channel and  Doppler effect due to mobility of UAVs. Furthermore, multipath fading in air-to-air communications needs to be characterized while considering UAVs' altitude as well as antennas' movement.   

\subsection{UAV Deployment}
In terms of open problems for UAV deployment, there is a need for new solutions to optimal 3D placement of UAVs while accounting for their unique features. For instance, one of the key open problems is the optimal 3D placement of UAVs in presence of terrestrial networks. For instance, there is a need to study how UAVs must be deployed in coexistence with cellular networks while considering mutual interference between such aerial and terrestrial systems. {\color{black} Other key open problems in deployment include: \vspace{0.03cm}

   1) \emph{Joint optimization of deployment and bandwidth allocation for low latency communications}: In order to minimize the maximum transmission latency of users which are served by drone-BSs, one problem is to jointly optimize the 3D locations of drone-BSs and bandwidth allocation. In particular, given  a number of drone-BSs, locations of users, and the total amount of bandwidth available for serving users,  one important open problem  is to find the optimal location of each drone-BS and its transmission bandwidth such that the maximum downlink transmission latency of the users is minimized. \vspace{0.05cm}

  2) \emph{Joint optimal 3D placement and cell association for flight time
  	minimization}: The flight time of a drone-BS that provides wireless services to users depends on many factors such as the load and number  of users connected to the drone-BS as well as the downlink transmission rate. In this problem, given the number of drone-BSs, the total flight time of drone-BSs needed for completely servicing users should be minimized by jointly optimizing the locations of drone-BSs and user-to-drone associations. \vspace{0.05cm}

  3) \emph{Obstacle aware deployment of UAVs for maximizing wireless coverage}: The coverage performance of drone-BSs that serve ground users is affected by obstacles. One key open problem here is to maximize the total coverage areas of drone-BSs by optimal placement of drone-BSs based on the locations of users and obstacles. In particular, given the locations of ground users and obstacles in the environment, the 3D positions of drone-BSs can be determined such that the maximum number of users are covered by drones. This is particularly useful if the drones operate at high frequency bands (e.g., at millimeter wave frequencies).}

 \subsection{\textcolor{black}{UAV Trajectory Optimization}}
 \textcolor{black}{While the potential mobility of UAVs provides promising opportunities, it introduces new challenges and technical problems.  In a UAV-assisted wireless network, the trajectory of UAVs needs to be optimized with respect to key performance metrics such as throughput, energy and spectral efficiency, and delay. Furthermore, trajectory optimization problems must account for the dynamic aspects and type of UAVs. While there has been a number of attractive studies on UAV trajectory optimization, there are still several open problems that include:  1) UAV trajectory optimization based on the mobility patterns of ground users for maximizing the coverage performance, 2) Obstacle aware trajectory optimization of UAVs considering users' delay constraints and UAVs' energy consumption,  3) Trajectory optimization for maximizing reliability and minimizing latency  in UAV-enabled wireless networks, and 4) Joint control, communication, and trajectory optimization of UAVs for flight time minimization. Finally, for cellular-connected UAV-UEs, optimizing trajectory while minimizing interference to the ground users and being cognizant of the downtilt of the antennas of the ground base stations is yet another open problem.} 
      
\subsection{Performance Analysis}
For performance analysis, there are numerous problems that can still be studied. For instance, one must completely characterize the performance of UAV-enabled wireless networks, that consist of both aerial and terrestrial users and base stations,  in terms of coverage and capacity. In particular, there is a need for tractable expressions for coverage probability and spectral efficiency in heterogeneous aerial-terrestrial networks. Moreover, fundamental  performance analysis needs to be done to capture inherent tradeoffs between spectral efficiency and and energy efficiency in UAV networks. Another open problem is to evaluate the performance of UAV-enabled wireless networks while incorporating the mobility of UAVs. The fundamental analysis of such mobile wireless networks involves capturing the spatial and temporal variations of various performance metrics in the network. For instance, there is a need to study how the trajectory of UAVs impacts their performance in terms of throughput, latency, and energy efficiency. Finally, the effect of dynamic scheduling on the performance of UAV communication systems can be analyzed. 

\subsection{Planning Cellular Networks with UAVs}
An efficient network planning with UAVs requires addressing a number of key problems. For example,  what is the minimum number of UAVs needed to provide a full coverage for given a geographical area that is partially covered by ground base stations. Solving such problems is particularity challenging when the geographical area of interest does not have a regular geometric shape (e.g., disk or square). Another design problem is the backhaul-aware deployment of UAVs while using them as aerial base stations. In this case, while deploying UAV-BSs, one must consider both the backhaul connectivity of UAVs and their users' quality-of-service.  Other important open problems include: 1) performing efficient frequency planning when both ground and aerial BSs and users exist, 2) developing new approaches to dynamically provision UAVs on the fly whenever they join network, and 3) designing robust and adaptive network planning techniques that can account for highly mobile drone-UEs. Last but not least, it is imperative to analyze the signaling overhead associated with the deployment of both UAV-BSs and UAV-UEs, while characterizing how that overhead can affect the performance.

\subsection{Resource Management in UAV Networks}
Resource management is another key research problem in UAV-based communication systems. In particular, there is a need for a framework that can dynamically manage various resources including bandwidth, energy, transmit power, UAV's flight time, and number of UAVs, among others. For instance, how to adaptively adjust the transmit power and trajectory  of a flying UAV that serves ground users. In this case, a key problem is to provide optimal bandwidth allocation mechanisms that can capture the impact of UAVs' locations, mobility, LoS interference, and traffic distribution of ground users.  Also, there is a need for designing efficient scheduling techniques to mitigate interference between aerial and terrestrial base stations in a UAV-assisted cellular network. In addition, one must analyze dynamic spectrum sharing in a heterogeneous network of both flying and ground base stations. Finally, adopting suitable frequency bands (e.g., WiFi, LTE bands) for UAV operations is of important design problems.

\subsection{Drone-UEs Scenarios}     
Naturally, flying drones that act as users within cellular networks can introduce new design challenges. In particular, while using drone-UEs in wireless networks, one must account for mobility, LoS interference, handover, energy constraints, and low-latency control of drones. In this regard, key open problems in drone-UEs communications include: 1) developing robust interference mitigation techniques for  massive drone-UEs deployment scenarios, 2) designing dynamic handover mechanisms to manage frequent handovers due to mobility, 3) providing accurate ground-to-air channel models for BSs-to-drone communications, 4) proposing new scheduling schemes while considering battery limitations of drones, 5) designing effective solutions that allow meeting URLLC requirements for drone-UEs, and 6) analyzing application-specific quality-of-service measures.

{\color{black}

\subsection{Lessons Learned}
 Despite the notable number of works on UAV-based wireless communications, there
	are many fundamental open problems that needs be studied. Key open problems in UAV networks exist in various areas such as comprehensive channel model for UAV communications, energy-aware deployment, analysis of signaling and overhead, reliable communications with path planning, low latency control, interference and handover management. 
}
\section{Analytical Frameworks to Enable UAV-based Communications}\label{sec:frameworks}
\begin{figure}[!t]
	\begin{center}
		\vspace{-0.2cm}
		\includegraphics[width=9.0cm]{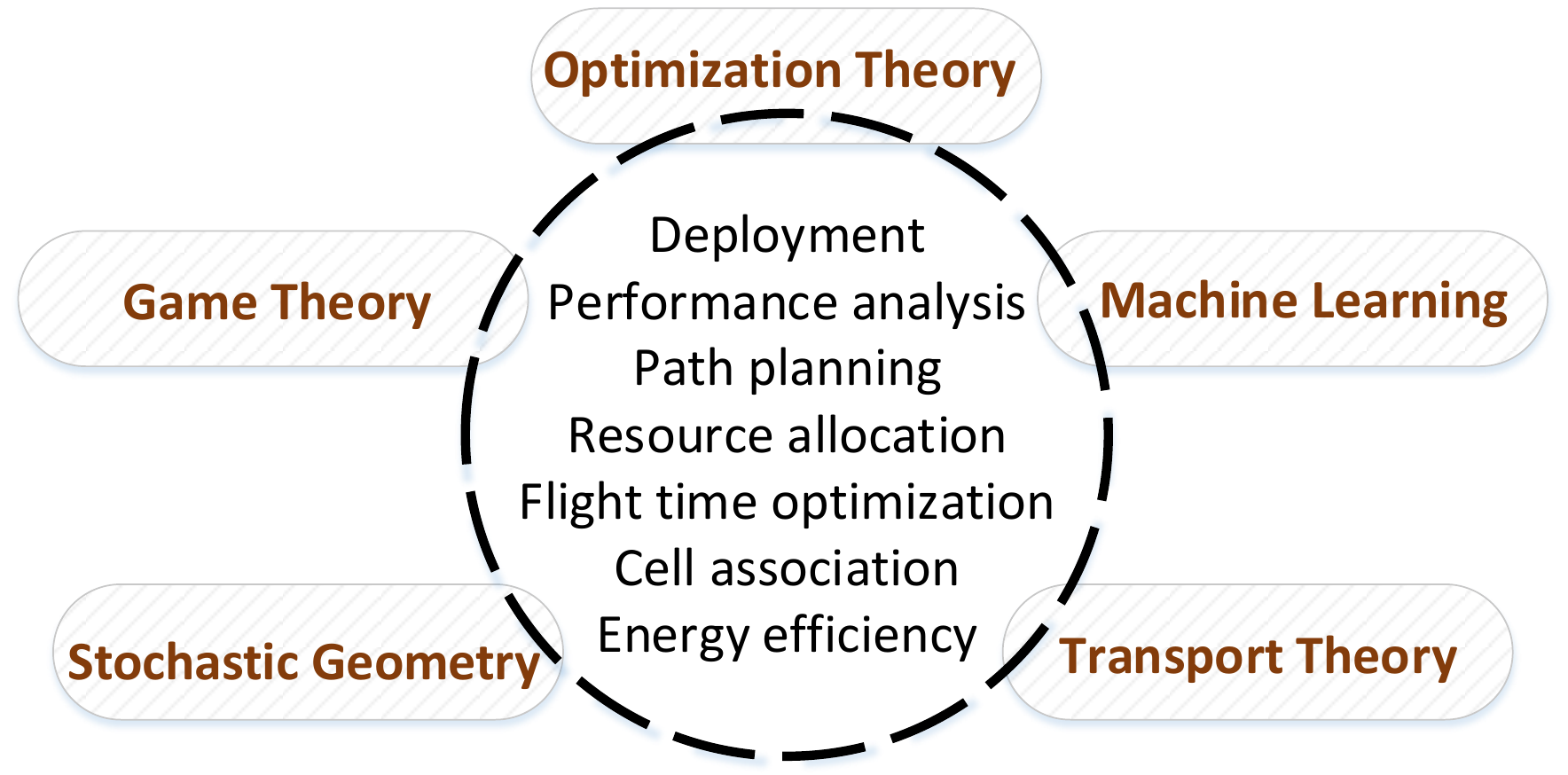}
		\vspace{-0.1cm}
		\caption{ \small Mathematical tools for designing UAV communication systems.}\vspace{-0.3cm}
		\label{Tools}
	\end{center}
\end{figure}

Having identified the research directions and their associated challenges and open problems, next, we turn our attention to the analytical frameworks needed to design, analyze, and optimize the use of UAVs for wireless networking purposes. Indeed, this research area is highly interdisciplinary and it will require drawing on tools from conventional fields such as communication theory, optimization theory, and network design, as well as emerging fields such as stochastic geometry, machine learning, and game theory, as listed in Figure \ref{Tools}.

\subsection{Centralized Optimization Theory for UAV Communication}
During the first phase of deployment of UAVs as flying base stations, despite their inherent autonomy, we envision that UAVs will initially rely on centralized control. This is particularly important for applications such as cellular network capacity enhancement, in which cellular operators may not be willing to relinquish control of their network during the early trials of a technology such as UAVs. In such scenarios, many of the identified research problems will very naturally involve the need to formulate and solve challenging centralized optimization problems. Such problems can be run at the level of a cloud (e.g., as is done in a cloud-assisted radio access network) \cite{CRAN} or at the level of a ground macrocell base station that is capable to control some of the UAVs.

It is worth noting that lessons learned from conventional terrestrial cellular network optimization problem can prove to be very handy in UAV communication. For example, classical approaches such as successive convex optimization \cite{sequentialnonlinear} can be used for optimizing the 3D location and trajectory of UAVs. However, many of the problems identified here will require more advanced optimization techniques. For example, when analyzing user association problems, one will naturally end up with challenging mixed integer programming problems, that cannot be solved using traditional algorithms, such as those used for convex optimization. In this regard, advanced mathematical tools such as optimal transport theory \cite{villani2003} can provide tractable solutions for a wide range of cell association problems that seek to optimize UAV's flight time, throughput, and energy-efficiency of UAV-enabled wireless networks.

\subsection{Optimal Transport Theory for UAV Networks}
Optimal transport theory \cite{villani2003} can enable deriving
tractable solutions for the notoriously difficult optimization problems that accompany the problems of user association, resource allocation, and flight time optimization in UAV-enabled wireless networks. By exploiting new ideas from probability theory and statistics, optimal transport theory enables capturing generic distributions of wireless devices, which, in turn, allows a much deeper fundamental analysis of network performance optimization than existing heuristic works. Optimal transport is a field in mathematics that studies scenarios in which
goods are transported between various locations. 

One popular example is the so-called ore mining problem. In this illustrative example, we are given a collection of mines mining iron ore, and a collection of factories which
consume the iron ore that the mines produce. The goal is to find the optimal way to transport (move) the ore from the mines to the factories, to minimize a certain cost function that captures key factors such as the costs of transportation, the location of the mines, and the productivity of the factories. Optimal transport theory aims to find an optimal
mapping between any two arbitrary probability measures. In particular, in a semi-discrete optimal
transport problem, a continuous probability density function must be mapped to a discrete
probability measure.  

Remarkably, such mathematical framework can be used to solve a number of complex problems in UAV communications. For instance, in a semi-discrete optimal transport case, the optimal transport map will optimally
partition the continuous distribution and assign each partition to one point in the discrete
probability measure. Clearly, such optimal
partitions can be considered as optimal cell association in UAV-to-user (in UAV base station scenarios) and BS-to-UAV (in drone-UE cases) communications. Therefore, within
the framework of optimal transport theory, one can address cell association problems for any
general spatial distribution of users. In fact, optimal transport theory enables the derivation of tractable solutions to variety of user association resource allocation, energy management, and flight optimization problems in UAV-enabled wireless networks. In particular, given any spatial distribution of ground users (that can be estimated  using UAV-based aerial imaging), one can exploit optimal transport theory to derive the optimal cell association and resource management schemes that lead to the maximum system performance in terms of energy efficiency, throughput, and delay under explicit flight time constraints of UAVs \cite {MozaffariFlightTime, Letter_OT}.

\subsection{Performance Analysis using Stochastic Geometry}
Stochastic geometry techniques have emerged as powerful tools for performance analysis of ad-hoc and cellular networks \cite{haenggi}. The key principle is to endow the locations devices, e.g., users and base stations, as a point process, and then evaluate key performance metrics such as coverage, rate, throughput, or delay. While stochastic geometry  has been utilized for the analysis of two-dimensional heterogeneous cellular networks, it can be potentially adopted to characterize the performance of 3D UAV networks \cite{VishnuJournal}. Nevertheless, one must use tractable and realistic point processes to model the locations of UAVs. For instance, the Binomial and Poisson cluster processes \cite{baccelli2010stochastic} are more suitable when UAVs are deployed at user hotspots, and the goal is to serve a massive number of users in a specific area. The processes with repulsion between points, e.g., Matern hard core process \cite{haenggi}, is more suitable for the a case in which UAVs are not allowed to be closer than a certain distance. Therefore, by exploiting tools from stochastic geometry and adopting a suitable point process model, the performance of UAV-enabled wireless networks can be characterized. This, in turn, can reveal the key design insights and inherent tradeoffs in UAV communications.  


\subsection{Machine Learning}
Machine learning enables systems to improve their performance by automatically learning from their environment and their past experience. Machine learning can be potentially leveraged to design and optimize UAV-based wireless communication systems \cite{MingzheTutorial,Ferd}. For instance, using reinforcement learning algorithms, drones can dynamically adjust their positions, flight
directions,  and motion control to service their ground users. In this case, drones are able to rapidly adapt to dynamic environments in a  self-organizing way, and autonomously optimize their trajectory.  In addition, by leveraging neural networks techniques and  performing data analytics,  one can predict the ground users' behavior and effectively deploy and operate drones. For example, machine learning tools enable predicting users' mobility and their load distribution that can be used to perform optimal deployment and path planning of drones. Such information about users' mobility pattern and traffic distribution is particularity useful in designing cache-enabled drone systems. Machine learning can also be used to learn the radio environment maps and to build a 3D channel model using UAVs. Such radio environment maps can be subsequently used to optimally deploy and operate UAV communication systems.

\subsection{Game Theory}
Distributed decision making will be an integral component of UAV networks. As such, along with the use of machine learning, game theory~\cite{C1,C2} will provide important foundations for distributed decision making in UAV-based wireless networks. Game theory is a natural tool to analyze resource management and trajectory optimization problems in which the decision is done at the level of each UAV. In such cases, each UAV will have its own, individual objective function that captures its own QoS. Here, the inherent coupling of the UAVs objective functions due to factors such as interference or collisions, strongly motivate the use of game-theoretic analysis for resource management. In a UAV-enabled network, distributed resource management problems will now involve different types of players (UAVs, BSs, UEs), as well as multi-dimensional strategy spaces that include energy, spectrum, hover/flight times, and 3D locations. This, in turn, will motivate the use of advanced game-theoretic mechanisms, such as the emerging notion of a multi-game~\cite{C3}, that go beyond classical game-theoretic constructs that are used for conventional terrestrial resource management problems \cite{GCBG2018}. In particular, multi-games allow capturing the fact that, in a UAV network, multiple games may co-exist, such as a game among UAVs and a game among terrestrial BSs, and, as such, analysis of such multi-game scenarios is needed.

Moreover, when UAVs are supposed to operate autonomously, it is imperative to jointly optimize their communication and control systems. Such an optimization must be distributed and done at the level of each individual, autonomous UAV, thus again motivating the use of game theory. Here,  stochastic differential games~\cite{TB} will be an important tool since they can naturally integrate both communication and control, whereby communication objectives can be included in utility functions while the control system dynamics can be posed as differential equation constraints. Moreover, the sheer scale of ultra dense cellular networks with a massive number of UAVs will require tools to analyze the asymptotic performance of the system. To this end, tools from mean-field game theory \cite{C4,C5,C6} are useful to perform such large-system analysis and gain insights on how energy efficiency, spectrum efficiency, and the overall network QoS can scale with the number of users.  

Moreover, cooperative behavior is another important aspect of UAV communications. For instance, how to dynamically form swarms of UAVs and enable their coordination is an important open problem. To address it, one can leverage tools from coalitional game theory, such as those developed in \cite{A,B,C} for wireless networks, in general, and in \cite{D,E}, for UAV systems, in particular. Other relevant game-theoretic tools include contract theory~\cite{C7}, to design incentive mechanisms and matching theory~\cite{C8} to study network planning problems. In addition, multiple synergies between machine learning, optimal transport theory, optimization theory, and game theory can be built and analyzed for a variety of problems in UAV communication systems.

{\color{black}
\subsection{Lessons Learned and Summary}

In Table \ref{BigTable}, we summarize the key challenges, open problems, important references, and analytical tools to analyze, optimize, and design UAV-enabled wireless networks.

 In summary, in order to address the fundamental challenges in UAV communication systems and efficiently use UAVs for  wireless networking applications, we need to leverage various mathematical tools. In this regard, the following mathematical tools can be utilized: 1) Optimization theory can be used for addressing problems related to deployment and path planning, 2) Stochastic geometry for performance analysis, 3) Optimal transport theory for cell association and load balancing problems, 4) Machine learning for motion control and channel modeling, and 5) Game theory for resource management and trajectory optimization problems.
}

\section{Concluding Remarks}
In this tutorial, we have provided a comprehensive study on the use of UAVs in wireless networks. We have investigated two main use cases of UAVs, namely, aerial base stations and cellular-connected users, i.e., UAV-UEs. For each use case of UAVs, we have explored key challenges, applications, and fundamental open problems. Moreover, we have presented the major state of \textcolor{black} {the} art pertaining to challenges in UAV-enabled wireless networks, along with insightful representative results. Meanwhile,  we have described mathematical tools and techniques needed for meeting UAV challenges as well as analyzing UAV-enabled wireless networks. Such an in-depth study on UAV communication and networking provides unique guidelines for optimizing, designing, and operating UAV-based wireless communication systems.  \vspace{0.24cm}

\def\baselinestretch{1.031}
\bibliographystyle{IEEEtran}

\bibliography{references}

\begin{thebibliography}{100}
\providecommand{\url}[1]{#1}
\csname url@samestyle\endcsname
\providecommand{\newblock}{\relax}
\providecommand{\bibinfo}[2]{#2}
\providecommand{\BIBentrySTDinterwordspacing}{\spaceskip=0pt\relax}
\providecommand{\BIBentryALTinterwordstretchfactor}{4}
\providecommand{\BIBentryALTinterwordspacing}{\spaceskip=\fontdimen2\font plus
\BIBentryALTinterwordstretchfactor\fontdimen3\font minus
  \fontdimen4\font\relax}
\providecommand{\BIBforeignlanguage}[2]{{%
\expandafter\ifx\csname l@#1\endcsname\relax
\typeout{** WARNING: IEEEtran.bst: No hyphenation pattern has been}%
\typeout{** loaded for the language `#1'. Using the pattern for}%
\typeout{** the default language instead.}%
\else
\language=\csname l@#1\endcsname
\fi
#2}}
\providecommand{\BIBdecl}{\relax}
\BIBdecl

\bibitem{HandbookUAV}
K.~P. Valavanis and G.~J. Vachtsevanos, \emph{Handbook of unmanned aerial
  vehicles}.\hskip 1em plus 0.5em minus 0.4em\relax Springer Publishing
  Company, Incorporated, 2014.

\bibitem{austin2011unmanned}
R.~Austin, \emph{Unmanned aircraft systems: {UAVS} design, development and
  deployment}.\hskip 1em plus 0.5em minus 0.4em\relax John Wiley \& Sons, 2011,
  vol.~54.

\bibitem{hanscomunmanned}
R.~W. Beard and T.~W. McLain, \emph{Small unmanned aircraft: Theory and
  practice}.\hskip 1em plus 0.5em minus 0.4em\relax Princeton university press,
  2012.

\bibitem{Micro}
M.~Asadpour, B.~V. den Bergh, D.~Giustiniano, K.~A. Hummel, S.~Pollin, and
  B.~Plattner, ``Micro aerial vehicle networks: an experimental analysis of
  challenges and opportunities,'' \emph{IEEE Communications Magazine}, vol.~52,
  no.~7, pp. 141--149, July 2014.

\bibitem{stansbury2008survey}
R.~S. Stansbury, M.~A. Vyas, and T.~A. Wilson, ``A survey of {UAS} technologies
  for command, control, and communication ({C3}),'' in \emph{Unmanned Aircraft
  Systems}.\hskip 1em plus 0.5em minus 0.4em\relax Springer, 2008, pp. 61--78.

\bibitem{puri2005survey}
A.~Puri, ``A survey of unmanned aerial vehicles ({UAV}) for traffic
  surveillance,'' \emph{Department of computer science and engineering,
  University of South Florida}, 2005.

\bibitem{IoTJournal}
M.~Mozaffari, W.~Saad, M.~Bennis, and M.~Debbah, ``Mobile unmanned aerial
  vehicles ({UAV}s) for energy-efficient {Internet of Things} communications,''
  \emph{IEEE Transactions on Wireless Communications}, vol.~16, no.~11, pp.
  7574--7589, Nov. 2017.

\bibitem{Irem}
R.~Yaliniz, A.~El-Keyi, and H.~Yanikomeroglu, ``Efficient 3-{D} placement of an
  aerial base station in next generation cellular networks,'' in \emph{Proc. of
  IEEE International Conference on Communications (ICC)}, Kuala Lumpur,
  Malaysia, May. 2016.

\bibitem{Bucaille}
I.~Bucaille, S.~Hethuin, A.~Munari, R.~Hermenier, T.~Rasheed, and S.~Allsopp,
  ``Rapidly deployable network for tactical applications: Aerial base station
  with opportunistic links for unattended and temporary events absolute
  example,'' in \emph{Proc. of IEEE Military Communications Conference
  (MILCOM)}, San Diego, CA, USA, Nov. 2013.

\bibitem{mozaffari2}
M.~{Mozaffari}, W.~Saad, M.~Bennis, and M.~Debbah, ``Unmanned aerial vehicle
  with underlaid device-to-device communications: Performance and tradeoffs,''
  \emph{IEEE Transactions on Wireless Communications}, vol.~15, no.~6, pp.
  3949--3963, June 2016.

\bibitem{HouraniOptimal}
A.~Hourani, K.~Sithamparanathan, and S.~Lardner, ``Optimal {LAP} altitude for
  maximum coverage,'' \emph{IEEE Wireless Communication Letters}, vol.~3,
  no.~6, pp. 569--572, Dec. 2014.

\bibitem{Mozaffari}
M.~{Mozaffari}, W.~Saad, M.~Bennis, and M.~Debbah, ``Drone small cells in the
  clouds: Design, deployment and performance analysis,'' in \emph{Proc. of IEEE
  Global Communications Conference (GLOBECOM)}, San Diego, CA, USA, Dec. 2015.

\bibitem{Letter}
{M. Mozaffari}, W.~Saad, M.~Bennis, and M.~Debbah, ``Efficient deployment of
  multiple unmanned aerial vehicles for optimal wireless coverage,'' \emph{IEEE
  Communications Letters}, vol.~20, no.~8, pp. 1647--1650, Aug. 2016.

\bibitem{zhang}
Y.~Zeng, R.~Zhang, and T.~J. Lim, ``Wireless communications with unmanned
  aerial vehicles: opportunities and challenges,'' \emph{IEEE Communications
  Magazine}, vol.~54, no.~5, pp. 36--42, May 2016.

\bibitem{bor}
I.~Bor-Yaliniz and H.~Yanikomeroglu, ``The new frontier in ran heterogeneity:
  Mui-tier drone-cells,'' \emph{IEEE Communications Magazine}, vol.~54, no.~11,
  pp. 48--55, 2016.

\bibitem{Rohde}
S.~Rohde and C.~Wietfeld, ``Interference aware positioning of aerial relays for
  cell overload and outage compensation,'' in \emph{Proc. of IEEE Vehicular
  Technology Conference (VTC)}, Quebec, QC, Canada, Sept. 2012.

\bibitem{yanmaz2018drone}
E.~Yanmaz, S.~Yahyanejad, B.~Rinner, H.~Hellwagner, and C.~Bettstetter, ``Drone
  networks: Communications, coordination, and sensing,'' \emph{Ad Hoc
  Networks}, vol.~68, pp. 1--15, 2018.

\bibitem{Sky}
Facebook, ``Connecting the world from the sky,'' Facebook Technical Report,
  2014.

\bibitem{GoogleLoon}
K.~Kamnani and C.~Suratkar, ``A review paper on {Google Loon} technique,''
  \emph{International Journal of Research In Science \& Engineering}, vol.~1,
  no.~1, pp. 167--171, 2015.

\bibitem{wu2018uav}
Q.~{Wu}, J.~Xu, and R.~Zhang, ``{UAV}-enabled aerial base station {(BS)
  III/III}: Capacity characterization of {UAV}-enabled two-user broadcast
  channel,'' \emph{available online: arxiv.org/abs/1801.00443}, 2018.

\bibitem{wu2018common}
Q.~{Wu} and R.~Zhang, ``Common throughput maximization in {UAV}-enabled {OFDMA}
  systems with delay consideration,'' \emph{available online:
  arxiv.org/abs/1801.00444}, 2018.

\bibitem{al2015internet}
A.~Al-Fuqaha, M.~Guizani, M.~Mohammadi, M.~Aledhari, and M.~Ayyash, ``Internet
  of things: A survey on enabling technologies, protocols, and applications,''
  \emph{IEEE Communications Surveys \& Tutorials}, vol.~17, no.~4, pp.
  2347--2376, 2015.

\bibitem{PA00}
T.~Park, N.~Abuzainab, and W.~Saad, ``Learning how to communicate in the
  {Internet of Things}: Finite resources and heterogeneity,'' \emph{IEEE
  Access}, vol.~4, Nov. 2016.

\bibitem{IoT2014}
A.~Zanella, N.~Bui, A.~Castellani, L.~Vangelista, and M.~Zorzi, ``{Internet of
  Things} for smart cities,'' \emph{IEEE Internet of Things Journal}, vol.~1,
  no.~1, pp. 22--32, Feb. 2014.

\bibitem{ferdowsi2017deep}
A.~Ferdowsi and W.~Saad, ``Deep learning-based dynamic watermarking for secure
  signal authentication in the {Internet of Things},'' in \emph{Proc. of IEEE
  International Conference on Communications (ICC)}, Kansas City, MO, USA, May
  2018.

\bibitem{Ding1}
G.~Ding, Q.~Wu, L.~Zhang, Y.~Lin, T.~A. Tsiftsis, and Y.~D. Yao, ``An amateur
  drone surveillance system based on the cognitive {Internet of Things},''
  \emph{IEEE Communications Magazine}, vol.~56, no.~1, pp. 29--35, Jan. 2018.

\bibitem{QualcomUAV}
``Paving the path to {5G}: Optimizing commercial {LTE} networks for drone
  communication,'' \emph{available online:
  https://www.qualcomm.com/news/onq/2016/09/06/paving-path-5g-optimizing-commercial-lte-networks-drone-communication.}

\bibitem{stewart2014google}
J.~Stewart, ``{Google} tests drone deliveries in project wing trials,''
  \emph{BBC World Service Radio}, 2014.

\bibitem{HouraniModeling}
A.~Hourani, S.~Kandeepan, and A.~Jamalipour, ``Modeling air-to-ground path loss
  for low altitude platforms in urban environments,'' in \emph{Proc. of IEEE
  Global Telecommunications Conference (GLOBECOM)}, Austin, TX, USA, Dec. 2014.

\bibitem{FAA}
D.~Gettinger and A.~H. Michel, ``Drone sightings and close encounters: An
  analysis,'' \emph{Center for the Study of the Drone, Bard College,
  Annandale-on-Hudson, NY, USA}, 2015.

\bibitem{fotouhi2018survey}
A.~Fotouhi, H.~Qiang, M.~Ding, M.~Hassan, L.~G. Giordano, A.~Garcia-Rodriguez,
  and J.~Yuan, ``Survey on uav cellular communications: Practical aspects,
  standardization advancements, regulation, and security challenges,''
  \emph{available online: arxiv.org/abs/1809.01752}, 2018.

\bibitem{stocker2017review}
C.~St{\"o}cker, R.~Bennett, F.~Nex, M.~Gerke, and J.~Zevenbergen, ``Review of
  the current state of {UAV} regulations,'' \emph{Remote sensing}, vol.~9,
  no.~5, p. 459, 2017.

\bibitem{AkramMagazin}
S.~Chandrasekharan, K.~Gomez, A.~Al-Hourani, S.~Kandeepan, T.~Rasheed,
  L.~Goratti, L.~Reynaud, D.~Grace, I.~Bucaille, T.~Wirth, and S.~Allsopp,
  ``Designing and implementing future aerial communication networks,''
  \emph{IEEE Communications Magazine}, vol.~54, no.~5, pp. 26--34, May 2016.

\bibitem{ALZ1}
M.~Alzenad, A.~El-Keyi, F.~Lagum, and H.~Yanikomeroglu, ``{3-D} placement of an
  unmanned aerial vehicle base station {(UAV-BS)} for energy-efficient maximal
  coverage,'' \emph{IEEE Wireless Communications Letters}, vol.~6, no.~4, pp.
  434--437, Aug. 2017.

\bibitem{ALZ2}
M.~Alzenad, A.~El-Keyi, and H.~Yanikomeroglu, ``{3-D} placement of an unmanned
  aerial vehicle base station for maximum coverage of users with different
  {QoS} requirements,'' \emph{IEEE Wireless Communications Letters}, vol.~7,
  no.~1, pp. 38--41, Feb. 2018.

\bibitem{Qin}
Q.~Wu, Y.~Zeng, and R.~Zhang, ``Joint trajectory and communication design for
  multi-uav enabled wireless networks,'' \emph{IEEE Transactions on Wireless
  Communications, Early access}, 2018.

\bibitem{Azari}
A.~M. Hayajneh, S.~A.~R. Zaidi, D.~C. McLernon, and M.~Ghogho, ``Drone
  empowered small cellular disaster recovery networks for resilient smart
  cities,'' in \emph{Proc. of IEEE International Conference on Sensing,
  Communication and Networking (SECON Workshops)}, June 2016.

\bibitem{VshalUAV}
V.~Sharma, R.~Sabatini, and S.~Ramasamy, ``{UAVs} assisted delay optimization
  in heterogeneous wireless networks,'' \emph{IEEE Communications Letters},
  vol.~20, no.~12, pp. 2526--2529, Dec. 2016.

\bibitem{Lyu}
J.~Lyu, Y.~Zeng, R.~Zhang, and T.~J. Lim, ``Placement optimization of
  {UAV}-mounted mobile base stations,'' \emph{IEEE Communications Letters},
  vol.~21, no.~3, pp. 604--607, March 2017.

\bibitem{Jeong}
S.~Jeong, O.~Simeone, and J.~Kang, ``Mobile edge computing via a {UAV}-mounted
  cloudlet: Optimal bit allocation and path planning,'' \emph{IEEE Transactions
  on Vehicular Technology, Early access}, 2017.

\bibitem{MozaffariFlightTime}
M.~Mozaffari, W.~Saad, M.~Bennis, and M.~Debbah, ``Wireless communication using
  unmanned aerial vehicles ({UAVs}): Optimal transport theory for hover time
  optimization,'' \emph{IEEE Transactions on Wireless Communications}, vol.~16,
  no.~12, pp. 8052--8066, Dec. 2017.

\bibitem{Complition}
Y.~Zeng, X.~Xu, and R.~Zhang, ``Trajectory optimization for completion time
  minimization in {UAV}-enabled multicasting,'' \emph{available online:
  arxiv.org/abs/1708.06478}, 2017.

\bibitem{Proactive}
P.~Yang, X.~Cao, C.~Yin, Z.~Xiao, X.~Xi, and D.~Wu, ``Proactive drone-cell
  deployment: Overload relief for a cellular network under flash crowd
  traffic,'' \emph{IEEE Transactions on Intelligent Transportation Systems},
  vol.~18, no.~10, pp. 2877--2892, Oct. 2017.

\bibitem{Flying1}
M.~A. Khan, A.~Safi, I.~M. Qureshi, and I.~U. Khan, ``Flying ad-hoc networks
  ({FANETs}): A review of communication architectures, and routing protocols,''
  in \emph{Proc. of IEEE First International Conference on Latest trends in
  Electrical Engineering and Computing Technologies (INTELLECT)}, Karachi,
  Pakistan, Nov. 2017.

\bibitem{Flying3}
W.~Zafar and B.~M. Khan, ``Flying ad-hoc networks: Technological and social
  implications,'' \emph{IEEE Technology and Society Magazine}, vol.~35, no.~2,
  pp. 67--74, June 2016.

\bibitem{bekmezci2013flying}
I.~Bekmezci, O.~K. Sahingoz, and {\c{S}}.~Temel, ``Flying ad-hoc networks
  ({FANETs}): A survey,'' \emph{Ad Hoc Networks}, vol.~11, no.~3, pp.
  1254--1270, 2013.

\bibitem{sahingoz2014networking}
O.~K. Sahingoz, ``Networking models in flying ad-hoc networks ({FANETs}):
  Concepts and challenges,'' \emph{Journal of Intelligent \& Robotic Systems},
  vol.~74, no. 1-2, pp. 513--527, 2014.

\bibitem{Low}
N.~H. Motlagh, T.~Taleb, and O.~Arouk, ``Low-altitude unmanned aerial
  vehicles-based internet of things services: Comprehensive survey and future
  perspectives,'' \emph{IEEE Internet of Things Journal}, vol.~3, no.~6, pp.
  899--922, Dec. 2016.

\bibitem{Airborne}
X.~Cao, P.~Yang, M.~Alzenad, X.~Xi, D.~Wu, and H.~Yanikomeroglu, ``Airborne
  communication networks: A survey,'' \emph{IEEE Journal on Selected Areas in
  Communications, Early access}, 2018.

\bibitem{karapantazis2005broadband}
S.~Karapantazis and F.~Pavlidou, ``Broadband communications via high-altitude
  platforms: a survey,'' \emph{IEEE Communications Surveys \& Tutorials},
  vol.~7, no.~1, pp. 2--31, 2005.

\bibitem{sekander2018multi}
S.~Sekander, H.~Tabassum, and E.~Hossain, ``Multi-tier drone architecture for
  {5G/B5G} cellular networks: Challenges, trends, and prospects,'' \emph{IEEE
  Communications Magazine}, vol.~56, no.~3, pp. 96--103, 2018.

\bibitem{hayat2016survey}
S.~Hayat, E.~Yanmaz, and R.~Muzaffar, ``Survey on unmanned aerial vehicle
  networks for civil applications: A communications viewpoint.'' \emph{IEEE
  Communications Surveys and Tutorials}, vol.~18, no.~4, pp. 2624--2661, 2016.

\bibitem{gupta2016survey}
L.~Gupta, R.~Jain, and G.~Vaszkun, ``Survey of important issues in {UAV}
  communication networks,'' \emph{IEEE Communications Surveys \& Tutorials},
  vol.~18, no.~2, pp. 1123--1152, 2016.

\bibitem{LTE_Sky}
B.~V.~D. Bergh, A.~Chiumento, and S.~Pollin, ``{LTE} in the sky: trading off
  propagation benefits with interference costs for aerial nodes,'' \emph{IEEE
  Communications Magazine}, vol.~54, no.~5, pp. 44--50, May 2016.

\bibitem{Ismail_survey}
W.~Khawaja, I.~Guvenc, D.~Matolak, U.-C. Fiebig, and N.~Schneckenberger, ``A
  survey of air-to-ground propagation channel modeling for unmanned aerial
  vehicles,'' \emph{available online: arxiv.org/abs/1801.01656}, 2018.

\bibitem{Samarakoon}
S.~Samarakoon, M.~Bennis, W.~Saad, M.~Debbah, and M.~Latva-aho, ``Ultra dense
  small cell networks: Turning density into energy efficiency,'' \emph{IEEE
  Journal on Selected Areas in Communications}, vol.~34, no.~5, pp. 1267--1280,
  May 2016.

\bibitem{Omid1}
O.~Semiari, W.~Saad, M.~Bennis, and Z.~Dawy, ``Inter-operator resource
  management for millimeter wave multi-hop backhaul networks,'' \emph{IEEE
  Transactions on Wireless Communications}, vol.~16, no.~8, pp. 5258--5272,
  Aug. 2017.

\bibitem{Omid2}
O.~Semiari, W.~Saad, and M.~Bennis, ``Joint millimeter wave and microwave
  resources allocation in cellular networks with dual-mode base stations,''
  \emph{IEEE Transactions on Wireless Communications}, vol.~16, no.~7, pp.
  4802--4816, July 2017.

\bibitem{Contract}
Y.~Zhang, L.~Song, W.~Saad, Z.~Dawy, and Z.~Han, ``Contract-based incentive
  mechanisms for device-to-device communications in cellular networks,''
  \emph{IEEE Journal on Selected Areas in Communications}, vol.~33, no.~10, pp.
  2144--2155, Oct. 2015.

\bibitem{ContextOmid}
O.~Semiari, W.~Saad, S.~Valentin, M.~Bennis, and H.~V. Poor, ``Context-aware
  small cell networks: How social metrics improve wireless resource
  allocation,'' \emph{IEEE Transactions on Wireless Communications}, vol.~14,
  no.~11, pp. 5927--5940, Nov. 2015.

\bibitem{Absolute}
I.~Bucaille, S.~Hethuin, A.~Munari, R.~Hermenier, T.~Rasheed, and S.~Allsopp,
  ``Rapidly deployable network for tactical applications: Aerial base station
  with opportunistic links for unattended and temporary events absolute
  example,'' in \emph{Military Communications Conference, MILCOM 2013-2013
  IEEE}.\hskip 1em plus 0.5em minus 0.4em\relax IEEE, 2013, pp. 1116--1120.

\bibitem{OffloadingLyu}
J.~Lyu, Y.~Zeng, and R.~Zhang, ``{UAV}-aided offloading for cellular hotspot,''
  \emph{IEEE Transactions on Wireless Communications}, vol.~17, no.~6, pp.
  3988--4001, June 2018.

\bibitem{ATDrone}
``{AT\&T} detail network testing of drones in football stadiums,''
  \emph{available online:
  https://www.androidheadlines.com/2016/09/att-detail-network-testing-of-drones-in-football-stadiums.html.}

\bibitem{Gomez}
K.~Gomez, A.~Hourani, L.~Goratti, R.~Riggio, S.~Kandeepan, and I.~Bucaille,
  ``Capacity evaluation of aerial {LTE} base-stations for public safety
  communications,'' in \emph{Proc. IEEE European Conference on Networks and
  Communications (EuCNC)}, June 2015.

\bibitem{PublicSafety}
G.~Baldini, S.~Karanasios, D.~Allen, and F.~Vergari, ``Survey of wireless
  communication technologies for public safety,'' \emph{IEEE Communications
  Surveys Tutorials}, vol.~16, no.~2, pp. 619--641, Second 2014.

\bibitem{Ismail}
A.~Merwaday and I.~Guvenc, ``{UAV} assisted heterogeneous networks for public
  safety communications,'' in \emph{Proc. of IEEE Wireless Communications and
  Networking Conference Workshops (WCNCW)}, March 2015.

\bibitem{orsino2017effects}
A.~Orsino, A.~Ometov, G.~Fodor, D.~Moltchanov, L.~Militano, S.~Andreev, O.~N.
  Yilmaz, T.~Tirronen, J.~Torsner, G.~Araniti \emph{et~al.}, ``Effects of
  heterogeneous mobility on {D2D}-and drone-assisted mission-critical {MTC} in
  {5G},'' \emph{IEEE Communications Magazine}, vol.~55, no.~2, pp. 79--87,
  2017.

\bibitem{Nam2013}
Y.~H. Nam, B.~L. Ng, K.~Sayana, Y.~Li, J.~Zhang, Y.~Kim, and J.~Lee,
  ``Full-dimension {MIMO} ({FD-MIMO}) for next generation cellular
  technology,'' \emph{IEEE Communications Magazine}, vol.~51, no.~6, pp.
  172--179, June 2013.

\bibitem{3GPP36897}
{3GPP}, ``Study on elevation beamforming/full-dimension ({FD}) {MIMO} for
  {LTE},'' \emph{TR 36.897}, May 2017.

\bibitem{lee3D}
W.~Lee, S.-R. Lee, H.-B. Kong, and I.~Lee, ``{3D} beamforming designs for
  single user {MISO} systems,'' in \emph{Proc. of IEEE Global Communications
  Conference (GLOBECOM)}, Atlanta, GA, USA, Dec. 2013.

\bibitem{namMIMO}
Y.-H. Nam, M.~S. Rahman, Y.~Li, G.~Xu, E.~Onggosanusi, J.~Zhang, and J.-Y.
  Seol, ``Full dimension {MIMO} for {LTE}-advanced and {5G},'' in \emph{Proc.
  of Information Theory and Applications Workshop ({ITA})}, San Diego, CA, USA,
  Feb. 2015.

\bibitem{sha}
M.~Shafi, M.~Zhang, P.~J. Smith, A.~L. Moustakas, and A.~F. Molisch, ``The
  impact of elevation angle on {MIMO} capacity,'' in \emph{Proc. of IEEE
  International Conference on Communications}, vol.~9, Istanbul, Turkey, June
  2006.

\bibitem{cheng}
X.~Cheng, B.~Yu, L.~Yang, J.~Zhang, G.~Liu, Y.~Wu, and L.~Wan, ``Communicating
  in the real world: {3D MIMO},'' \emph{IEEE Wireless Communications magazine},
  vol.~21, no.~4, pp. 136--144, 2014.

\bibitem{Li}
Y.~Li, X.~Ji, D.~Liang, and Y.~Li, ``Dynamic beamforming for three-dimensional
  {MIMO} technique in {LTE}-advanced networks,'' \emph{International Journal of
  Antennas and Propagation}, vol. 2013, 2013.

\bibitem{3GPP36777}
{3GPP}, ``Enhanced {LTE} support for aerial vehicles,'' \emph{TR 36.777}, May
  2017.

\bibitem{CommControl}
M.~Mozaffari, W.~Saad, M.~Bennis, and M.~Debbah, ``Communications and control
  for wireless drone-based antenna array,'' \emph{IEEE Transactions on
  Communications}, vol.~67, no.~1, pp. 820--834, Jan. 2019.

\bibitem{3GPP38811}
{3GPP}, ``Study on nr to support non-terrestrial networks,'' \emph{TR 38.811},
  Jan. 2018.

\bibitem{IsmailmmW}
N.~Rupasinghe, Y.~Yapici, I.~Guvenc, and Y.~Kakishima, ``Non-orthogonal
  multiple access for {mmWave} drones with multi-antenna transmission,''
  \emph{available online: arxiv.org/abs/1711.10050}, 2017.

\bibitem{tork}
E.~Torkildson, H.~Zhang, and U.~Madhow, ``Channel modeling for millimeter wave
  {MIMO},'' in \emph{Proc. of Information Theory and Applications Workshop
  (ITA), 2010}.\hskip 1em plus 0.5em minus 0.4em\relax IEEE, 2010, pp. 1--8.

\bibitem{IoTVision}
J.~Gubbi, R.~Buyya, S.~Marusic, and M.~Palaniswami, ``{Internet of Things}
  ({IoT}): A vision, architectural elements, and future directions,''
  \emph{Future generation computer systems}, vol.~29, no.~7, pp. 1645--1660,
  2013.

\bibitem{dawy}
Z.~Dawy, W.~Saad, A.~Ghosh, J.~G. Andrews, and E.~Yaacoub, ``Toward massive
  machine type cellular communications,'' \emph{IEEE Wireless Communications},
  vol.~24, no.~1, pp. 120--128, Feb. 2017.

\bibitem{lien}
S.-Y. Lien, K.-C. Chen, and Y.~Lin, ``Toward ubiquitous massive accesses in
  3gpp machine-to-machine communications,'' \emph{IEEECommunications Magazine},
  vol.~49, no.~4, pp. 66--74, April. 2011.

\bibitem{pang}
Y.~Pang, Y.~Zhang, Y.~Gu, M.~Pan, Z.~Han, and P.~Li, ``Efficient data
  collection for wireless rechargeable sensor clusters in harsh terrains using
  {UAV}s,'' in \emph{Proc. of IEEE Global Communications Conference
  (GLOBECOM)}, Austin, TX, USA, Dec. 2014.

\bibitem{MehdiM2M}
M.~N. Soorki, M.~Mozaffari, W.~Saad, M.~H. Manshaei, and H.~Saidi, ``Resource
  allocation for machine-to-machine communications with unmanned aerial
  vehicles,'' in \emph{IEEE Globecom Workshops (GC Wkshps)}, Dec. 2016.

\bibitem{ProactiveCaching2016}
J.~Qiao, Y.~He, and S.~Shen, ``Proactive caching for mobile video streaming in
  millimeter wave {5G} networks,'' \emph{IEEE Transactions on Wireless
  Communications}, vol.~15, no.~10, pp. 7187--7198, Oct. 2016.

\bibitem{Tran2016Octopus}
T.~X. Tran and D.~Pompili, ``Octopus: A cooperative hierarchical caching
  strategy for cloud radio access networks,'' in \emph{Proc. of IEEE
  International Conference on Mobile Ad Hoc and Sensor Systems ({MASS})},
  Brasilia, Brazil, Oct. 2016, pp. 154--162.

\bibitem{guo2015cooperative}
Y.~Guo, L.~Duan, and R.~Zhang, ``Cooperative local caching under heterogeneous
  file preferences,'' \emph{IEEE Transactions on Communications}, vol.~65,
  no.~1, pp. 444--457, Jan. 2017.

\bibitem{bacstug2015cache}
E.~Bastug, M.~Bennis, M.~Kountouris, and M.~Debbah, ``Cache-enabled small cell
  networks: {M}odeling and tradeoffs,'' \emph{EURASIP J. Wireless Commun.
  Netw.,Special Issue Tech. Adv. Design Deployment Future Heterogeneous Netw.},
  vol. 2015, no.~1, Feb 2015.

\bibitem{ye2016tradeoff}
Z.~Ye, C.~Pan, H.~Zhu, and J.~Wang, ``Tradeoff caching strategy of outage
  probability and fronthaul usage in {C}loud-{RAN},'' \emph{available online:
  arxiv.org/abs/1611.02660}, Nov. 2016.

\bibitem{mingzhe}
{M. Chen}, M.~Mozaffari, W.~Saad, C.~Yin, M.~Debbah, and C.~S. Hong, ``Caching
  in the sky: Proactive deployment of cache-enabled unmanned aerial vehicles
  for optimized quality-of-experience,'' \emph{IEEE Journal on Selected Areas
  in Communications}, vol.~35, no.~5, pp. 1046--1061, May 2017.

\bibitem{Ding2}
H.~Wang, G.~Ding, F.~Gao, J.~Chen, J.~Wang, and L.~Wang, ``Power control in
  {UAV}-supported ultra dense networks: Communications, caching, and energy
  transfer,'' \emph{available online: arxiv.org/abs/1712.05004}, 2017.

\bibitem{Ramy}
R.~Amer, W.~Saad, H.~ElSawy, M.~Butt, and N.~Marchetti, ``Caching to the sky:
  Performance analysis of cache-assisted {CoMP} for cellular-connected
  {UAVs},'' in \emph{Proc. of IEEE Wireless Communications and Networking
  Conference (WCNC), Wireless Networks Track}, Marrakech, Morocco,, 2019.

\bibitem{bamburry2015drones}
D.~Bamburry, ``Drones: Designed for product delivery,'' \emph{Design Management
  Review}, vol.~26, no.~1, pp. 40--48, 2015.

\bibitem{mozaffari2018beyond}
M.~Mozaffari, A.~T.~Z. Kasgari, W.~Saad, M.~Bennis, and M.~Debbah, ``Beyond
  {5G} with {UAVs}: Foundations of a {3D} wireless cellular network,''
  \emph{IEEE Transactions on Wireless Communications}, vol.~18, no.~1, pp.
  357--372, Jan. 2019.

\bibitem{densification}
N.~Bhushan, J.~Li, D.~Malladi, R.~Gilmore, D.~Brenner, A.~Damnjanovic, R.~T.
  Sukhavasi, C.~Patel, and S.~Geirhofer, ``Network densification: the dominant
  theme for wireless evolution into {5G},'' \emph{IEEE Communications
  Magazine}, vol.~52, no.~2, pp. 82--89, Feb. 2014.

\bibitem{Ge}
X.~Ge, S.~Tu, G.~Mao, C.~X. Wang, and T.~Han, ``{5G} ultra-dense cellular
  networks,'' \emph{IEEE Wireless Communications}, vol.~23, no.~1, pp. 72--79,
  Feb. 2016.

\bibitem{MmWave}
Z.~Gao, L.~Dai, D.~Mi, Z.~Wang, M.~A. Imran, and M.~Z. Shakir, ``Mmwave
  massive-{MIMO}-based wireless backhaul for the {5G} ultra-dense network,''
  \emph{IEEE Wireless Communications}, vol.~22, no.~5, pp. 13--21, Oct. 2015.

\bibitem{WirelessBackhaul}
U.~Siddique, H.~Tabassum, E.~Hossain, and D.~I. Kim, ``Wireless backhauling of
  {5G} small cells: challenges and solution approaches,'' \emph{IEEE Wireless
  Communications}, vol.~22, no.~5, pp. 22--31, Oct. 2015.

\bibitem{ursula}
U.~Challita and W.~Saad, ``Network formation in the {Sky}: Unmanned aerial
  vehicles for multi-hop wireless backhauling,'' in \emph{Proc. of IEEE Global
  Telecommunications Conference (GLOBECOM)}, Singapore, Dec. 2017.

\bibitem{ferdowsi2017colonel}
A.~Ferdowsi, W.~Saad, and N.~B. Mandayam, ``{Colonel Blotto} game for secure
  state estimation in interdependent critical infrastructure,'' \emph{available
  online: arxiv.org/abs/1709.09768}, 2017.

\bibitem{GesbertMap}
J.~Chen, U.~Yatnalli, and D.~Gesbert, ``Learning radio maps for {UAV}-aided
  wireless networks: A segmented regression approach,'' in \emph{Proc. of IEEE
  International Conference on Communications (ICC)}, Paris, France, May 2017.

\bibitem{zaj}
A.~Zaji{\'c}, \emph{Mobile-to-mobile wireless channels}.\hskip 1em plus 0.5em
  minus 0.4em\relax Artech House, 2012.

\bibitem{Zheng}
Y.~Zheng, Y.~Wang, and F.~Meng, ``Modeling and simulation of pathloss and
  fading for air-ground link of {HAP}s within a network simulator,'' in
  \emph{Proc. of IEEE International Conference on Cyber-Enabled Distributed
  Computing and Knowledge Discovery (CyberC)}, Beijing, China, Oct. 2013.

\bibitem{Holis}
J.~Holis and P.~Pechac, ``Elevation dependent shadowing model for mobile
  communications via high altitude platforms in built-up areas,'' \emph{IEEE
  Transactions on Antennas and Propagation}, vol.~56, no.~4, pp. 1078--1084,
  April 2008.

\bibitem{yun2015ray}
Z.~Yun and M.~F. Iskander, ``Ray tracing for radio propagation modeling:
  principles and applications,'' \emph{IEEE Access}, vol.~3, pp. 1089--1100,
  2015.

\bibitem{Matolak}
D.~W. Matolak, ``Air-ground channels amp; models: Comprehensive review and
  considerations for unmanned aircraft systems,'' in \emph{Proc. of IEEE
  Aerospace Conference}, Big Sky, MT, USA, Mar. 2012.

\bibitem{Matolak2017}
D.~W. Matolak and R.~Sun, ``Air–ground channel characterization for unmanned
  aircraft systems—part i: Methods, measurements, and models for over-water
  settings,'' \emph{IEEE Transactions on Vehicular Technology}, vol.~66, no.~1,
  pp. 26--44, Jan. 2017.

\bibitem{FengModelling}
Q.~Feng, E.~K. Tameh, A.~R. Nix, and J.~McGeehan, ``Modelling the likelihood of
  line-of-sight for air-to-ground radio propagation in urban environments,'' in
  \emph{Proc. of IEEE Global Telecommunications Conference (GLOBECOM)}, San
  Diego, CA, USA, Nov. 2006.

\bibitem{Channel3D}
K.~Daniel, M.~Putzke, B.~Dusza, and C.~Wietfeld, ``Three dimensional channel
  characterization for low altitude aerial vehicles,'' in \emph{Proc. of IEEE
  International Symposium on Wireless Communication Systems}, Sep. 2010, pp.
  756--760.

\bibitem{UAVChannel2}
E.~Yanmaz, R.~Kuschnig, and C.~Bettstetter, ``Channel measurements over
  802.11a-based {UAV}-to-ground links,'' in \emph{Proc. IEEE GLOBECOM Workshops
  (GC Wkshps)}, Dec. 2011.

\bibitem{UAVChannel3}
K.~Sasloglou, I.~A. Glover, V.~Gazis, P.~Kikiras, K.~Mathioudakis, and
  I.~Andonovic, ``Empirical channel models for optimized communications in a
  network of unmanned ground vehicles,'' in \emph{Proc. IEEE International
  Symposium on Signal Processing and Information Technology}, Dec. 2013.

\bibitem{Bettstetter}
E.~Yanmaz, R.~Kuschnig, and C.~Bettstetter, ``Achieving air-ground
  communications in 802.11 networks with three-dimensional aerial mobility,''
  in \emph{Proc. of IEEE INFOCOM}, Turin, Italy, April 2013.

\bibitem{Kalantari}
E.~Kalantari, H.~Yanikomeroglu, and A.~Yongacoglu, ``On the number and {3D}
  placement of drone base stations in wireless cellular networks,'' in
  \emph{Proc. of IEEE Vehicular Technology Conference}, 2016.

\bibitem{Jacob}
H.~Shakhatreh, A.~Khreishah, J.~Chakareski, H.~B. Salameh, and I.~Khalil, ``On
  the continuous coverage problem for a swarm of {UAV}s,'' in \emph{Proc. of
  IEEE 37th Sarnoff Symposium}, Sep. 2016, pp. 130--135.

\bibitem{Azari2}
M.~M. Azari, F.~Rosas, K.~C. Chen, and S.~Pollin, ``Joint sum-rate and power
  gain analysis of an aerial base station,'' in \emph{Proc. of IEEE GLOBECOM
  Workshops}, Dec. 2016.

\bibitem{Haya}
A.~M. Hayajneh, S.~A.~R. Zaidi, D.~C. McLernon, and M.~Ghogho, ``Optimal
  dimensioning and performance analysis of drone-based wireless
  communications,'' in \emph{Proc. of IEEE GLOBECOM Workshops}, Dec. 2016.

\bibitem{jia}
S.~Jia and Z.~Lin, ``Modeling unmanned aerial vehicles base station in
  ground-to-air cooperative networks,'' \emph{IET Communications}, 2017.

\bibitem{ITUR}
ITU-R, ``Rec. p.1410-2 propagation data and prediction methods for the design
  of terrestrial broadband millimetric radio access systems,'' \emph{Series,
  Radiowave propagation}, 2003.

\bibitem{kosmerl}
J.~Kosmerl and A.~Vilhar, ``Base stations placement optimization in wireless
  networks for emergency communications,'' in \emph{Proc. of IEEE International
  Conference on Communications (ICC)}, Sydney, Australia, June. 2014.

\bibitem{Kalantari2}
E.~Kalantari, M.~Z. Shakir, H.~Yanikomeroglu, and A.~Yongacoglu,
  ``Backhaul-aware robust {3D} drone placement in {5G+} wireless networks,'' in
  \emph{Proc. of IEEE International Conference on Communications Workshops (ICC
  Workshops)}, May 2017, pp. 109--114.

\bibitem{E}
W.~Saad, Z.~Han, T.~Ba\c{s}ar, M.~Debbah, and A.~Hj{\o}rungnes, ``A selfish
  approach to coalition formation among unmanned air vehicles in wireless
  networks,'' in \emph{Proc. of the International Conference on Game Theory for
  Networks (GameNets)}, 2009, pp. 259--267.

\bibitem{Daniel}
K.~Daniel and C.~Wietfeld, ``Using public network infrastructures for {UAV}
  remote sensing in civilian security operations,'' DTIC Document, Tech. Rep.,
  Mar. 2011.

\bibitem{zhan2006}
P.~Zhan, K.~Yu, and A.~L. Swindlehurst, ``Wireless relay communications using
  an unmanned aerial vehicle,'' in \emph{Proc. IEEE 7th Workshop on Signal
  Processing Advances in Wireless Communications}, Cannes, France, July 2006.

\bibitem{de}
E.~P. De~Freitas, T.~Heimfarth, I.~F. Netto, C.~E. Lino, C.~E. Pereira, A.~M.
  Ferreira, F.~R. Wagner, and T.~Larsson, ``{UAV} relay network to support wsn
  connectivity,'' in \emph{Proc. of International Congress on Ultra Modern
  Telecommunications and Control Systems and Workshops (ICUMT)}.\hskip 1em plus
  0.5em minus 0.4em\relax IEEE, 2010, pp. 309--314.

\bibitem{orfanus}
D.~Orfanus, E.~P. de~Freitas, and F.~Eliassen, ``Self-organization as a
  supporting paradigm for military {UAV} relay networks,'' \emph{IEEE
  Communications Letters}, vol.~20, no.~4, pp. 804--807, 2016.

\bibitem{gaspar2000upper}
Z.~G{\'a}sp{\'a}r and T.~Tarnai, ``Upper bound of density for packing of equal
  circles in special domains in the plane,'' \emph{Periodica Polytechnica.
  Civil Engineering}, vol.~44, no.~1, p.~13, 2000.

\bibitem{dou}
K.~Do{\u{g}}an{\c{c}}ay, ``{UAV} path planning for passive emitter
  localization,'' \emph{IEEE Transactions on Aerospace and Electronic Systems},
  vol.~48, no.~2, pp. 1150--1166, 2012.

\bibitem{Rucco}
A.~Rucco, A.~P. Aguiar, and J.~Hauser, ``Trajectory optimization for
  constrained {UAVs}: A virtual target vehicle approach,'' in \emph{Proc. IEEE
  International Conference on Unmanned Aircraft Systems (ICUAS)}, June 2015.

\bibitem{CooperativePath}
J.~S. Bellingham, M.~Tillerson, M.~Alighanbari, and J.~P. How, ``Cooperative
  path planning for multiple {UAV}s in dynamic and uncertain environments,'' in
  \emph{Proc. IEEE Conference on Decision and Control}, Dec. 2002.

\bibitem{FlightDemonstrations}
J.~How, Y.~Kuwata, and E.~King, ``Flight demonstrations of cooperative control
  for {UAV} teams,'' in \emph{AIAA 3rd" Unmanned Unlimited" Technical
  Conference, Workshop and Exhibit}, 2004, p. 6490.

\bibitem{Autonomous}
J.~Tisdale, Z.~Kim, and J.~K. Hedrick, ``Autonomous {UAV} path planning and
  estimation,'' \emph{IEEE Robotics Automation Magazine}, vol.~16, no.~2, pp.
  35--42, June 2009.

\bibitem{chandler2000uav}
P.~Chandler, S.~Rasmussen, and M.~Pachter, ``{UAV} cooperative path planning,''
  in \emph{AIAA Guidance, Navigation, and Control Conference and Exhibit},
  2000, p. 4370.

\bibitem{Jiang}
F.~Jiang and A.~L. Swindlehurst, ``Optimization of {UAV} heading for the
  ground-to-air uplink,'' \emph{IEEE Journal on Selected Areas in
  Communications}, vol.~30, no.~5, pp. 993--1005, June 2012.

\bibitem{zengThroughput}
Y.~Zeng, R.~Zhang, and T.~J. Lim, ``Throughput maximization for {UAV}-enabled
  mobile relaying systems,'' \emph{IEEE Transactions on Communications},
  vol.~64, no.~12, pp. 4983--4996, Dec. 2016.

\bibitem{fran}
C.~D. Franco and G.~Buttazzo, ``Energy-aware coverage path planning of
  {UAVs},'' in \emph{Proc. of IEEE International Conference on Autonomous Robot
  Systems and Competitions (ICARSC)}, Vila Real, Portugal, April 2015, pp.
  111--117.

\bibitem{gro}
E.~I. Gr{\o}tli and T.~A. Johansen, ``Path planning for {UAVs} under
  communication constraints using splat! and milp,'' \emph{Journal of
  Intelligent \& Robotic Systems}, vol.~65, no. 1-4, pp. 265--282, 2012.

\bibitem{tis}
J.~Tisdale, Z.~Kim, and J.~K. Hedrick, ``Autonomous {UAV} path planning and
  estimation,'' \emph{IEEE Robotics \& Automation Magazine}, vol.~16, no.~2,
  pp. 35--42, 2009.

\bibitem{H}
Z.~Han, A.~L. Swindlehurst, and K.~Liu, ``Optimization of {MANET} connectivity
  via smart deployment/movement of unmanned air vehicles,'' \emph{IEEE
  Transactions on Vehicular Technology}, vol.~58, no.~7, pp. 3533--3546, Dec.
  2009.

\bibitem{MozaffariGPS}
M.~Mozaffari, A.~Broumandan, K.~O'Keefe, and G.~Lachapelle, ``Weak {GPS} signal
  acquisition using antenna diversity,'' \emph{Navigation}, vol.~62, no.~3, pp.
  205--218, 2015.

\bibitem{3GPP}
{3GPP}, ``Study on {RAN} improvements for machine type communication,''
  \emph{TR 37.868}, Sept. 2011.

\bibitem{SudheeshLetter}
P.~Sudheesh, M.~Mozaffari, M.~Magarini, W.~Saad, and P.~Muthuchidambaranathan,
  ``Sum-rate analysis for high altitude platform ({HAP}) drones with tethered
  balloon relay,'' \emph{IEEE Communications Letters}, vol.~22, no.~6, pp.
  1240--1243, 2018.

\bibitem{Performance2011}
A.~I. Alshbatat and L.~Dong, ``Performance analysis of mobile ad hoc unmanned
  aerial vehicle communication networks with directional antennas,''
  \emph{International Journal of Aerospace Engineering}, vol. 2010, 2011.

\bibitem{guo}
W.~Guo, C.~Devine, and S.~Wang, ``Performance analysis of micro unmanned
  airborne communication relays for cellular networks,'' in \emph{Proc. of IEEE
  International Symposium on Communication Systems, Networks \& Digital Signal
  Processing (CSNDSP)}, Manchester, UK, July 2014, pp. 658--663.

\bibitem{RelayCommunications}
P.~Zhan, K.~Yu, and A.~L. Swindlehurst, ``Wireless relay communications with
  unmanned aerial vehicles: Performance and optimization,'' \emph{IEEE
  Transactions on Aerospace and Electronic Systems}, vol.~47, no.~3, pp.
  2068--2085, July 2011.

\bibitem{VishnuJournal}
V.~V. Chetlur and H.~S. Dhillon, ``Downlink coverage analysis for a finite
  {3-D} wireless network of unmanned aerial vehicles,'' \emph{IEEE Transactions
  on Communications}, vol.~65, no.~10, pp. 4543--4558, Oct. 2017.

\bibitem{Spectrum}
C.~Zhang and W.~Zhang, ``Spectrum sharing for drone networks,'' \emph{IEEE
  Journal on Selected Areas in Communications}, vol.~35, no.~1, pp. 136--144,
  Jan. 2017.

\bibitem{mumtaz}
S.~Mumtaz, S.~Huq, K.~Mohammed, A.~Radwan, J.~Rodriguez, and R.~L. Aguiar,
  ``Energy efficient interference-aware resource allocation in {LTE-D2D}
  communication,'' in \emph{Proc. of IEEE International Conference on
  Communications (ICC)}, Sydney, Australia, June. 2014.

\bibitem{haenggi}
M.~Haenggi, \emph{Stochastic geometry for wireless networks}.\hskip 1em plus
  0.5em minus 0.4em\relax Cambridge University Press, 2012.

\bibitem{lee}
N.~Lee, X.~Lin, J.~G. Andrews, and R.~Heath, ``Power control for {D2D}
  underlaid cellular networks: Modeling, algorithms, and analysis,'' \emph{IEEE
  Journal on Selected Areas in Communications}, vol.~33, no.~1, pp. 1--13, Feb.
  2015.

\bibitem{XX}
X.~Xu, W.~Saad, X.~Zhang, X.~Xu, and S.~Zhou, ``Joint deployment of small cells
  and wireless backhaul links in next-generation networks,'' \emph{IEEE
  Communications Letters}, vol.~19, no.~12, pp. 2250--2253, Dec. 2015.

\bibitem{horwath2007experimental}
J.~Horwath, N.~Perlot, M.~Knapek, and F.~Moll, ``Experimental verification of
  optical backhaul links for high-altitude platform networks: Atmospheric
  turbulence and downlink availability,'' \emph{International Journal of
  Satellite Communications and Networking}, vol.~25, no.~5, pp. 501--528, 2007.

\bibitem{fidler}
F.~Fidler, M.~Knapek, J.~Horwath, and W.~R. Leeb, ``Optical communications for
  high-altitude platforms,'' \emph{IEEE Journal of selected topics in quantum
  electronics}, vol.~16, no.~5, pp. 1058--1070, 2010.

\bibitem{HalimBackhaul}
M.~Alzenad, M.~Z. Shakir, H.~Yanikomeroglu, and M.~Alouini, ``{FSO}-based
  vertical backhaul/fronthaul framework for {5G+} wireless networks,''
  \emph{IEEE Communications Magazine}, vol.~56, no.~1, pp. 218--224, Jan. 2018.

\bibitem{Vishal}
V.~Sharma, M.~Bennis, and R.~Kumar, ``{UAV}-assisted heterogeneous networks for
  capacity enhancement,'' \emph{IEEE Communications Letters}, vol.~20, no.~6,
  pp. 1207--1210, June 2016.

\bibitem{OTUAV}
{M. Mozaffari}, W.~Saad, M.~Bennis, and M.~Debbah, ``Optimal transport theory
  for power-efficient deployment of unmanned aerial vehicles,'' in \emph{Proc.
  of IEEE International Conference on Communications (ICC)}, May 2016.

\bibitem{Letter_OT}
M.~Mozaffari, W.~Saad, M.~Bennis, and M.~Debbah, ``Optimal transport theory for
  cell association in {UAV}-enabled cellular networks,'' \emph{IEEE
  Communications Letters}, vol.~21, no.~9, pp. 2053--2056, Sep. 2017.

\bibitem{FarajLetter}
F.~Lagum, I.~Bor-Yaliniz, and H.~Yanikomeroglu, ``Strategic densification with
  {UAV-BSs} in cellular networks,'' \emph{IEEE Wireless Communications Letters,
  Early access}, 2017.

\bibitem{Boris_Back}
B.~Galkin, J.~Kibi{\l}da, and L.~A. DaSilva, ``Backhaul for low-altitude {UAV}s
  in urban environments,'' May 2018.

\bibitem{Talebi1}
A.~{Taleb Zadeh Kasgari}, W.~Saad, and M.~Debbah, ``Brain-aware wireless
  networks: Learning and resource management,'' in \emph{Proc. of IEEE Asilomar
  Conference on Signals, Systems and Computers}, Pacific Grove, CA, USA, Nov.
  2017.

\bibitem{Talebi2}
A.~{Taleb Zadeh Kasgari} and W.~Saad, ``Stochastic optimization and control
  framework for {5G} network slicing with effective isolation,'' in \emph{Proc.
  of Annual Conference on Information Sciences and Systems (CISS)}, Princeton,
  USA, Mar. 2018.

\bibitem{PantisanoSpectrum}
F.~Pantisano, M.~Bennis, W.~Saad, and M.~Debbah, ``Spectrum leasing as an
  incentive towards uplink macrocell and femtocell cooperation,'' \emph{IEEE
  Journal on Selected Areas in Communications}, vol.~30, no.~3, pp. 617--630,
  April 2012.

\bibitem{Uragun}
B.~Uragun, ``Energy efficiency for unmanned aerial vehicles,'' in \emph{Proc.
  of IEEE 10th International Conference on Machine Learning and Applications
  and Workshops (ICMLA)}, vol.~2, Honolulu, HI, USA, Dec. 2011, pp. 316--320.

\bibitem{ZhangEnergy}
Y.~Zeng and R.~Zhang, ``Energy-efficient {UAV} communication with trajectory
  optimization,'' \emph{IEEE Transactions on Wireless Communications}, vol.~16,
  no.~6, pp. 3747--3760, June 2017.

\bibitem{Cooperative}
T.~X. Tran, A.~Hajisami, and D.~Pompili, ``Cooperative hierarchical caching in
  {5G} cloud radio access networks,'' \emph{IEEE Network}, vol.~31, no.~4, pp.
  35--41, July 2017.

\bibitem{Zorbas}
D.~Zorbas, T.~Razafindralambo, F.~Guerriero \emph{et~al.}, ``Energy efficient
  mobile target tracking using flying drones,'' \emph{Procedia Computer
  Science}, vol.~19, pp. 80--87, June. 2013.

\bibitem{ant}
S.~R. Anton and D.~J. Inman, ``Performance modeling of unmanned aerial vehicles
  with on-board energy harvesting,'' in \emph{SPIE Smart Structures and
  Materials+ Nondestructive Evaluation and Health Monitoring}.\hskip 1em plus
  0.5em minus 0.4em\relax International Society for Optics and Photonics, 2011,
  pp. 79\,771H--79\,771H.

\bibitem{ZhangLetter}
J.~Lyu, Y.~Zeng, and R.~Zhang, ``Cyclical multiple access in {UAV}-aided
  communications: A throughput-delay tradeoff,'' \emph{available online:
  arxiv.org/abs/1608.03180}, 2016.

\bibitem{shar}
M.~S. Sharawi, D.~N. Aloi, O.~Rawashdeh \emph{et~al.}, ``Design and
  implementation of embedded printed antenna arrays in small {UAV} wing
  structures,'' \emph{IEEE Transactions on Antennas and Propagation}, vol.~58,
  no.~8, pp. 2531--2538, 2010.

\bibitem{Ceran}
E.~T. Ceran, T.~Erkilic, E.~Uysal-Biyikoglu, T.~Girici, and K.~Leblebicioglu,
  ``Optimal energy allocation policies for a high altitude flying wireless
  access point,'' \emph{Transactions on Emerging Telecommunications
  Technologies}, vol.~28, no.~4, 2017.

\bibitem{Mingzhe_LTE}
{M. Chen}, W.~Saad, and C.~Yin, ``Liquid state machine learning for resource
  allocation in a network of cache-enabled {LTE-U UAVs},'' in \emph{Proc. of
  Global Communications Conference (GLOBECOM)}, Singapore, Dec. 2017.

\bibitem{zeng2018energy}
Y.~Zeng, J.~Xu, and R.~Zhang, ``Energy minimization for wireless communication
  with rotary-wing {UAV},'' \emph{available online: arxiv.org/abs/1804.02238},
  2018.

\bibitem{chen2017virtual}
M.~{Chen}, W.~Saad, and C.~Yin, ``Virtual reality over wireless networks:
  quality-of-service model and learning-based resource management,''
  \emph{available online: arxiv.org/abs/1703.04209}, 2017.

\bibitem{JacobVR}
J.~Chakareski, ``Aerial {UAV-IoT} sensing for ubiquitous immersive
  communication and virtual human teleportation,'' in \emph{2017 IEEE
  Conference on Computer Communications Workshops (INFOCOM WKSHPS)}, May 2017,
  pp. 718--723.

\bibitem{chen2017echo}
M.~Chen, W.~Saad, and C.~Yin, ``Echo state learning for wireless virtual
  reality resource allocation in {UAV}-enabled {LTE-U} networks,'' in
  \emph{Proc. of the IEEE International Conference on Communications (ICC)},
  Kansas city, USA, May 2018.

\bibitem{zhang2011signalling}
H.~Zhang, W.~Ma, W.~Li, W.~Zheng, X.~Wen, and C.~Jiang, ``Signalling cost
  evaluation of handover management schemes in lte-advanced femtocell,'' in
  \emph{Proc. of IEEE Vehicular Technology Conference (VTC Spring)}, May 2011.

\bibitem{godor2015survey}
G.~G{\'o}dor, Z.~Jak{\'o}, {\'A}.~Knapp, and S.~Imre, ``A survey of handover
  management in {LTE}-based multi-tier femtocell networks: Requirements,
  challenges and solutions,'' \emph{Computer Networks}, vol.~76, pp. 17--41,
  2015.

\bibitem{Handover1}
R.~Arshad, H.~Elsawy, S.~Sorour, T.~Y. Al-Naffouri, and M.~Alouini, ``Handover
  management in {5G} and beyond: A topology aware skipping approach,''
  \emph{IEEE Access}, vol.~4, pp. 9073--9081, 2016.

\bibitem{Coexistence}
M.~M. Azari, F.~Rosas, A.~Chiumento, and S.~Pollin, ``Coexistence of
  terrestrial and aerial users in cellular networks,'' in \emph{Proc. of IEEE
  Global Telecommunications Conference (GLOBECOM) Workshops}, Singapore, Dec.
  2017.

\bibitem{azari2017reshaping}
M.~M. Azari, F.~Rosas, and S.~Pollin, ``Reshaping cellular networks for the
  sky: The major factors and feasibility,'' \emph{available online:
  arxiv.org/abs/1710.11404}, 2017.

\bibitem{lin2017sky}
X.~Lin, V.~Yajnanarayana, S.~D. Muruganathan, S.~Gao, H.~Asplund, H.-L.
  Maattanen, S.~Euler, Y.-P.~E. Wang \emph{et~al.}, ``The sky is not the limit:
  {LTE} for unmanned aerial vehicles,'' \emph{available online:
  arxiv.org/abs/1707.07534}, 2017.

\bibitem{challita2018cellular}
U.~Challita, W.~Saad, and C.~Bettstetter, ``Cellular-connected {UAVs} over
  {5G}: Deep reinforcement learning for interference management,'' \emph{IEEE
  Transactions on Wireless Communications, accepted and to appear}, 2019.

\bibitem{garcia2018essential}
A.~Garcia-Rodriguez, G.~Geraci, D.~L{\'o}pez-P{\'e}rez, L.~G. Giordano,
  M.~Ding, and E.~Bj{\"o}rnson, ``The essential guide to realizing
  {5G}-connected {UAVs} with massive {MIMO},'' \emph{available online:
  arxiv.org/abs/1805.05654}, 2018.

\bibitem{OmidMatching}
O.~Semiari, W.~Saad, S.~Valentin, M.~Bennis, and B.~Maham, ``Matching theory
  for priority-based cell association in the downlink of wireless small cell
  networks,'' in \emph{Proc. of IEEE International Conference on Acoustics,
  Speech and Signal Processing (ICASSP)}, May 2014, pp. 444--448.

\bibitem{PantisanoInterf}
F.~Pantisano, M.~Bennis, W.~Saad, M.~Debbah, and M.~Latva-aho, ``Interference
  alignment for cooperative femtocell networks: A game-theoretic approach,''
  \emph{IEEE Transactions on Mobile Computing}, vol.~12, no.~11, pp.
  2233--2246, Nov. 2013.

\bibitem{ZhaoInterf}
N.~Zhao, F.~Cheng, F.~R. Yu, J.~Tang, Y.~Chen, G.~Gui, and H.~Sari, ``Caching
  uav assisted secure transmission in hyper-dense networks based on
  interference alignment,'' \emph{IEEE Transactions on Communications, Early
  access}, 2018.

\bibitem{FengPath}
Q.~Feng, J.~McGeehan, E.~K. Tameh, and A.~R. Nix, ``Path loss models for
  air-to-ground radio channels in urban environments,'' in \emph{Proc. of IEEE
  Vehicular Technology Conference (VTC)}, Melbourne, Vic, Australia, May 2006.

\bibitem{dan}
K.~Daniel, M.~Putzke, B.~Dusza, and C.~Wietfeld, ``Three dimensional channel
  characterization for low altitude aerial vehicles,'' in \emph{Proc. of IEEE
  International Symposium on Wireless Communication Systems (ISWCS),}, York,
  UK, Sep. 2010.

\bibitem{vin}
P.~J. Vincent, M.~Tummala, and J.~McEachen, ``An energy-efficient approach for
  information transfer from distributed wireless sensor systems,'' in
  \emph{Proc. of IEEE International Conference on System of Systems Engineering
  (IEEE/SMC)}.\hskip 1em plus 0.5em minus 0.4em\relax IEEE, 2006, pp. 6--pp.

\bibitem{WangTrajectory}
H.~Wang, G.~Ren, J.~Chen, G.~Ding, and Y.~Yang, ``Unmanned aerial vehicle-aided
  communications: Joint transmit power and trajectory optimization,''
  \emph{IEEE Wireless Communications Letters, Early access}, 2018.

\bibitem{CRAN}
M.~Peng, Y.~Sun, X.~Li, Z.~Mao, and C.~Wang, ``Recent advances in cloud radio
  access networks: System architectures, key techniques, and open issues,''
  \emph{IEEE Communications Surveys Tutorials}, vol.~18, no.~3, pp. 2282--2308,
  thirdquarter 2016.

\bibitem{sequentialnonlinear}
A.~V. Fiacco and G.~P. McCormick, \emph{Nonlinear programming: sequential
  unconstrained minimization techniques}.\hskip 1em plus 0.5em minus
  0.4em\relax Siam, 1990, vol.~4.

\bibitem{villani2003}
C.~Villani, \emph{Topics in optimal transportation}.\hskip 1em plus 0.5em minus
  0.4em\relax American Mathematical Soc., 2003, no.~58.

\bibitem{baccelli2010stochastic}
F.~Baccelli and B.~B{\l}aszczyszyn, ``Stochastic geometry and wireless
  networks: Volume {II} applications,'' \emph{Foundations and Trends in
  Networking}, vol.~4, no. 1--2, pp. 1--312, 2010.

\bibitem{MingzheTutorial}
M.~{Chen}, U.~Challita, W.~Saad, C.~Yin, and M.~Debbah, ``Machine learning for
  wireless networks with artificial intelligence: A tutorial on neural
  networks,'' \emph{available online: arxiv.org/abs/1710.02913}, 2017.

\bibitem{Ferd}
{U. Challita, A. Ferdowsi, M. Chen, and W. Saad}, ``Machine learning for
  wireless connectivity and security of cellular-connected {UAVs},'' \emph{IEEE
  Wireless Communications Magazine, Special Issue on Integrating {UAVs} into
  {5G} and Beyond, to appear}, 2018.

\bibitem{C1}
Z.~Han, D.~Niyato, W.~Saad, T.~Ba{\c{s}}ar, and A.~Hj{\o}rungnes, \emph{Game
  theory in wireless and communication networks: theory, models, and
  applications}.\hskip 1em plus 0.5em minus 0.4em\relax Cambridge University
  Press, 2012.

\bibitem{C2}
G.~Bacci, S.~Lasaulce, W.~Saad, and L.~Sanguinetti, ``Game theory for networks:
  A tutorial on game-theoretic tools for emerging signal processing
  applications,'' \emph{IEEE Signal Processing Magazine}, vol.~33, no.~1, pp.
  94--119, 2016.

\bibitem{C3}
K.~Hamidouche, W.~Saad, and M.~Debbah, ``A multi-game framework for harmonized
  {LTE-U} and {WiFi} coexistence over unlicensed bands,'' \emph{IEEE Wireless
  Communications}, vol.~23, no.~6, pp. 62--69, 2016.

\bibitem{GCBG2018}
A.~Ferdowsi, A.~Sanjab, W.~Saad, and T.~Ba{\c{s}}ar, ``{Generalized Colonel
  Blotto} game,'' in \emph{Proc. IEEE American Control Conference}, Milwaukee,
  WI, USA, June 2018.

\bibitem{TB}
T.~Ba\c{s}ar and G.~J. Olsder, \emph{Dynamic noncooperative game theory}.\hskip
  1em plus 0.5em minus 0.4em\relax Siam, 1999, vol.~23.

\bibitem{C4}
J.~Apaloo, \emph{Advances in Dynamic and Mean Field Games: Theory,
  Applications, and Numerical Methods}.\hskip 1em plus 0.5em minus 0.4em\relax
  Birkhauser, 2018.

\bibitem{C5}
S.~Samarakoon, M.~Bennis, W.~Saad, M.~Debbah, and M.~Latva-Aho, ``Ultra dense
  small cell networks: Turning density into energy efficiency,'' \emph{IEEE
  Journal on Selected Areas in Communications}, vol.~34, no.~5, pp. 1267--1280,
  2016.

\bibitem{C6}
K.~Hamidouche, W.~Saad, M.~Debbah, and H.~V. Poor, ``Mean-field games for
  distributed caching in ultra-dense small cell networks,'' in \emph{Proc. IEEE
  American Control Conference (ACC)}, Boston, MA, USA, July 2016, pp.
  4699--4704.

\bibitem{A}
W.~Saad, Z.~Han, M.~Debbah, A.~Hj{\o}rungnes, and T.~Ba\c{s}ar, ``Coalitional
  game theory for communication networks,'' \emph{IEEE Signal Processing
  Magazine}, vol.~26, no.~5, pp. 77--97, Sep. 2009.

\bibitem{B}
W.~Saad, Z.~Han, M.~Debbah, and A.~Hj{\o}rungnes, ``A distributed coalition
  formation framework for fair user cooperation in wireless networks,''
  \emph{IEEE Transactions on Wireless Communications}, vol.~8, no.~9, pp.
  4580--4593, Sep. 2009.

\bibitem{C}
W.~Saad, Z.~Han, M.~Debbah, A.~Hj{\o}rungnes, and T.~Ba\c{s}ar, ``Coalitional
  games for distributed collaborative spectrum sensing in cognitive radio
  networks,'' in \emph{Proc. of the IEEE International Conference on Computer
  Communications (INFOCOM)}, Rio de Janeiro, Brazil, April 2009, pp.
  2114--2122.

\bibitem{D}
W.~Saad, Z.~Han, T.~Ba\c{s}ar, M.~Debbah, and A.~Hj{\o}rungnes, ``Hedonic
  coalition formation for distributed task allocation among wireless agents,''
  \emph{IEEE Transactions on Mobile Computing}, vol.~10, no.~9, pp. 1327--1344,
  Sep. 2011.

\bibitem{C7}
Y.~Zhang, L.~Song, W.~Saad, Z.~Dawy, and Z.~Han, ``Contract-based incentive
  mechanisms for device-to-device communications in cellular networks,''
  \emph{IEEE Journal on Selected Areas in Communications}, vol.~33, no.~10, pp.
  2144--2155, 2015.

\bibitem{C8}
Y.~Gu, W.~Saad, M.~Bennis, M.~Debbah, and Z.~Han, ``Matching theory for future
  wireless networks: fundamentals and applications,'' \emph{IEEE Communications
  Magazine}, vol.~53, no.~5, pp. 52--59, 2015.

\end{thebibliography}
\end{document}